\documentclass[5p,final]{elsarticle}
\usepackage{amsfonts}
\usepackage{amssymb}
\usepackage{amsmath}  
\usepackage{amsthm}
\usepackage{geometry}
\usepackage{graphicx}
\usepackage{enumitem} 
\usepackage{algorithm}
\usepackage{algcompatible}
\usepackage{algpseudocode}
\usepackage[colorlinks=true, allcolors=blue]{hyperref}
\usepackage[utf8]{inputenc} 
\usepackage{url}
\usepackage{cleveref}
\usepackage[english]{babel} 
\usepackage{subfiles}
\usepackage{xfrac}
\usepackage{multirow}
\usepackage{balance}
\usepackage{booktabs}
\usepackage[normalem]{ulem}
\usepackage{caption}
\usepackage{subcaption}
\usepackage{lipsum}
\usepackage{todonotes}
\usepackage{float}
\usepackage{placeins}
\usepackage{widetext}
\usepackage{pgfplots}
\usepackage{tikz}
\usetikzlibrary{shapes.geometric, arrows}
\usepackage{tikz-3dplot}
\usepackage[squaren]{SIunits}
\usepackage{cuted}
\usetikzlibrary{decorations.pathreplacing,positioning, arrows.meta}
\usepackage{listofitems}
\usepackage[toc,page]{appendix}
\newcommand{\nocontentsline}[3]{}
\newcommand{\tocless}[2]{\bgroup\let\addcontentsline=\nocontentsline#1{#2}\egroup}

\newtheorem{statement}{Statement}

\newcommand{\bc}{\begin{center}}
\newcommand{\ec}{\end{center}}
\newcommand{\be}{\begin{equation}}
\newcommand{\ee}{\end{equation}}
\newcommand{\bea}{\begin{eqnarray}}
\newcommand{\eea}{\end{eqnarray}}
\newcommand{\beq}{\begin{eqnarray*}}
\newcommand{\eeq}{\end{eqnarray*}}
\newcommand{\bv}{\left( \begin{array}{c} }
\newcommand{\ev}{\end{array} \right) }

\newcommand{\R}{\mathbb{R}}

\newcommand{\E}{\mathbb{E}}

\newcommand{\de}{\partial}
\newcommand{\mcl}[1]{\mathcal{#1}}

\newcommand{\mlm}{\lim_{\substack{\Delta x \rightarrow 0 \\ \Delta t \rightarrow 0}}}

\pgfplotsset{compat=1.18}

\usepackage{natbib}
\def \FigPath       {"Figures/"}
\def \PathSingles   {\FigPath "Singles/"}

\def \PathKExpInterp{\FigPath "K_exp_interp/"}
\def \PathKExpMO    {\FigPath "K_exp_MO/"}
\def \PathKInterp   {\FigPath "K_interp/"}
\def \PathKMO       {\FigPath "K_MO/"}
\def \PathSFDC      {\FigPath "SF_DC/"}
\def \PathSFDD      {\FigPath "SF_DD/"}
\def \PathSFDT      {\FigPath "SF_DT/"}
\def \PathSFDV      {\FigPath "SF_DV/"}
\def \PathTikz      {\FigPath "Tikz/"}
\def \PathVK        {\FigPath "V_K/"}

\begin{document}
\title{Non-uniformly sampled simulated price impact of an order-book.}

\author[uct-sta]{Derick Diana}
\ead{derick.diana.dd@gmail.com}
\author[uct-sta]{Tim Gebbie}
\ead{tim.gebbie@uct.ac.za}
\address[uct-sta]{Department of Statistical Sciences, University of Cape Town, Rondebosch, South Africa}
\begin{abstract}
We extend a Discrete Time Random Walk (DTRW) numerical scheme to simulate the anomalous diffusion of financial market orders in a simulated order book. Here using random walks with Sibuya waiting times to include a time-dependent stochastic forcing function with non-uniformly sampled times between order book events in the setting of fractional diffusion. This models the fluid limit of an order book by modelling the continuous arrival, cancellation and diffusion of orders in the presence of information shocks. We study the impulse response and stylised facts of orders undergoing anomalous diffusion for different forcing functions and model parameters. Concretely, we demonstrate the price impact for flash limit-orders and market orders and show how the numerical method generates kinks in the price impact. We use cubic spline interpolation to generate smoothed price impact curves. The work promotes the use of non-uniform sampling in the presence of diffusive dynamics as the preferred simulation method. 
\end{abstract}


\maketitle


\section{Introduction} \label{sec:intro}

Does one need a complex mechanistic agent-based model to reasonably recover and simulate continuous-time double auction stock market price dynamics, or can this be reasonably achieved in the continuum limit? We consider this question by extending a model first used to describe the fluid limit of the latent-order book to modelling the diffusion of orders in a lit order book \cite{TLDLJB2011}.  This model was further extended and explored, first in the context of anomalous price impact \cite{TLDLJB2011,MTBPRL2014} and then for fractional diffusion \cite{BB2018}. Here we use the model to combine price discovery and the dynamics of the limit order book that includes an external driving force and anomalous diffusion, the intensities and rates describing the arrival and cancellation of orders, with a finite difference numerical scheme that allows for non-uniformly sampled update times. 

Describing stock market behaviour through the lens of agent-based models and complexity can be seen to begin with the seminal work of \citet{BPS1997}. In their paper two kinds of agent are simulated; one being ``intelligent", and the other simply being a ``noise" trader. They found that when only noise traders are used the behaviour of the system corresponds to reaction diffusion processes of the form $A+B\rightarrow0$. Here there are two type of agents, $A$ and $B$, which are represented as particles. They are injected into different ends of a tube, they then bump into each other, diffuse, and can annihilate when they bump and interact with agents of a different type. The annihilation process can then be interpreted as a successful trade between a buyer and seller. However, including the ability for ``noise" agents to imitate others makes such processes unsuitable. This is important because a combination of both types of agents with the added ability for agents to copy each other's prices (as an examples of herding behaviour) results in a models which better match empirical data in terms of Hurst exponents of observed price paths. 

\citet{TT1999} considered this formulation of the stock market problem and were able to extend it, for the same agent types, combinations and behaviours, to include: unbiased diffusion, biased diffusion, and biased diffusion with copying. They further explained how imbalances in the number of buying and selling agents can cause prices changes. A key insight was to show that biasing agents to move towards the current price reproduces a random walk. Finally, they derive scaling functions which describe the Hurst exponents in all three models, and again find the combination model to match real world data. They provided the first heuristic method that clearly linked price fluctuations to diffusive motion of individual agents using the representation introduced by \citet{BPS1997}, but where the model could start to conform with the measured fractal scaling observed in real financial markets. 

These developments suggested that the dynamics of financial markets could be understood and clarified from first principles by extending kinetic theory to the analysis of stock markets. However, a key stumbling block remained. How to empirically validate microscopic models using data arising from mesoscopic observations when most of the prevailing models were focused on validating macroscopic and mesoscopic models. 
Theoretical models avoided establishing empirical links between the microscopic models and observed financial markets by direct comparison of trader dynamics, the kinetic theory, price dynamics, and their ability recover observed mesoscopic and macroscopic features. \citet{KSTT2018,KSTT2018-2} provided and quantified a conceptual framework that directly linked the microscopic kinetic theory of traders to mesoscopic observations.     

\citet{KSTT2018,KSTT2018-2} used hierarchical theory to derive a Boltzmann-like and Langevin-like equation to describe the behaviour of trend following agents by treating the problem using statistical mechanics and physical analogies. Their theory consisted of three scales. First, a ``micro" scale in which individual traders participate in the market (they use a physics analogy to say that this corresponds to individual particles in motion in a gas). We do not directly consider this scale in this paper. Next, a ``meso" scale in which the agent participation results in distribution of bids and offers in the lit order book (in their analogy this is the velocity distribution of all the gas molecules), and we directly model this by tracking the number of bids and asks at discrete grid points. Finally, a ``macro" scale in which one observes the resulting market prices (in their analogy this is the path one can observe by following a single particle - or tracer - through the gas) and we measure this by solving for the intercept amongst our grid points.  

They further analysed the Langevin-like equation to derive three different behaviours which can occur based on how strongly the agents follow recent price trends. They show that when agents only weakly follow recent price trends, then the model has strongly diffusive behaviour and the returns distribution is Gaussian. At the other extreme, when agents strongly follow recent price trends, the returns distribution becomes exponential and the drift behaviour dominates. Finally, in the case somewhere between the two extremes, the returns distribution is exponential once more. In this paper, we will have strongly normally distributed returns which means that our model should be operating in the weakly trend following case.

The unobservable latent-order book has been described as the primary source of orders driving market price discovery and captures the intentions of all traders engaged at the microscopic level with trading activity that is observed at the mesoscopic and macroscopic level. The latent order book literature is well established \cite{TLDLJB2011,DBMB2014,MTBPRL2014,BOUCHAUD200957}. The initial departure point, as motivated by \citet{DBMB2014}, for this type of approach to model simulation (and calibration) is one that is informed by the zero-intelligence agent argument \cite{SFGK2003,FPZI2005}. The original model of \citet{TLDLJB2011} described the latent-order book in the continuous-time fluid limit where the dynamics are captured by a reaction-diffusion equation. The reaction-diffusion model $A + B \to \emptyset$ can then be generalised to study the impact of an excess of A (or B) at the order-flow reaction front \cite{MTBPRL2014}. Here we applied the same model to the problem of modelling a lit order book in a steady-state configuration, where the order density source term is made to linearly track the price (then the source term is stationary in the frame of the order book) it will then suppress the dynamics at the boundary. This is simulated using the stochastic finite difference numerical scheme developed by \citet{ANGSTMANN2016508}.

Key drivers of aggregate emergent market dynamics in the continuous-time domain can be largely independent of the direct specification of individual agent-behaviours. There are many fairly well understood examples of agent behaviour derived models \cite{KT2002,FJ2002,PTSPSJ2006,CIP2009,BOUCHAUD200957} where some of the pressing issues relating to the combination of computational efficiency of the models and trust relate to the complex challenges of calibration \cite{PG2016,P2020,FA2013}. However, if the observable features can be successfully described in terms of the dynamics relating to order-flow and information arrival directly, and hence described in terms of a order book itself \cite{LD2010, TLDLJB2011} then it may make sense to sacrifice model complexity and mechanistic agent realism for parsimony and predictability. This will allow models to focus on mesoscopic properties as averaged representations of the underlying microscopic mechanisms. This is an idea that is at the heart of kinetic theory where the thermodynamic proprieties arising from averaging processes are what are modelled and mapped to observables and experiments.

Using this type of simplified model gives one computational advantages that can then allow us to numerically more fully explore the parameter space of the model, and the resulting stylised facts, towards building model configurations and numerical schemes that can produce reasonably realistic simulated price responses and dynamics with computational speed. This anticipates the idea that complex interactions of many heterogeneous financial market agents competing within a financial market can be reasonably approximated with simple source terms, a stochastic driving force, within the broad context of diffusion models to capture zero-intelligence dynamics. Thus, even when many heterogeneous agents are engaging in purposeful, competitive, and strategic decision making within a hierarchically organised system, when averaged, this may be well approximated by diffusive dynamics arising from a financial market kinetic theory with it associated statistical mechanics. 

One key issue here is that this argues that mesoscopic 
properties emerge from the microscopic interactions of trading agents, and that the observable mesoscopic properties are the result of averaged microscopic interactions. This implies that the relationships between buying and selling, and why these are correlated, should have a microscopic foundation. However, prices that emerge from trading can remain uncorrelated with auto-correlations that die out faster than those in the order-flow itself.  

\citet{G2008} considered the two schemes proposed in \cite{LF2004} and \cite{BGPW2003}. These attempted to explain why, despite the fact that buys and sells are correlated, prices remain uncorrelated. He then shows that these two models can be seen as the same model with different interpretations for the terms in the models. He further proposes that if there are groups of market participants who collectively condition liquidity on past order flow such as to make this model hold, then price impact will be power law. Alternatively, if the individual initiating a transaction does this conditioning themselves, then one obtains a price impact which is logarithmic. This is an important insight, and should be recovered by any reasonable kinetic theory of a financial market. 

Here we show how the shape of the price impact is in fact dependent on the discretisation procedure used to simulated the averaged reaction-diffusion model of order flow interactions, and as such, that there are model hyper-parameters that can be used to tune the model to fit observations, that are only weakly dependent on the model parameters themselves. In particular, the dependency on the choice of sampling, and how this can mislead the fits between log and power law price impact.    

The main contribution of this work is to provide a concrete example of a mesoscopic averaged kinetic theory inspired model of order flow in a general simulation framework, and to then produce price impact curves from the models price response to shocks and sequences of volume impulses. This is used to demonstrate the variety of stylised facts that are possible, but also the limitations of this class of models. Concretely, the work extends the initial numerical simulation and calibration prototype developed by \citet{Gant2022a,Gant2022b} where the diffusive model was calibrated to data using a simplified formulation of the \citet{ANGSTMANN2016508,AHJM2016} update scheme, to a more realistic simulation configuration. 

Specifically we now explored the anomalous diffusion extension in terms of the simulated price impact (Section \ref{sec:lob}), along with various stylised facts, in terms of the the extended model parameters (Section \ref{ssec:DtDx} and Table \ref{tab:parameters}). Then we extend this to include the ability to generate non-uniformly sampled mid-price data paths (Section \ref{ssec:updateequationarrivaltimes}). A key contribution is that we show and explain the impact of the numerical scheme on the form of the price impact (\ref{app:PriceImpactKinks}), and then provide an interpolation method to smooth these artefact's (Section \ref{ssec:priceimpactlimit}). We now consider the model. 

\section{Limit Order Book (LOB)} \label{sec:lob}

The latent order book proposed by \citet{TLDLJB2011} approximates the limit order book in the fluid limit (hydrodynamic limit \cite{GD2018}) using the evolution of the order-book density imbalance where $x$ refers to the logarithm of the price and $t$ is some calendar time:
\begin{equation}
\varphi(x,t) = \rho_B(x,t) - \rho_A(x,t),
\end{equation}
given be the difference between density of bids to buy $\rho_B(x,t)$ and offers to sell $\rho_A(x,t)$. From the combined order book volume density imbalance $\varphi(x,t)$ at price $x$ and time $t$ we can find the mid-price $p(t)$ at time $t$:\footnote{Non-trivial order-flow currents $j_a \neq j_b$ imply a separation $x_b \equiv \max( \{x: \rho_b(x)>0\})$ and $x_a \equiv \min( \{x: \rho_a(x)>0\})$ such that one obtains a spread $s(t)=x_a - x_b$ where the transaction price $p(t)$ is not in general equal to the mid-price $m(t) \equiv (x_b + x_a)/2$; we would then need to solve two coupled equations for a moving boundary problem, rather than the combined steady-state equation \cite{MTBPRL2014}. Here $j_a=j_b$ by construction.}
\begin{equation}
p(t) := \{{x} : \varphi(x,t)=0 \}.
\end{equation} \label{eq:midprice}
The order-imbalance drives trading, and the volume of trading $Q$ over a given time period {\it e.g.} $t \in [0,T]$, can be found by using the equivalent of Ohm's law for a financial market, and then integrating over time: 
\begin{equation}\label{eq:volumetraded}
    Q = D \int_0^T \left.{\partial_x \varphi(x,t)}\right|_{p_t}dt.
\end{equation}
We assume the volume density evolves as a reaction-diffusion equation \cite{TLDLJB2011,MTBPRL2014,DBMB2015QF,BB2018} in terms of the removal (annihilation) rate $a(x,t)$, a stochastic driving force $f(x,t)$ and a source (creation) term $c(x,t)$ (which can be dependent on all available past information). We further assume that the diffusion of orders is anomalous with diffusion parameter $D_{\alpha}$ with corresponding Riemann-Liouville operator $D^{1-\alpha}_{t}$ for times $t$ with parameter $\alpha$. Then the  accumulated volume that has not been removed, the order survival function, is:\footnote{We note that the form of $\theta D^{1-\alpha}_t \theta^{-1}$, together with Equation [\ref{eq:theta}], results in a tempered fractional derivative (see \cite{FMJ2015}) which may suppress the Sibuya waiting times in \ref{app:AnomalousDerivation}.}
\begin{equation}
    \theta(x,t) = e^{- \int_0^{t} a(x,\tau)d\tau}.
    \label{eq:theta}
\end{equation} 
Now in this model new information is expected to trigger trading, this is some forcing function. We can choose the forcing function in terms of a time dependent stochastic process $V_t$ so that $V_t = 2 \beta D_{\alpha} f(t)$ with inverse temperature $\beta$ (\ref{app:force}). These ideas can be combined to then provide a reaction-diffusion equation \cite{TLDLJB2011,MTBPRL2014,DBMB2015QF,BB2018} with processes for order addition and removal to describe the anomalous diffusion of limit orders as new information arrives:
\begin{align}\label{eq:MainPDELongNotation}
\partial_t \varphi(x,t) =&
D_\alpha \partial_{xx}\Bigl[\theta(t) D^{1-\alpha}_{t,t'} \Bigl( \frac{\varphi(x,t')}{\theta(t')} \Bigr)\Bigr] \nonumber \\
&+ V_t \partial_x \left[\theta(t) D^{1-\alpha}_{t,t'} \left( \frac{\varphi(x,t')}{\theta(t')} \right)\right] \nonumber
\\ &- a \varphi(x,t) + c(x,t).
\end{align}
The inclusion of $t'$ is to emphasize the variable of integration of the Riemann-Liouville operator $D^{1-\alpha}_{t,t'}$, which integrates the objects passed to it, up to time $t$. Equation [\ref{eq:MainPDE}] then provides a streamlined version of the above by using a new partial derivative notation and leaving out function parameters:
\begin{equation}\label{eq:MainPDE}
\varphi_t =
D_\alpha \left[\theta D^{1-\alpha}_{t} \left( \frac{\varphi}{\theta} \right)\right]_{xx} \negthickspace  \negthickspace  \negthickspace
+ V_t \left[\theta D^{1-\alpha}_{t} \left( \frac{\varphi}{\theta} \right)\right]_x 
- a \varphi + c.
\end{equation}
We see that orders will in general evolve under an anomalous diffusion process\footnote{$\lim_{\alpha \to 1}$ recovers $\varphi_t = D \varphi_{xx} + V_t \varphi_x - a \varphi +c$.}, the idea is that this captures zero-intelligence trader activity. There is a stochastic forcing term that will bias the diffusion which represents new information, and the volume of trading that is associated with this is the response of the market to the arrival of this new information. In addition, orders can be cancelled. 

Finally, there is a source term that represents the overall propensity for the supply and demand of orders within the market. The source term $c$ will characterises the liquidity properties of the market itself and is a mesoscopic representation of the aggregate propensity for microscopic trader activity.

Interpreting this model and hence implementing a consistent numerical scheme is surprisingly nuanced. It should be emphasised that the approach adopted here is comparable {\it in the appropriate limits}, but not microscopically equivalent to all or any of the many other reaction-diffusion type market models. Firstly, we start by saying that new information arrives at the current time, on the lattice (see the first term on the RHS of Equation \ref{eq:Update1}); however, for example, \citet{BB2018} include a convolution over the entire past of the information arrival -- this is microscopically a different model. This results in their fractional differential equation having a fractional derivative acting on their source term, whereas ours (Equation \ref{eq:MainPDELongNotation}) does not, highlighting the differences between the models once again.

Second, \citet{BB2018} start with a continuum representation {\it a-priori}, and then apply the diffusion limits as an initial approximation along with a truncated waiting time distribution to find a propagator representation, before transforming back from the dual space. This can lead to nuanced differences in interpretation, particularly when considering boundary and initial conditions in the context of a numerical scheme. 

Here we claim more generality \cite{ANGSTMANN2016508,AHJM2016,ADHJL2015}. We also claim a consistent link between the microscopic foundational model used in the simulations and the stated continuum reaction-diffusion equation and its associated boundary and initial conditions because we take the diffusion limit after the integral transforms and without needing to truncate the path when showing that the numerical scheme (Equation \ref{eq:FinalUpdateEquation}) is consistent with the model (Equation \ref{eq:MainPDE}).  

\subsection{Order volume conservation} \label{ssec:mass}

In equilibrium the total volume area of orders $\varphi$ must remain constant through time. This volume is:
\begin{equation}\label{eq:VolumeEquation}
    \phi(t) = \int_0^L \varphi(x,t) dx.
\end{equation}
Using this,  Equation [\ref{eq:MainPDE}] can be rewritten. The derivatives with respect to $x$ can pass through the operator $D^{1-\alpha}_{t}$ because this is a linear operator, and has only a $t$ dependence. We rewrite [\ref{eq:MainPDELongNotation}] and use that $V_t = 2 \beta D_{\alpha} f(t)$:
\begin{align}
    \partial_t \varphi(x,t) &= D_\alpha \Bigl[\theta(t) D^{1-\alpha}_{t,t'} 
    \Bigl(\frac{1}{\theta(t')} \left({\partial_{xx} \varphi(x,t') - }\right. \nonumber \\ &\left.{2 \beta f(t) \partial_x \varphi(x,t')}\right)  \Bigr) \Bigr] - a \, \varphi(x,t) +c(x,t). \label{eq:7}
\end{align}
We integrate both sides of Equation [\ref{eq:7}] over $x\in [0,L]$. The integrals over $\varphi$ result in $\phi$ as in [\ref{eq:VolumeEquation}]. We note that the integral with respect to $x$ commutes with the operators $\de_t$ and $D^{1-\alpha}_{t,t'}$, and then
simplify and streamline the notation, to find:
\begin{align}
\phi_t = D_\alpha \left[{\theta D^{1-\alpha}_{t} \left[{\frac{1}{\theta} \left({\varphi_x \bigr|^L_0 -  2 \beta f \varphi|^L_0} \right)} \right] } \right] - a \phi + C. \nonumber 
\end{align}
Here $C$ are the orders created from the source terms accumulated over the volume domain. The boundary conditions we need to ensure balance for some $t' < t$ are: $\varphi_x|_0 = 2 \beta f(t)\varphi|_0$, and
$\varphi_x|_L = 2 \beta f(t)\varphi|_L$. 

It is then required that for all time:
\begin{align}
        \varphi_x|_0 =\varphi|_0 =0, \mbox{and }
        \varphi_x|_L =\varphi|_L =0. 
        \label{eq:boundary}
\end{align}
This eliminates the dissipative term, and the remaining PDE is: $\phi_t = - a \phi + C$. So that $\phi(0) = 0$ and $C= 0$ for all times $t$. This suggest that anti-symmetric source functions are prudent choices.

\subsection{Order arrival and removal} \label{ssec:source}

\begin{figure}[ht]
\centering
\captionsetup{type=figure}
\includegraphics[width=0.45\textwidth]{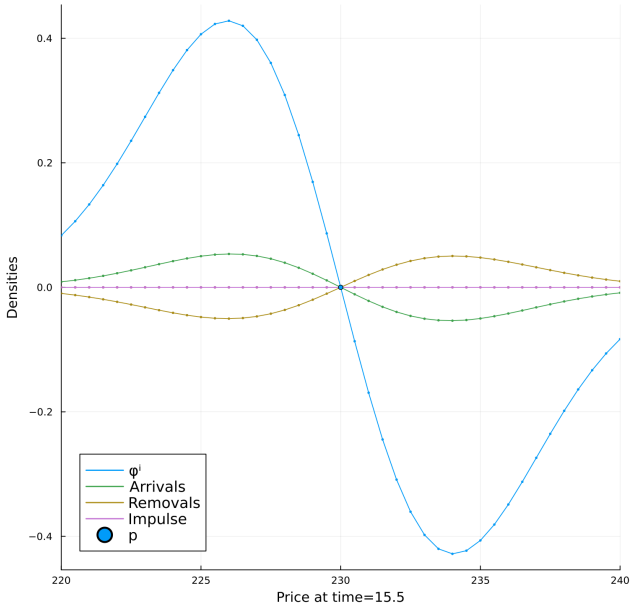}
\caption{The state of the observable (lit) Limit-Order Book (LOB) in equilibrium is shown in blue where the cross-over point from negative to positive resting orders is the prevailing mid-price $p$ for the asset associated with this market. One can see that the density of resting orders for different price levels. The market is in equilibrium so that the scaled versions of the density of order arrivals at different price levels (green) are balanced by order cancellations (gold). Lastly any trade shocks will be shown in pink (See Fig. \ref{fig:VisualizeImpactSpike} for plots that demonstrate the resulting shock dynamics).}
\label{fig:StationaryVisualisationExp}
\end{figure}

We further subdivide the creation terms conceptually into the following components:
\begin{equation}
c(x,p,t)=s(x,p,t)+ \delta(x,p,t).
\end{equation}
Here $s(x,p,t)$ represents the market source of volume and $\delta(x,p,t)$ will be an impulse trade representing some instantaneous, isolated, exogenously introduced trade that shocks the system. This impulse trade will be used in later sections to probe the market response (See Section \ref{ssec:priceimpacts}). The market source term will either be: i.) latent demand from investors outside of the observed limit order book \cite{TLDLJB2011,DBMB2015QF,BB2018}, or ii.) lit market demand that is entirely within the limit order book (Figure \ref{fig:StationaryVisualisationExp}). 

The equilibrium supply and demand will be captured by the {\it lit market source}. This is defined to be the situation with zero volume at the boundaries \cite{Gant2022a}. Here at logarithmic price $x$, mid-price $p(t)$, with intensity parameter $\kappa$, and scale parameter $\mu$:
\begin{equation}\label{eq:source}
s(x,t) = - \kappa \bigl(\mu (x - p(t))\bigr) e^{-\bigl(\mu (x - p(t))\bigr)^2}.
\end{equation} 
This is convenient, not only because of its interpretation, but also because with Robin boundary conditions the integral under the density of resting orders $\varphi$ will be constant. Both the lit and latent order book sources can be used to describe equilibrium dynamics of the order book. This represents the system that will be perturbed with additional orders and information shocks. In this paper we initially restrict ourselves to studying the lit order book. 

To describe non-equilibrium supply and demand we differentiate between two different order types: {\it market orders} which remove volume from the order book at the best possible price with demand either buying or selling volume from the order book, and {\it limit orders} which add volume to the order book, either buying or selling volume. Both will asymmetrically introduce volume that was not already existing in the order book. We can add a third order type to capture the idea of a trading schedule, the {\it meta-order} as introduced by \citet{TLDLJB2011} which is a parent order with many individual child orders executed at different times. This type of order is not discussed or considered any further here and we consider only instantaneous orders.

There is an important caveat, we cannot in general assume that all order types are mutually exclusive. Effective orders can be problematic. \citet{SFGK2003} point out, for example, that when a parent limit order of one type (buy or sell) crosses the current mid-price (termed a ``crossing order") and meets a limit order of the opposite type, these orders may be interpreted as a market order within this model. The limit order in question is effectively a market order. However, if the crossing order is larger than the opposite limit order, then one obtains a new limit order of the opposite sign. This is not generally expected behaviour of a market orders in many models where it is assumed this is an effect of little consequence and can largely be ignored. This is not the case, \citet{JCG2021} considered a Hawkes point-process which generates order types from a selection of types including, limit and market orders, which they feed to a matching engine. They showed that, depending on how such interactions (crossing orders and other market mechanisms) are handled, one's ability to infer the underlying process is affected and the differences can be statistically significant. 

\subsection{Time and time-scales} \label{ssec:timescales}

In the continuous time fluid limit setting the time variable $t$ references order book updates, this suggests that there is a transaction for every order-book update, and that every order book update conforms to a calendar time measurement. This does not faithfully capture the physics of an actual financial market. There are several inter-related timescales in real financial markets \cite{CHANG2021126329}.

\begin{figure}
\input{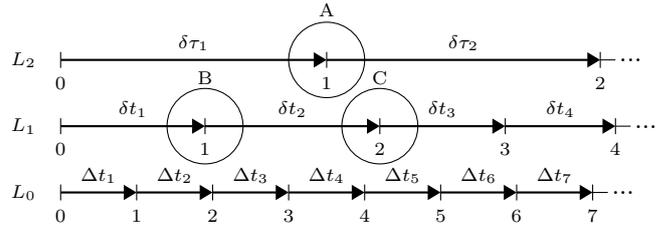}
\captionsetup{type=figure}
\caption{A visualisation of the interacting time scales between our model (level $L_0$) and the physical system (levels $L_1$ and $L_2$). Here $L_1$ represents the order book update events, and $L_2$ the trade events. The horizontal continuum lines represent the continuous passing of time in the model system which Eqn. \ref{eq:MainPDE} describes. Into this continuum, we insert events separated by a sequence $\{\Delta t_m\}_{m=1}^N$ which are the times at which our chosen algorithm allows us to observe this system (either update Eqn. [\ref{eq:FinalUpdateEquation}] or[\ref{eq:UpdateEquationExpArrival}] - see Sec. \ref{ssec:sampling} and \ref{ssec:priceimpactlimit}) giving us observations at $t_n= \sum_{m=1}^n \Delta t_m$. This is level $L_0$. 
We also sample the system at times $\ell_n =\sum_{m=1}^n \delta t_n$ to get the set of order book updates events ($L_1$) and similarly at $\tau_n=\sum_{m=1}^n \tau_m$ to get the set of trade events ($L_2$). Interpolation may be required for $L_1$ and $L_2$ events, but we don't consider this case.
}
\label{fig:TimeScales}
\end{figure}

First, there is a difference between event time and calendar time. Event time is the integer indexed time between events, in our setting this will typically be either: i.) an order book update event, or ii.) a transaction event (here index by $\ell$ by convention). Calendar events are measure in calendar time {\it i.e.} millisecond, seconds, minutes and so on; and not typically indexed and measured as a counting process. Here we use the parameter $t$, and index this with a counting process index $n$. Second, in financial markets there can be many more order book events than transactions because order book updates do not always lead to transaction events. In this model both order book events and trade events are dense on the real line. This becomes important when we want to relate the time-scales in the continuum limit model with time-scales in the discrete non-uniformly sampled asynchronous physical world of real financial markets. To simulate the continuous time fluid limit models we will need to discretise the model. Some care is prudent in both interpreting, and then mapping this interpretation to the material world where measurements take place. In the end, we implement this in a simple manner that identifies trade event times and sampled order book events, in the observable world, as sampling events that sample the continuous time model.  

In practice, there are two different discretisation scales: 1.) $\delta t$ the average time between order-book update events, this is the time $t$ we find in the Equation [\ref{eq:FinalUpdateEquation}], and 2.) $\delta \tau$ the average time between observed trades in the market. This is the sampling time we will expect in Equation [\ref{eq:midprice}]. In practice, price impact is computed to measure the changes in the order book due to transactions of different volumes and prices. This is measured by recording the mid-price just proceeding, and then immediately after the $\ell^{\mathrm{th}}$ transaction event, to then compute the mid-price change associated with the volume of the transaction in question. Price response is then the evolution of transaction induced changes to the order book as a function of a delay, in the trade event count, relative to a particular set of trades of a certain size. This can be understood to be the gradient of the order density in the continuum limit.

It is typical that price impact (and price response more generally) are measured in terms of event time counts; where $\ell$ counts transaction events subsequent to the trade in question. This is because mid-prices are those measured immediately after a trade event and compared to those immediately before \cite{LeeReady1991,LimCoggins2005,HARVEY2017416}. Price impact is measure in terms of changes in the mid-price relative to time indexed by trade counts and not calendar time. This means that the trade event index $\ell$ has an averaging waiting time $\delta \tau$ between the transactions, where $\delta t \neq \delta \tau$ and $\delta \tau \ge \delta t$. However, in this model we have the simulation time $t$ of the simulation environment, and then the lattice size $\Delta t$ representing the order book time changes in the simulation environment. This is the background simulation time (See Figure \ref{fig:visualisationofexpscheme}). The simulation time is dense on the real line so can be thought of the situation where the calendar times, the trade times, and the order book event times are coincident.

This simulation time needs to be mapped to the observed calendar time of the market, and the observable lattice size needs to be related to the known order book update times (here indexed by an integer event count $k$), and these in turn to the trade times. This requires sampling. Particularly if a model is to be calibrated to real world data, to move from the model to the physically measurable quantities {\it i.e.} the quotes and hence mid-prices, the time at which trades occur, and the times at which order book updates occur to the model update times $\Delta t$. 

\begin{figure}[ht]
\centering
\includegraphics[width=0.45\textwidth]{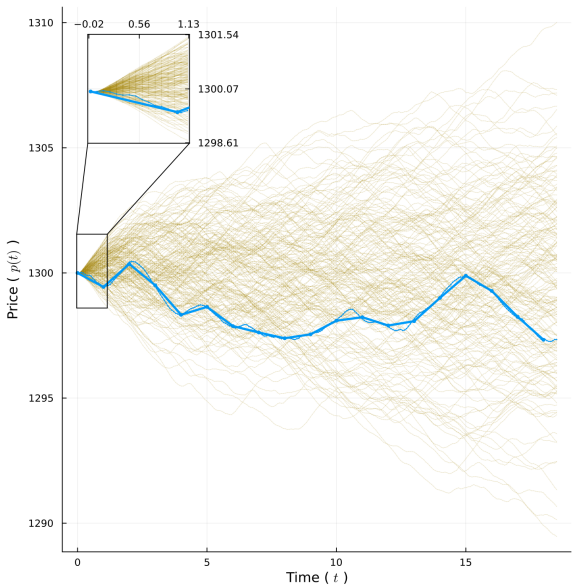}
\caption{An example of simulated mid-price paths arising from the model as discussed in Sec. \ref{ssec:simulatinglimitordrebook}. The parameters are those from Tab. \ref{tab:parameters}. The figure shows an indicative example of an underlying simulated mid-price path in the simulation background lattice $L_0$ which represents order book events (a thin blue line with small points). This path is then sampled at trade event times that would conceptually coincide with those observed in financial markets (and whose rate is set by calibration). These are the observable sampled event times (the thick line points). The order book events are sampled to coincide with the sampled price paths in $L_2$ (see Sec. \ref{ssec:timescales}). 
This figure shows paths simulated for a chosen set of random realisations of the stochastic potential $V_t$ which represent the arrival of random information. The inset figure shows two sampled order book events (large blue dots) at $\ell=0$ and $\ell=1$ (in order book event time). There are many small dots, both yellow and blue, at the background simulation times that represent the underlying, intermediate order book events. The underlying background lattice time is $t$ such that the sampled order book events are integer counts $\ell$ sampled from $t$.}
\label{fig:manypaths}
\end{figure}

We choose to sample and map the continuous simulation time the model speaks of to a set of sampled order book events. Figures \ref{fig:IncreasingVarianceOfSpikeSmallDx}, \ref{fig:manypaths} and all the stylised facts figures \ref{fig:StylizedFactsDifferentDiffusions}- \ref{fig:StylizedFactsDifferentDx} emphasise this choice. Figure \ref{fig:IncreasingVarianceOfSpikeSmallDx} is explained in Section \ref{ssec:measurediffusion}, but one may view the curvature in the top-left inset of Figure \ref{fig:IncreasingVarianceOfSpikeSmallDx} as being a measure of how sub-diffusive the system is. Each dot represents an $L_0$ (background) event, and we must choose how to map these events to $L_1$ (observable) order book events where the model describes the order book. 

Continuous time is on the $x$-axis and the scheme requires many more $L_0$ events to simulate the same length of continuous time the larger $\alpha$ gets (see \ref{app:complexity}). If one were to simply state that all $L_0$ events are also $L_1$ order book events, then in order book event time these plots become more and more stretched out, and may not represent the sub-diffusion behaviour one is trying to model. Instead one must choose order book events from $L_0$ so that in the order book sampling time the plots show in Figure \ref{fig:IncreasingVarianceOfSpikeSmallDx} would look similar. 

A natural way to proceed is to identify times at which we sample prices in the observed market with trade events in $L_2$ (the trade events that are observed), and to then make the choice that $L_1$ and $L_2$ events are commensurate, and so these larger sampling dots are also $L_1$ events. This ensure that the all the order book sampling events are trade sampling events. Then to sample at integer values of the continuous model time that preserve the diffusive properties to the required numerical precision at the aggregate level, which is the choice made in this paper. The larger points in Figure \ref{fig:IncreasingVarianceOfSpikeSmallDx} represent then observable events which we would call order book events which coincide in both $L_1$ and $L_2$ (and are equivalent to counts $\ell$). We note that one could use half-integer or smaller for a slight decrease in computation time. This is despite knowing that there are more order book update events than trade events in real financial markets. This becomes important once one tries to link the model to real world data using some sort of calibration process, or when trying to compare stylised facts from observed data with that from simulated data.

Another approach to interpreting the model is to argue that we collapse $L_1$ into $L_0$ - that the simulation iterations represent the actual order book updates and their associated mid-price changes. While sampling the trade events whose realised prices are the prevailing mid-prices. In this approach we would, again, start sampling at trade event times that are observable in actual markets, and to then sample mid-prices from the underlying model assuming that the realised trade price are the same as the mid-prices. In this picture the model is then iterating order book events in $L_0$ at a higher rate than the trade event sampling rate. The events in $L_0$ are then the necessary order book updates required to generate the appropriate diffusion when sampled at event times $\ell$. Either way, we end up with the same model. 

The key is to realise that at every simulation update in this model there are in fact explicit mid-price changes and implicit trading, this is because orders annihilate each other at the interface between bids and offers at every time in the model. We do not directly observe trades in the model, we can only observe the mid-prices from the order book itself.  

\subsection{Sampling the order book} \label{ssec:sampling}

The physical system we are modelling is observed and measured by recording Trade And Quote (TAQ) data. This includes all the order book events (Quote updates) and the transaction events (Trade updates). These are both in general non-uniformly sampled (as shown as level $L_1$ and $L_2$ in Figure \ref{fig:TimeScales}). In the continuous time setting of the fluid limit it may seem natural to assume that all order book events and trade events conform at time $t$ and are dense on the real line. This is physically how the model has been constructed. However, in order to calibrate the model, which is an approximation, to the physical world, we will need to down sample from the continuous-time model (as sampled from $t$ in level $L_0$ in Figure \ref{fig:TimeScales}) as discussed previously in Section \ref{ssec:timescales}.

Concretely, we will use two very specific sampling time constructions. The first, which we will call {\it uniform time sampling}, using Figure \ref{fig:TimeScales}, is where we choose the model simulation time to be uniformly $\Delta t$. Then we can choose the order book update time $\delta t$, and trade times $\delta \tau$, to be the same multiple $\gamma_1$ of model time\footnote{Here the uniform sampling model uses $\delta \tau = \delta t = 8 \Delta t$ so that $\gamma_1=8=\gamma_2$ so that $\gamma_{2,1}=1$.}.

The second construction, which we will call {\it non-uniform sampling}, is when we set the trade event times $\delta \tau_m$ to be non-uniformly sampled, perhaps from some random process. Here at each calendar time $t_m$ there is a new waiting time $\delta \tau_m$. This process is taken to be modelling the waiting times between measured trade events, and we can the choose these to align the order book update event waiting times between observed trades ($\delta t_m = \delta \tau_m$). We can then fix these so that they all align with the underlying simulated order book update times $\Delta t_m$ in the underlying model. However, this requires that the underlying model be extended to accommodate a non-uniform time lattice while being consistent with the diffusion process. We provide such an extension in Section \ref{ssec:updateequationarrivaltimes}.

In general, the average number of order book events per trade event is a free-parameter that we need to choose, or that is set via a calibration process. In practice, we will always work backwards from the physical system we are measuring, {\it i.e.} from the average waiting time between trades, and thus between mid-price measurement events. This is because mid-prices that are measured to compute the price impact in the physical systems, are always those immediately after and before a trade event \cite{LeeReady1991,LimCoggins2005,HARVEY2017416}. That is why in this work we align the trade time events $L_2$ with the sampled order book update events in $L_1$.

Thus, there are three important time ratio's in this work and these are defined by the average sampling time ratio's using the setup visualised in Figure \ref{fig:TimeScales}:
\begin{enumerate}
    \item Trade event time vs. simulation time: 
 \begin{equation} \label{eq:gamma1}
        \gamma_1 =\frac{\E[\delta \tau_m]}{\E[\Delta t_m]} =\frac{\delta \tau}{\Delta t}.
\end{equation}
    \item Order-book event update time vs. simulation time:
            \begin{equation} \label{eq:gamma2}
        \gamma_2 = \frac{\E[\delta t_m]}{\E[\Delta t_m]} = \frac{\delta t}{\Delta t}.
\end{equation}
    \item Trade event time vs. order-book event time:  
    \begin{equation} \label{eq:gamma12}
        \gamma_{2,1}= \frac{\E[\delta t_m]}{\E[\Delta \tau_m]} = \frac{\delta t}{\delta \tau}.
\end{equation}
\end{enumerate}
These are important when we construct the price impact because they help encode how the mesoscopic processes will map into the macroscopic observable. Crucially, we will show that the shape of the price impact curves are dependent on choices of these parameters. 

In this paper each simulation is conceptually taken to represents 8 hours of calendar time trading where $\delta \tau = \E[\delta \tau_m]$ is the average waiting time between trades over that period. The average time between order book events over the same typical 8 hours of simulated trading is then $\delta t = \E[\delta t_m]$. The discretisation procedure then starts with $\Delta t = \delta t$ and that $\Delta t = \E[\Delta t_{m}]$. This then allows the waiting time $\Delta t_m$ on the unobserved model lattice to be the same as that between transaction events. This will then set the non-uniform nature of the order book price updates necessary to ensure consistency with the fractional diffusion. We exploit this relationship (See Equation [\ref{eq:latticesize}]) throughout this work. 

\section{Anomalous Stochastic Finite Difference} \label{sec:stochastic}

\subsection{Simulating the limit order book} 
\label{ssec:simulatinglimitordrebook}

\citet{ANGSTMANN2016508} provide a numerical scheme that can solve Equation [\ref{eq:MainPDE}] but exclude the first derivative term that couples to the random force {\it i.e.}:
\begin{equation}
\varphi_t = c - a \varphi + D_\alpha \left[\theta D^{1-\alpha}_{t} \left( \frac{\varphi}{\theta} \right)\right]_{xx}. \label{eq:ReducedPDE}
\end{equation} 
Their scheme motivates the use of an update equation on a discrete price and time a grid with log price points $x_i=i \Delta x$ and times $t_n=n \Delta t$ to give the volume density on the lattice $\varphi^i_n = \varphi(x_i,t_n)$. The update equations allows us to move through the price and time grid in the positive time direction. The order arrival survival function $\theta$ from some times $t \in [t_m,t_n]$ where $t_m<t_n$ is approximated by a discretized annihilation operator representing accumulated removed orders using a Poisson process with removal rates $a_{\Delta}(x_i,t_n)$ on the grid\footnote{Here $y_{\Delta}$ for some y is as $y(x,t)= \mlm y_{\Delta}(x,t)$.}:
\begin{equation}
    \theta^i_{m,n} = \exp\left({-\int_{t_m}^{t_n} a_\Delta (x_{i},t')dt'}\right). 
\end{equation}
The accumulated created orders for a single $\Delta t$ time step accumulate over $t \in [t_{n-1},t_n]$ and is found using a creation operator $c_{\Delta}(x_i,t_n)$: 
\begin{equation}
    C^i_{n-1,n} = \int_{t_{n-1}}^{t_n}  \negthickspace   \negthickspace   \negthickspace   c_\Delta(x_i,t')dt'.
\end{equation}
The update formula uses a Discrete Time Random Walk (DTRW) \cite{ANGSTMANN2016508}. This encodes left, right, and self jumps at some time $t_n$ with respective probabilities $P^{(+1)}_n$, $P^{(-1)}_n$, and  $P^{(0)}_n$ to generate the diffusion:
\begin{align}
\varphi^{i}_{n}&=  C^i_{m,n} + \theta^i_{n-1,n} \varphi^i_{n-1} + \sum_{m=0}^{n-1} K_{n-m} \biggl[ P^{(+1)}_{n-1} \theta^{i-1}_{m,n-1} \varphi^{i-1}_{m} \nonumber \\
&+ P^{(-1)}_{n-1} \theta^{i+1}_{m,n-1} \varphi^{i+1}_{m} + (P^{(0)}_{n-1}-1) \theta^i_{m,n-1} \varphi^i_m
\biggr].\label{eq:UpdateRule0}
\end{align} 
Here $K_{n-m}$ is a memory kernel for a process with Sibuya waiting times \cite{ANGSTMANN2016508} and is defined as
\begin{equation}
K_{n-m} = \prod_{k=1}^{n-m}\left(1-\frac{2-\alpha}{k} \right) + \delta_{1,n-m}. \label{eq:MemKerRecursive}
\end{equation}
This is zero at $m=n$, unity at $m=n-1$ and prior time information is accumulated so that when $\alpha=1$ this becomes: $\delta_{m,n-1}$.

With no driving force the jump probabilities can be chosen as $P^{(+1)}_{n}=\frac{r}{2}$, $P^{(-1)}_{n}=\frac{r}{2}$, and  $P^{(0)}_{n}=1-r$ where there is a overall probability of a jump is $r$. The update equation can then be reduced to:
\begin{equation}
\begin{aligned}
\varphi^i_n =& C^i_{n-1,n} + \theta^i_{n-1,n} \varphi^i_{n-1}  + \sum_{m=0}^{n-1} K_{n-m} 
\biggl[ \tfrac{r}{2}  \theta^{i-1}_{m,n-1} \varphi^{i-1}_{m} \nonumber \\
&+\tfrac{r}{2} \theta^{i+1}_{m,n-1} \varphi^{i+1}_{m} -r \theta^{i}_{m,n-1} \varphi^i_m\biggr].\\
\end{aligned} \label{eq:UpdateRule1}
\end{equation}
To include a discrete driving force $F_n = f(t_n)$ we update the probability of right, left, and self jumps respectively: $P^{(+1)}_{n}=\tfrac{1}{2}(r-F_n)$, 
$P^{(-1)}_{n}=\tfrac{1}{2}(r+F_n)$, and $P^{(0)}_{n}=1-r$. Here the probability of self jumps is unchanged. The stochastic force is taken to be $F_n= \tfrac{V_t}{2 \beta D_{\alpha}}$ for some inverse temperature $\beta$, and $V_t$ as a stochastic process. The form of the stochastic force is discussed in \ref{app:force}. 

The update rule itself follows from the discrete master equation formulation of a reaction-diffusion system to describe the order volume density $\varphi^i_n$ on the lattice at price $x_i$ and time $t_n$. \ref{app:AnomalousDerivation} shows that the update rule given in Equation [\ref{eq:UpdateRule0}] satisfies Equation [\ref{eq:ReducedPDE}].  The equivalence with Equation [\ref{eq:ReducedPDE}] is demonstrated by using the update equation to formulate a finite difference method on the lattice by taking a discrete Laplace transform to dual variables, inverse Laplace transform these back to continuous price and time variables, and then taking appropriate limits to derive the advertised partial differential equation. 

Now, if we make a simplifying choice and set the annihilation rate to simply be a constant: $a(x,t)=\nu$ (in Equation [\ref{eq:theta}]), and use some source function $s(x,t)$ to approximate the creation operator, we find Equation [\ref{eq:FinalUpdateEquation}]. This is the update equation we will use to simulate the order book:
\begin{widetext}
\begin{equation}
\varphi^i_n = \sum_{m=0}^{n-1} K_{n-m} e^{-(n-m-1)\nu \Delta t} \biggl[ {\tfrac{1}{2}(r-F_{n-1}) \varphi^{i-1}_{m} +\tfrac{1}{2}(r+F_{n-1})  \varphi^{i+1}_{m}-r \varphi^{i}_{m}}\biggl] + e^{-\nu \Delta t} \varphi^{i}_{n-1}
+ s(x_i,t_{n-1}) \Delta t.
\label{eq:FinalUpdateEquation}
\end{equation}
\end{widetext}
Here $K_{n-m}$ is pre-computed using the recursive formula\footnote{The recursion is run until a value less than some tolerance $\epsilon=0.01$. The value of $n$ for which this occurs is a cut off: $n_c$. The sums in recursion then start from $m=(n-1)-n_c$. This means only the last $n_c$ steps are used in each calculation which saves computation time.} using Equation [\ref{eq:FinalUpdateEquation}]. The lower bound may be impractical to use in computations, so we label this lower bound $m_0$, and treat this as a model hyper-parameter. This allows one to get the correct diffusive behaviour over specific time intervals (see section \ref{ssec:measurediffusion}).

In Figure \ref{fig:manypaths}, we display an example of a collection of many simulated price paths that result from this scheme. The thin blue line point which occur in simulation background lattice $L_0$ (see  Section \ref{ssec:timescales}) for a chosen random realisation of $\{V_t\}$, and as a thick line points which occur on the sampled price path in $L_1$. These prices correspond to those sampled at the order book events and represent events one observes in our current formulation of the model. These sampled events are then those that are then mapped to the observed events in a real world realisation of a financial market. We also include in yellow the $L_0$ points for many other realisations. The paths are generated using the parameters in Table \ref{tab:parameters}.

\subsection{Initial and boundary conditions}

From Equation [\ref{eq:boundary}] our approach to finding appropriate initial and boundary conditions for the anomalous diffusion is to simulate the system using simple diffusive initial and boundary conditions
for the case of vanishing boundaries, until the system reaches equilibrium. These then are used as the initial conditions for the simulation of the anomalous diffusion system. 
This is discussed further in \ref{app:boundarycons}.

\subsection{Lattice parameters} \label{ssec:DtDx}

We assume that the diffusion limit exists:
\begin{equation}
    D_{\alpha} = \mlm \frac{r}{2}\frac{\Delta x^2}{\Delta t^{\alpha}}.
\end{equation}
We can then use this to set the lattice price grid increments in terms of the lattice time increments\footnote{This is just an uncertainty principle (where for $\alpha=1)$: $\langle \Delta x \tfrac{\Delta x}{\Delta t} \rangle = \tfrac{2D}{r}$ and $\langle \Delta t \tfrac{\Delta x^2}{\Delta t^2} \rangle = \tfrac{2D}{r}$ limit the minimum allowed grid combinations on the lattice.}: 
\begin{equation}
    \Delta x = \sqrt{\frac{2D_{\alpha}}{r}} \Delta t^{\frac{\alpha}{2}}. \label{eq:latticesize}
\end{equation}
We select the time increment $\Delta t$, and then use this to set the price increments $\Delta x$. In order to simulate non-uniformly sampled time grids we assume that the time changes are exponentially distributed: $\Delta t_m \sim \mathrm{Exp}(\lambda)$ at some time $t_m$. The default time increment for the simulations is with uniform sampling times $\Delta t$ where $\Delta t= \E[\Delta t_m] = \frac{1}{\lambda}$ \footnote{Here $\lambda = 8.0$ to have a time increment of $\Delta t= 0.125$ where $D=0.5$, $r=0.5$ and $\Delta x = 0.5$ when $\alpha=1.0$}. We modify the update equation so that the updates take place on a background grid $(x_i,t_n)$ that has uniformly sampled price changes but non-uniformly sampled time changes, but where the jumps take place using price changes dependent following Equation [\ref{eq:latticesize}] (See \ref{ssec:updateequationarrivaltimes}).

In practice we set the minimum and maximum price levels first {\it i.e.} $x_0$ and $x_M$ for some choice of $M$; the average number of grid points. Using this we find a $\Delta x$ to then find $\Delta t$ from Equation [\ref{eq:latticesize}]. This $\Delta t$ is then used to set the intensity parameter $\lambda = \Delta t^{-1}$. We then generate $M$ waiting times $\Delta t_n$. The background lattice $(x_i,t_n)$ is then built using $\Delta x$ and $\Delta t_m$ and Equation [\ref{eq:FinalUpdateEquation}] is modified and updated as in Equation [\ref{eq:UpdateEquationExpArrival}]. This is then demonstrated in Figure \ref{fig:StylizedFactsDifferentDx} where we compare the stylised facts for paths generated on a uniform time lattice with those on the non-uniformly sampled time lattice.

\subsection{Anomalous diffusion dependency} \label{ssec:measurediffusion}

\begin{figure}[ht]
\centering
\includegraphics[width=0.45\textwidth]{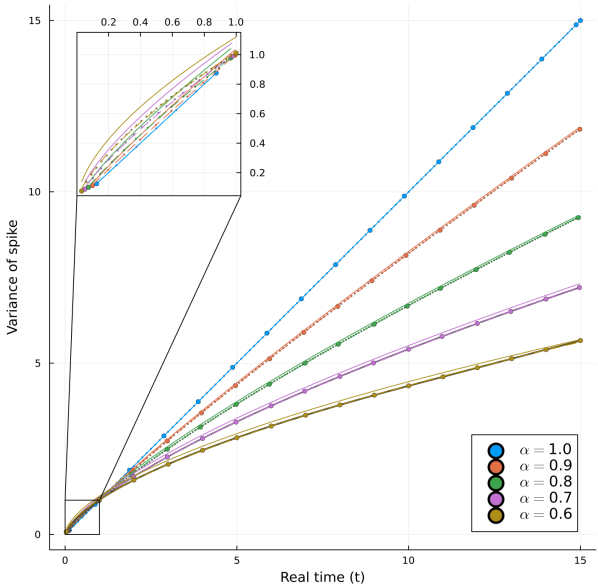}
\caption{We plot the simulated variance $\sigma_t$ of instantaneous pulses (corresponding to limit orders, see Sec. \ref{ssec:priceimpactlimit}) which occur at time 0 (vertical axis) as a function of time $t$ (horizontal axis) for $L/M=0.5$ (compared to $L/M=0.2$ in Fig. \ref{fig:IncreasingVarianceOfSpikeSmallDx}) as part of the discussion in Sec. \ref{ssec:measurediffusion}. We do this with different fractional diffusion rates $\alpha$. The theoretical prediction for $\sigma_t$ is $\sigma_t = (2D)/(\Gamma(1+\alpha)) \times t^\alpha$ which is shown as solid lines. We then do a numeric fit of the form $\sigma_t = a \times t^b$  where $a$ and $b$ are free parameters. This is plotted with a stippled line. The values of theoretical and numeric fits are shown in Tab. \ref{tab:measuredspikevarparameters}. We note that the behaviour is not as accurate as it is in Fig. \ref{fig:IncreasingVarianceOfSpikeSmallDx}, however, making $\Delta x$ larger has a very large saving in computation time (see Fig. \ref{fig:computcomplex}.)}
\label{fig:IncreasingVarianceOfSpike}
\end{figure}

\begin{figure}
\centering
\includegraphics[width=0.45\textwidth]{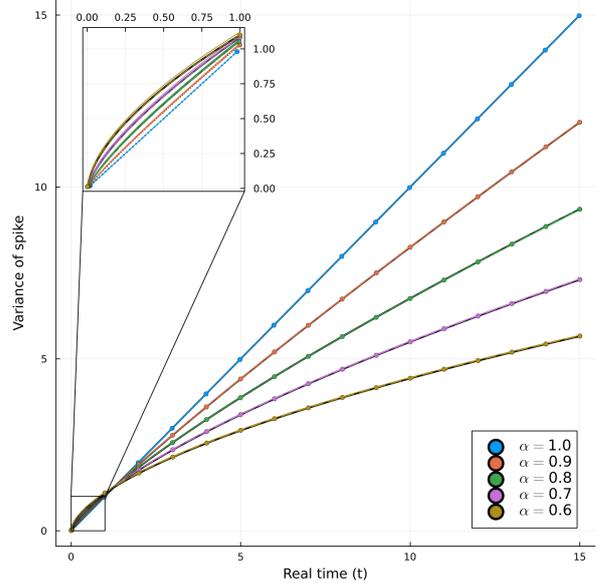}
\caption{We plot the simulated variance $\sigma_t$ of instantaneous pulses (corresponding to limit orders, see Sec. \ref{ssec:priceimpactlimit}) which occur at time 0 (vertical axis) as a function of time $t$ (horizontal axis) but for $L/M=0.2$ (compared to $L/M=0.5$ as in Fig. \ref{fig:IncreasingVarianceOfSpike}) as part of the discussion in Sec. \ref{ssec:measurediffusion}. We do this with different fractional diffusion rates $\alpha$. The theoretical prediction for $\sigma_t$ is $\sigma_t = (2D)/(\Gamma(1+\alpha)) \times t^\alpha$ which is shown as solid lines. We then do a numeric fit of the form $\sigma_t = a \times t^b$  where $a$ and $b$ are free parameters. This is plotted with a stippled line. The values of theoretical and numeric fits are shown in Tab. \ref{tab:measuredspikevarparameterssmalldx}. This verifies that the model is working correctly.}
\label{fig:IncreasingVarianceOfSpikeSmallDx}
\end{figure}

To simulate the system we first give the system time to relax to equilibrium, and then we introduce a single impulse event at trade time count\footnote{A number chosen to be large enough that in the slowest of the simulated systems we consider, it will reach equilibrium.} $\ell=200$. We then record the variances of the spike as it spreads out (diffuses) starting from the moment at which the spike is placed (which we assign $t=0$) and allowing the system to evolve for 20 units of trade time. We do this for different values of $\alpha$, $\Delta x$ and $m_0$ (see Section \ref{ssec:simulatinglimitordrebook}). This is shown in Figure \ref{fig:IncreasingVarianceOfSpike} and Figure \ref{fig:IncreasingVarianceOfSpikeSmallDx}. Both figures are for $\alpha$ drawn from the set \{1.0,0.9,0.8,0.7,0.6\}. 

Figure \ref{fig:IncreasingVarianceOfSpike} has $\Delta x=0.5$ and $m_0$ chosen so that the system remembers approximately the previous 12 trade events when $\alpha=0.6$. For larger values of $\alpha$, the memory then lasts longer. 

Figure \ref{fig:IncreasingVarianceOfSpikeSmallDx} has $\Delta x=0.2$ and $m_0$ chosen to be large enough that the system remembers more than 15 trade events in all cases. The fitted values are shown in tables \ref{tab:measuredspikevarparameters} and \ref{tab:measuredspikevarparameterssmalldx} respectively. We also show an inset with a region closer to the origin. We note that, as reported in \citet{ANGSTMANN2016508}, for smaller values of $\Delta x$, the result is closer to the theoretical answer. We include Figure \ref{fig:IncreasingVarianceOfSpikeSmallDx} to confirm that our numerical scheme is working as expected. 

In addition, the $\alpha=0.6$ line in Figure \ref{fig:IncreasingVarianceOfSpike} shows that once system can no longer remember its entire history (after 13 trade events) the diffusion no longer matches the theoretical result. Throughout this paper, we will use this value of $m_0$ so that the system remembers at least 13 previous order book trade events when $\alpha=0.6$. Finally, we note the appearance of more data points in Figure \ref{fig:IncreasingVarianceOfSpike} as compared with Figure \ref{fig:IncreasingVarianceOfSpikeSmallDx}. This follows from Equation \ref{eq:latticesize} where using a smaller $\Delta x$ results in many more time steps being required. 

\begin{table}[h!]
\begin{center}
\begin{tabular}{c c | c c} 
\hline
\multicolumn{2}{c|}{Theoretical} & \multicolumn{2}{c}{Estimated (Measured)} \\ 
\multicolumn{2}{c|}{$\sigma_t = \sigma_0 t^{\alpha}$} & 
\multicolumn{2}{c}{$\tilde \sigma_t = \hat \sigma_0 t^{\hat \alpha}$}\\
$\alpha$ & $\sigma_0$ & $\hat \alpha$  & $\hat \sigma_0$ \\ 
\hline
1.00 & 1.00 & 1.00000 $\pm$ [0.00000] & 1.00000 $\pm$ [0.00000] \\
0.90 & 1.04 & 0.90693 $\pm$ [0.00019] & 1.01610 $\pm$ [0.00045] \\
0.80 & 1.07 & 0.81310 $\pm$ [0.00031] & 1.02744 $\pm$ [0.00075] \\
0.70 & 1.10 & 0.71829 $\pm$ [0.00036] & 1.03360 $\pm$ [0.00087] \\
0.60 & 1.12 & 0.62951 $\pm$ [0.00039] & 1.02130 $\pm$ [0.00092] \\
\hline
\end{tabular}
\end{center}
\caption{Table of theoretical and measured parameters corresponding to Fig. \ref{fig:IncreasingVarianceOfSpike} which shows the increase in variance of a spike placed at a single point at a single time (limit orders, see Sec. \ref{ssec:priceimpactlimit}). Uncertainties are quoted to 2 significant figures and the fit was done using Julia's {\tt LsqFit} library. This gives a sense of the accuracy of the numerical implementation for our chosen parameters.}
\label{tab:measuredspikevarparameters}
\end{table}

\begin{table}
\begin{center}
\begin{tabular}{c c | c c} 
\hline
\multicolumn{2}{c|}{Theoretical} & \multicolumn{2}{c}{Estimated (Measured)} \\ 
\multicolumn{2}{c|}{$\sigma_t = \sigma_0 t^{\alpha}$} & 
\multicolumn{2}{c}{$\tilde \sigma_t = \hat \sigma_0 t^{\hat \alpha}$}\\
$\alpha$ & $\sigma_0$ & $\hat \alpha$  & $\hat \sigma_0$ \\ 
\hline
1.00 & 1.00 & 1.00000 $\pm$ [0.00000] & 1.00000 $\pm$ [0.00000] \\
0.90 & 1.04 & 0.90150 $\pm$ [0.00002] & 1.03471 $\pm$ [0.00004] \\
0.80 & 1.07 & 0.80262 $\pm$ [0.00002] & 1.06450 $\pm$ [0.00006] \\
0.70 & 1.10 & 0.70339 $\pm$ [0.00002] & 1.08811 $\pm$ [0.00005] \\
0.60 & 1.12 & 0.60386 $\pm$ [0.00001] & 1.10422 $\pm$ [0.00001] \\
\hline
\end{tabular}
\end{center}
\caption{Table of theoretical and measured parameters corresponding to Fig. \ref{fig:IncreasingVarianceOfSpikeSmallDx} which shows the increase in variance of a spike placed at a single point at a single time (limit orders, see Sec. \ref{ssec:priceimpactlimit}). Uncertainties are quoted to 1 significant figure and the fit was done using Julia's {\tt LsqFit} library. This shows that the bids and offers are diffusing in line with the theoretical model parameters and validates the numerical implementation.}
\label{tab:measuredspikevarparameterssmalldx}
\end{table}

\begin{figure}[h!]
\input{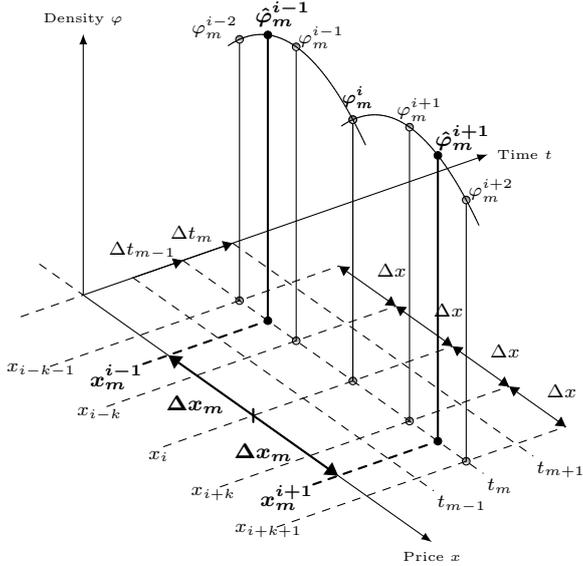}
\caption{The lattice specification used to support the simulation grid (see Sec. \ref{ssec:updateequationarrivaltimes}) shows how points not in the underlying background lattice are approximated in update Eqn. [\ref{eq:UpdateEquationExpArrival}] using Eqn. [\ref{eq:onjumplattice0}]. The horizontal plane represents the background lattice $(x_i,t_m)$. The vertical axis gives the order densities at a given lattice point: $\varphi^i_m$. The separation between times are $\Delta t_{m-1}$ and $\Delta t_{m}$; these can be non-uniform. The separation between prices in the background lattice are all uniformly $\Delta x$. The diagram shows a particle jumping to the current point $(x_i,t_m)$ where the jump distance is not a multiple of $\Delta x$ but is $\Delta x_m$. The source order densities are found using a three point approximation from Eqn.[\ref{eq:onjumplattice}] for $\hat \varphi^{i-1}_m$ and $\hat \varphi^{i+1}_m$.}
\label{fig:visualisationofexpscheme}
\end{figure}

\subsection{Non-uniform sampling update equation} 
\label{ssec:updateequationarrivaltimes}

Consider the final update equation [\ref{eq:FinalUpdateEquation}]. In this equation, we have implicitly assumed that $\Delta t$ and $\Delta x$ are constant during the simulation. This defines a uniform lattice. Now let us rather suppose that we generate a sequence of changes in time, which we will denote $\left\{\Delta t_m\right\}_{m=1}^M$; these refer to the waiting times (in simulation time) between simulation events. Here interpreted as price update events that then coincide with order book update events. 

The simulation starts at $t_0$ and continues until time $t_M$. These time steps are no longer uniform. Step $n-1$ to $n$ will have some $\Delta t_n$ drawn from some distribution {\it e.g.} an exponential. Here we will declare a target average arrival delay, and then generate the sequence of times around it, and then use that to determine the price (spatial) grid size at each time. We can do this because the diffusion is built on a random walk process as are the stochastic forces and these are all price independent probabilities, but there can be time dependence for the stochastic force. 

Now, using the equation for lattice sizes, Equation [\ref{eq:latticesize}], as well as a fixed value for $D_{\alpha}$, $\alpha$ and $r$, we generate, for each $n$, a corresponding lattice spacing sequence $\left\{\Delta x_k\right\}_{k=1}^M$. We have two sequences of lattice sizes. Here we denote $\Delta x$ to be the average grid width: $\Delta x = \E[\Delta x_k]$. 

A choice of $\Delta t_{n-1}$, and thus $\Delta x_{n-1}$, determines the price gap from where the left and right jumps to price $x^i_{n-1}$ will occur at time $t_{n-1}$ to time $t_n$ {\it i.e.} this unique pair is associated with the $n^{\mathrm{th}}$ event and then determines $x^{i-1}_{n-1}$ and $x^{i+1}_{n-1}$ unique to that event. Here these points are: $x^{i\pm1}_{n-1}=x_i \pm \Delta x_{n-1}$. This will then imply that we have similar event specific values for all the price points $x^i_n$. 

However, the $i^\mathrm{th}$ the price lattice point at time $t_n$ will not necessarily be equivalent to the price value at the $i^\mathrm{th}$ lattice price point at time $t_m$, if it exists at all. This means that we need to introduce a background lattice. Here we now have two price (spatial) coordinate systems, that of the background $(x_i,t_n)$, and that associated with the jumps: $x^i_n$. To proceed with the simulation we first generate the background lattice, this will have non-uniformly sampled times but uniformly sampled prices. We use the average grid size $\Delta x$ to generate the background lattice: $(x_i,t_n)$. 

Here we assume there is some approximation function $\varphi_\Delta (x,t)$ for the densities on the background lattice: $\varphi^i_n = \varphi_{\Delta}(x_i,t_n)$. Here the grid spacing is $\Delta x$ for the prices but $\Delta t_n$ for the temporal spaces which are now non-uniform. At some time $t_{n-1}$ and with some time increment $\Delta t_{n-1} = t_{n} - t_{n-1}$ we can then use the diffusion constraint in Equation [\ref{eq:latticesize}] to find the unique price increment at time $t_{n-1}$, {\it i.e.} $\Delta x_{n-1}$. Then we can find the prices from which the right and left jumps will occur that are consistent with the diffusion to the order of the approximations:
\begin{equation}
 \hat \varphi^{i\pm 1}_{m} = \varphi_{\Delta}(x^{i \pm 1}_{m}, t_{m}) = \varphi_{\Delta}(x_i \pm \Delta x_{m}, t_{m}).
\end{equation}
The probabilities of left and right jumps do not depend on the sequence $\left\{\Delta x_m\right\}_{m=1}^M$, they only depend on the most recent entry at $n-1$. In contrast, $\varphi$ makes use of the entire history of the sequence $\left\{\Delta x_m\right\}_{m=1}^M$ where at each time $t_m$ we have the unique $\Delta t_m$ and hence its unique $\Delta x_m$ relative to the background points $x_i$. We rewrite Equation [\ref{eq:FinalUpdateEquation}] as in Equation [\ref{eq:UpdateEquationExpArrival}] where $\varphi^i_n$ is on the background lattice:
\begin{widetext}
\begin{equation}
\varphi^i_n = \sum_{m=0}^{n-1} K_{n-m} e^{-\nu (t_{n-1}-t_m)} \left[{\tfrac{1}{2}\left({r+F_{n-1}}\right) \hat \varphi^{i-1}_{m} + \tfrac{1}{2}\left({r-F_{n-1}}\right) \hat \varphi^{i+1}_{m} - r \varphi^{i}_{m} }\right] + e^{-\nu \Delta t_{n-1}} \varphi^{i}_{n-1} + s(x^i_n,t_{n-1}) \Delta t_{n-1}. 
\label{eq:UpdateEquationExpArrival}
\end{equation}
\end{widetext}
We need to approximate $\hat \varphi^{i\pm 1}_{m}$ from each $x^{i \pm 1}_{n-1}$, and $x_i$ at each time $t_m$. The point $x_i$ is in the background lattice $(x_i,t_n)$ and are uniformly sampled using $\Delta x$. While points $x^i_{n-1}$ for each $t_{n-1}$ are points in the jump lattice. This means that the price jump grid points do not necessarily align at different times: $x_{i \pm 1}\neq x^{i\pm1}_{m}$ because $x_i \pm \Delta x \neq x_i \pm \Delta x_m$. We can identify the points $x_i$ in the background grid with points $x^i_n$:
\begin{equation}
\hat \varphi^{i\pm1}_m \approx \varphi_{\Delta}(x_i \pm \Delta x_m,t_m). \label{eq:onjumplattice0}
\end{equation}
Here at some other time $t_m$ we search the background lattice for points $(x_k,t_m)$ such that $x_k \le x^{i \pm 1}_m \le x_{k+1}$ and then use a first order mid-point approximation to find the densities: 
\begin{equation}
   \hat \varphi^{i \pm 1}_m \approx \varphi^{k}_m +  \tfrac{( x^{i \pm 1}_m  - x_k)}{2 \Delta x} \left[{\varphi^{k+1}_m - \varphi^{k-1}_m} \right]. \label{eq:onjumplattice}
\end{equation}
This allows for the situations where $\Delta t_m > \Delta x$, $\Delta t_m < \Delta x$, and when $\Delta x_m = \Delta x$ we recover $\hat \varphi^{i \pm 1}_m = \varphi^{k}_m$ from the background\footnote{If $x_k=x_{i\pm1}$ then  $\hat \varphi^{i \pm 1}_{m} \approx \varphi^{i\pm 1}_{m} + \tfrac{(\Delta x -  \Delta x_{m})}{2 \Delta x} \left[{\varphi^{i}_{m} - \varphi^{i \pm 2}_{m}} \right]$.}. We can now use that $\Delta t_m \sim \mathrm{Exp}(\lambda)$ and the intensity $\lambda$ is chosen such that $\Delta t= \E[\Delta t_m] = \frac{1}{\lambda}$. 

\begin{table*}
\begin{center}
\begin{tabular}{l l r l l l} 
\toprule
\multicolumn{3}{c}{Parameters} & Value & Units & Description \\ 
\toprule
\multirow{11}{*}{\rotatebox{90}{Model}} &
\multirow{8}{*}{\rotatebox{90}{Fixed}}
 & $L$ & 200 & $[\$]$ & Maximum price $x_{\mathrm{max}}$ s.t. $x \in [0,L]$ and $p(t) \in [p(0)-\sfrac{L}{2},p(0)+\sfrac{L}{2}]$.\\ 
 & & $M$ & 400 & $[1]$ & Number of price grid elements s.t. $\Delta x = \sfrac{L}{M}$ (Eqn. \ref{eq:latticesize}) \\ 
 & & $D_{\alpha}$ & 0.5 &  $[\$^{2} T^{-\alpha}]$ & Diffusion constant (Eqn. [\ref{eq:MainPDE}]).\\ 
  & & $\nu$ & 0.5 & $[\$ T^{-1}]$ & The limit order removal rate (Eqn. [\ref{eq:MainPDE}]).\\ 
 & & $r$ & 0.5 & [1] & Probability of self jumps (Eqn. [\ref{eq:FinalUpdateEquation}]).\\ 
 & & $p(0)$ & 1300 & $[\$]$ & Initialised state of the order book intercept {\it i.e.}  $\varphi(p(0),0)=0$. \\ 
 & & $\kappa$ & 1.0 & $[\# T^{-1}]$  & Limit order arrival rate source function scaling (Eqn.[\ref{eq:source}]).\\ 
 & & $\mu$ & 0.1 & $[ \$^{-1} ]$ & Price axis source function scaling (Eqn. [\ref{eq:source}]).\\ 
 \cmidrule(rl){2-6}
 & \multirow{3}{*}{\rotatebox{90}{Free}}
 & $\alpha$ & $\in$ \{0.6,0.8,1.0\} & $[1]$ & Fractional diffusion parameter (Eqns [\ref{eq:MainPDE}] and [\ref{eq:FinalUpdateEquation}]).\\
 & & $\sigma$ & $\in$ \{0.5,1.0,1.5\} & $[\%]$ & Stochastic potential sampling distribution variance: $V_t \sim \mathcal{N}(0,\sigma)$\\
 & & $\rho$ & $\in$ \{0.0,0.8,0.9\} & $[\%]$ & Stochastic potential self correlation: $V_t \sim \rho V_{t-1}$\\
 \hline
 \multirow{3}{*}{\rotatebox{90}{Hyper}} &
\multirow{3}{*}{\rotatebox{90}{Free}}
 & $\gamma_1$ & $\in$ $[1,\infty)$ & $[\%]$ & Timescale ratio of order book times to simulation times. (Eqn. [\ref{eq:gamma1}]) \\
 & & $\gamma_{2}$ & $\in$ $[1,\infty)$ & $[\%]$ & Timescale ratio of trade times to simulation times. (Eqn. [\ref{eq:gamma12}]) \\
  & & $m_0$ & $\in$ [0,M] & $[1]$ & Minimum memory cut-off in anomalous diffusion: $\sum_{m=0}^{m_0} \cdot$ (Eqn. [\ref{eq:FinalUpdateEquation}]) \\
\bottomrule
\end{tabular}
\end{center}
\caption{Fixed and variable parameters are given with values for the fixed parameters, and ranges for variable parameters. Prices are measured in currency of account [\$], shares as counts [\#], time [$T$] is measured in simulation time as seconds $s$, and variance in units denoted [\%]. The symbol $[1]$ denotes a unit-less quantity. The impact of non-uniform time (see Sec. \ref{ssec:updateequationarrivaltimes}) is shown in Fig. \ref{fig:StylizedFactsDifferentDx}. The stylised facts plots showing the impact of free parameter $(\alpha,\sigma,\rho)$ combinations is given in \ref{app:appendix:additional} in figs. \ref{fig:StylizedFactsDifferentDiffusions}, \ref{fig:StylizedFactsDifferentVariances}, and \ref{fig:StylizedFactsDifferentCorrelation}.}
\label{tab:parameters}
\end{table*}

\begin{figure}[ht]
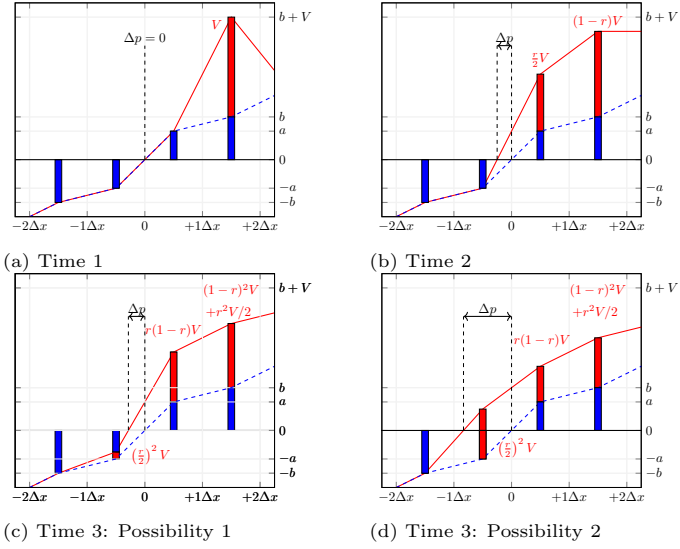

\begin{tabular}[l]{ll}
\begin{subfigure}[c]{0.24\textwidth}
     \resizebox{0.95\textwidth}{!}{
        \input{\PathTikz tikz_time1}}
\caption{Time 1} \label{fig:kt1}
\end{subfigure}
&\begin{subfigure}[c]{0.24\textwidth}
   \resizebox{0.95\textwidth}{!}{
        \input{\PathTikz tikz_time2}}
\caption{Time 2} \label{fig:kt2}
\end{subfigure}\\
\begin{subfigure}[c]{0.24\textwidth}
     \resizebox{0.95\textwidth}{!}{
        \input{\PathTikz tikz_time3-1}}
\caption{Time 3: Possibility 1} \label{fig:kt31}
\end{subfigure} 
&\begin{subfigure}[c]{0.24\textwidth}
     \resizebox{0.95\textwidth}{!}{
        \input{\PathTikz tikz_time3-2}}
\caption{Time 3: Possibility 2} \label{fig:kt32}
\end{subfigure}
\end{tabular}
\caption{The evolution of an impulse on the simulation grid is shown. The blue bars represent the order density of the system $\varphi$ before it is changed by a flash limit order (see Sec \ref{ssec:priceimpactlimit}) of volume $V$. The red bars show the impulse and its spread through simulation time. The blue dotted line approximates the equilibrium initial configuration of the system and is used to compute the initial mid-price using Eqn. [\ref{eq:midprice}]. The difference between the vertical dashed black lines represent the price impact by showing the mid-price change $\Delta p$ as time progresses. The new mid-price is calculated from the intercept of the red line (joining the red volume bars for the altered system) with the zero volume axis. At Time=1: Fig. \ref{fig:kt1} shows the impulse of volume $V$ at log-price position $\tfrac{3}{2}\Delta x$. At Time=2: Fig. \ref{fig:kt2} is one simulation time step later, and has the impulse spread according to update Eqn. [\ref{eq:FinalUpdateEquation}]. The stochastic force is ignored, and a fraction $\tfrac{r}{2}$ of orders have moved in either direction. There is now price impact and $\Delta p>0$. There are two possible steps that follow: i.) (Fig. \ref{fig:kt31}) Here the volume $V$ was small enough to ensure that the intercept remains in the region $[-\tfrac{1}{2}\Delta x, \tfrac{1}{2}\Delta x]$, or ii.) (Fig. \ref{fig:kt32}) The volume $V$ of the order is now large enough so that the intercept is now in the range $[-\tfrac{3}{2}\Delta x,-\tfrac{1}{2}\Delta x]$. The value at which this crossover occurs is the critical volume we label: $V_c$. See \ref{app:PriceImpactKinks}.}
\label{fig:KinkVisualisation}
\end{figure}


\section{Stylised facts} \label{sec:stylisedfact}

\subsection{Generic stylised facts} \label{ssec:stylisedfacts}

In Figure \ref{fig:StylizedFactsDifferentDx}, we generate stylised facts using the algorithm from Section \ref{ssec:simulatinglimitordrebook} for order book sampling events which happen at uniform time intervals, and the algorithm from Section \ref{ssec:updateequationarrivaltimes} which allows for order book events to happen at non-uniform times. In making this plot, we use parameters $\alpha=0.8$, $\beta=0.9$ and $\sigma=1.0$.  We first describe what each of the sub-figures represent.

In Figure \ref{fig:StylizedFactsDifferentDx:a}, we show the trajectory of $p(t)$ as a function of calendar time, that is, events in $L_0$. It is important to note the inset in Figure \ref{fig:StylizedFactsDifferentDx:a}, which shows a zoomed in view of the price path. Here there are small circles which represent events in $L_0$ (see Section \ref{ssec:timescales}) which we have previously described as unobserved. This is still the case and they are only shown here to give a sense of the number of steps required to compute these stylised facts. The large circles appearing in this inset are events in $L_1$, that is, sampled order book events. 

These sampled order book events conceptually coincide with trade events that we expect to measure from the real world observations and are solely used to produce all results in all other figures \ref{fig:StylizedFactsDifferentDx:a} to \ref{fig:StylizedFactsDifferentDx:f}. In \ref{fig:StylizedFactsDifferentDx:b} we show the log returns over the same time as in Figure \ref{fig:StylizedFactsDifferentDx:a}. In Figure \ref{fig:StylizedFactsDifferentDx:c}, we show the distribution of the log of the returns from Figure \ref{fig:StylizedFactsDifferentDx:b} so that it aligns with this. In Figure \ref{fig:StylizedFactsDifferentDx:d} we show the corresponding QQ-plots. In Figure \ref{fig:StylizedFactsDifferentDx:e} we show auto-correlation functions for (from the outside going in): the sign of the orders, the log of the returns, and finally the absolute value of the log of the returns. Order signs are determined by the tick rule. 

Finally, in Figure \ref{fig:StylizedFactsDifferentDx:f} we investigate the tail behaviour of the returns found from Figure \ref{fig:StylizedFactsDifferentDx:c}. We  use the method provided by \citet{C2001} to model threshold excess\footnote{We use a library specifically implemented around the description in \cite{C2001}: ``Extremes.jl"}. First, we select a threshold above which the mean excess plot is linear within uncertainty. This process was automated by allowing our code to select as a threshold the first value above which a straight line could be fit that lies entirely within the uncertainty bars of the mean excess plot. However, this did not always result in the best possible cut offs. \citet{C2001} recommends applying this rule less strictly in order to get more extreme values. To do this, we include a tolerance factor $\zeta$ which widens the uncertainty bars by a factor of $\zeta$ before allowing the algorithm to fit the line. If $\zeta$ is in $[-1,0]$, than the program simply takes the top $-\zeta$ proportion of the data to be extreme values. The mean excess plots, chosen $\zeta$ and straight line fits are shown as insets in Figure \ref{fig:StylizedFactsDifferentDx:f} and the tolerance factor used is shown in the top right of each inset. Figure \ref{fig:StylizedFactsDifferentDx:f} itself shows the resulting return level plot for the extreme values and includes a generalised Pareto fit shown as a solid line which can be used to check the fit's accuracy.

We now consider the qualitative features obtained 
in each subplot of Figure \ref{fig:StylizedFactsDifferentDx} - first analysing the generic features observed throughout this work by using the uniform algorithm in Section \ref{ssec:sfuniform} and then contrasting the effects of using the non-uniform algorithm in Section \ref{ssec:sfnonuniform}. 

\subsubsection{Uniform sampling}\label{ssec:sfuniform}

We consider the blue path in Figure \ref{fig:StylizedFactsDifferentDx} which corresponds to using the uniform sampling algorithm and the default parameters. 
Firstly, looking at \ref{fig:StylizedFactsDifferentDx:e}, we note that self correlations between the random information arrival with $V_t$ and $V_{t-1}$ via the correlation parameter $\rho$ has successfully introduced correlations in the ACF plots. This is seen in the trade signs (order flow), the log returns and the absolute value of the log returns. We further note that the auto-correlations in the order flow die off much faster than those in the absolute value of log returns - which matches the behaviour of real markets. 
Secondly, the QQ-plot \ref{fig:StylizedFactsDifferentDx:d}, together with the mean excess plots in \ref{fig:StylizedFactsDifferentDx:f}, indicate that the returns distribution \ref{fig:StylizedFactsDifferentDx:c} has light tails, which is not observed in real markets (see discussion in Section \ref{sec:conclusion}). 
Finally, from \ref{fig:StylizedFactsDifferentDx:b} we can see that there is no volatility clustering. 

\subsubsection{Non-uniform sampling}\label{ssec:sfnonuniform}

\begin{figure*}[ht!]
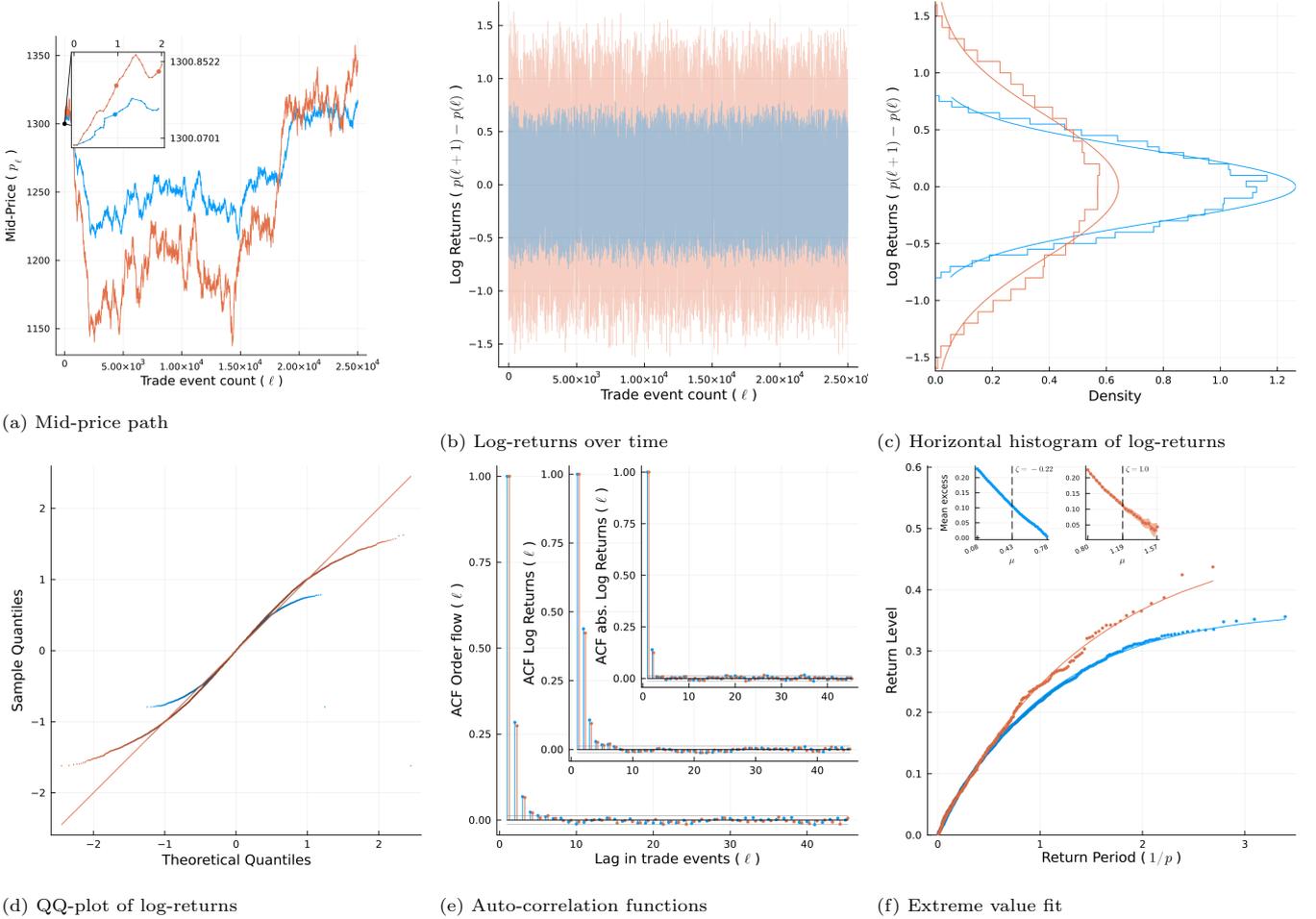

\centering
\begin{subfigure}[c]{0.32\textwidth}
\includegraphics[width=1.0\textwidth]{\PathSFDT SF_DT_1-mp.png} 
\caption{Mid-price path} \label{fig:StylizedFactsDifferentDx:a}
\end{subfigure}
\begin{subfigure}[c]{0.32\textwidth}
\includegraphics[width=1.0\textwidth]{\PathSFDT SF_DT_2-lr.png}
\caption{Log-returns over time} \label{fig:StylizedFactsDifferentDx:b}
\end{subfigure}
\begin{subfigure}[c]{0.32\textwidth}
\includegraphics[width=1.0\textwidth]{\PathSFDT SF_DT_3-hlr.png}
\caption{Horizontal histogram of log-returns} \label{fig:StylizedFactsDifferentDx:c}
\end{subfigure}
\begin{subfigure}[c]{0.32\textwidth}
\includegraphics[width=1.0\textwidth]{\PathSFDT SF_DT_4-qqlr.png}
\caption{QQ-plot of log-returns} \label{fig:StylizedFactsDifferentDx:d}
\end{subfigure}
\begin{subfigure}[c]{0.32\textwidth}
\includegraphics[width=1.0\textwidth]{\PathSFDT SF_DT_5-acfs.png}
\caption{Auto-correlation functions} \label{fig:StylizedFactsDifferentDx:e}
\end{subfigure}
\begin{subfigure}[c]{0.32\textwidth}
\includegraphics[width=1.0\textwidth]{\PathSFDT SF_DT_6-expl.png}
\caption{Extreme value fit} \label{fig:StylizedFactsDifferentDx:f}
\end{subfigure}
\caption{The impact of using the non-uniform algorithm (see Sec, \ref{ssec:updateequationarrivaltimes}) rather than the uniform algorithm (see Sec, \ref{ssec:simulatinglimitordrebook}) on the stylised facts are shown while using the default free parameter tuple $(\alpha=0.8,\rho=0.9,\sigma=1.0)$ (see Sec, \ref{ssec:stylisedfacts}) for both simulations. The red path is the default path which uses the uniform sampling algorithm and appears in all stylised facts plots for comparison (see figs.\ref{fig:StylizedFactsDifferentDiffusions}, \ref{fig:StylizedFactsDifferentVariances} and \ref{fig:StylizedFactsDifferentCorrelation}.) and it has $\Delta t_m = \lambda \approx 0.0743 \; \forall m$ (this is chosen by the algorithm given $\Delta x = 0.5$). The blue path uses the non-uniform sampling algorithm for comparison (this corresponds to the discussion in Sec. \ref{ssec:sfnonuniform}.) This algorithm results in $\Delta t_m \sim \textrm{Exp}(1/\lambda)$ such that $\E [\Delta t_m] = \lambda$. Each path has approximately 25,000 events. Using non-uniform sampling reduces the variance of the returns distribution, and is seen in Fig. \ref{fig:StylizedFactsDifferentDx:c}. In addition, the shape of the paths, shown in Fig. \ref{fig:StylizedFactsDifferentDx:a}, remain comparable - it is not difficult to believe they have been drawn from the same seed. The QQ-plot in Fig. \ref{fig:StylizedFactsDifferentDx:d} together with the mean excess plots in Fig. \ref{fig:StylizedFactsDifferentDx:f} show that the returns distributions have light tails. The self-correlations in the information arrivals $\rho$ still creates auto-correlation in the displayed ACFs, and is shown in Fig. \ref{fig:StylizedFactsDifferentDx:e}; the self-correlations is not affected by the non-uniform scheme. We still see the same behaviour of the ACFs where the auto-correlations in the order flow die off much faster than those in the absolute value of log returns. Finally, there is still no volatility clustering with either scheme. }
\label{fig:StylizedFactsDifferentDx}
\end{figure*}

We now consider the red path in Figure \ref{fig:StylizedFactsDifferentDx} which uses the non-uniform time sampling from Section \ref{ssec:updateequationarrivaltimes} (update Equation [\ref{eq:UpdateEquationExpArrival}].)

The non-uniform path shares the same generic features of the uniform path mentioned above (Section \ref{ssec:sfuniform}) with minor alterations. Firstly, the non-uniform time algorithm has reduced the variance of the original path as can be seen in Figure \ref{fig:StylizedFactsDifferentDx:a}. In addition, the distribution of extreme events has changed slightly but still has good Pareto fit. Note that we set $\xi=-0.22$ in the inset, choosing the cutoff manually\footnote{The automatic fitting algorithm from Sec. \ref{ssec:stylisedfacts} is not used.} such that a straight line could plausibly be fit above the cutoff and within uncertainty. The auto-correlations have not changed significantly compared to the uniform scheme. In summary, this means we are able to extract the same stylised facts using the non-uniform scheme as that of the uniform scheme. This is an important verification of the suitability of the non-uniform algorithm for market simulation work.

\subsection{Simulated price impacts}\label{ssec:priceimpacts}

An order schedule is a 3-tuple $(v_k,x_k,t_k)$ for the $k$th order, with volume $v_k$, target price $x_k$, and at time $t_k$. Price impact and response can be computed for any order schedule.
In the following sections we consider the price impacts $\mathcal{J}$ for two different order schedules that represent two order types: a flash limit order (Section \ref{ssec:priceimpactlimit}) and a market order (Section \ref{ssec:priceimpactmarket}). We use cubic spline interpolation to determine the price $p$ at any time together with the non-uniform sampling scheme (Section \ref{ssec:updateequationarrivaltimes}.). 

Each plot will encompass changing two factors. Firstly, various delays $\Delta n \in \mathbb{N}^{+}$ will be tried - that is, the number of time steps which have been allowed to pass in $L_0$ (see Section \ref{ssec:timescales}) since the placement of the order before measuring the price impact. Secondly, the volume of the order $Q$. We thus end up with the expression $\mathcal{J}(\Delta n,Q) \equiv p(n+\Delta n)|_Q - p(n)$ where $n$ is the time in $L_0$ of the arrival of the order. When plotting this expression as a function of $Q$, various lines defined by $\mathcal{J}_{\Delta n} \equiv \{(Q,\mathcal{J}(Q,\Delta n)):\forall Q\}$ are plotted and this is the conventional meaning of price impact. Each $\mathcal{J}_{\Delta n}$ corresponds to the conventional price impact line one would observe for different choices of $\gamma_1$ since here we have $\gamma_1=\Delta n$.
In each case we use 10 different sample paths (that is, different paths $V_t$) to introduce some uncertainty. Additionally, we set $\alpha=1.0$ leave all other parameters the same as shown in table $\ref{tab:parameters}$. 

\subsubsection{Simulated flash limit order}\label{ssec:priceimpactlimit}

Here we consider the placement of either a buying (positive) or selling order (negative) at a single reservation price. Analytically, this is an impulse limit order and the below term would be added to $C$ in Equation [\ref{eq:MainPDE}]: 
\begin{equation}
    s_{_\text{LO}}(x,t) = V \delta(x-x_0) \delta(t - t_0). 
\end{equation}
This may be interpreted as an order of volume $Q$ placed at the reservation price $x_0$ at time $t_0$. This will change the overall volume availability and so kick the system out of equilibrium.  We can then observe how the system returns to equilibrium. This is modelling the impact of flash orders on the overall order book. 

\begin{figure*}[ht!]
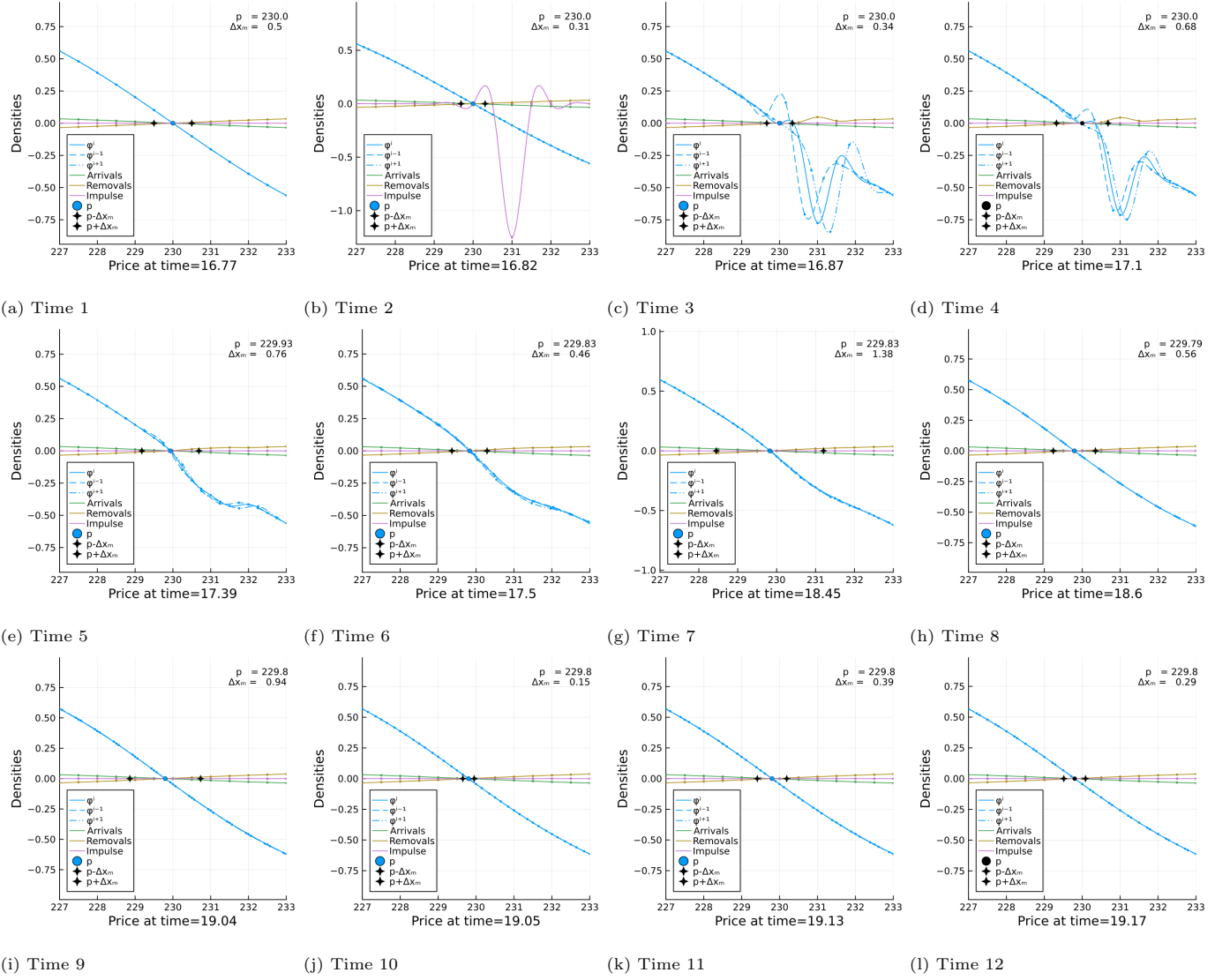

\centering
\begin{subfigure}[c]{0.24\textwidth}
\includegraphics[width=1.0\textwidth]{\PathKExpInterp K_exp_interp-1.png}
\caption{Time 1}\label{fig:VisualizePriceImpactSpikeExpInterp:a}
\end{subfigure}
\begin{subfigure}[c]{0.24\textwidth}
\includegraphics[width=1.0\textwidth]{\PathKExpInterp K_exp_interp-2.png}
\caption{Time 2}\label{fig:VisualizePriceImpactSpikeExpInterp:b}
\end{subfigure}
\begin{subfigure}[c]{0.24\textwidth}
\includegraphics[width=1.0\textwidth]{\PathKExpInterp K_exp_interp-3.png}
\caption{Time 3}\label{fig:VisualizePriceImpactSpikeExpInterp:c}
\end{subfigure}
\begin{subfigure}[c]{0.24\textwidth}
\includegraphics[width=1.0\textwidth]{\PathKExpInterp K_exp_interp-4.png}
\caption{Time 4}
\end{subfigure}
\begin{subfigure}[c]{0.24\textwidth}
\includegraphics[width=1.0\textwidth]{\PathKExpInterp K_exp_interp-5.png}
\caption{Time 5}
\end{subfigure}
\begin{subfigure}[c]{0.24\textwidth}
\includegraphics[width=1.0\textwidth]{\PathKExpInterp K_exp_interp-6.png}
\caption{Time 6}
\end{subfigure}
\begin{subfigure}[c]{0.24\textwidth}
\includegraphics[width=1.0\textwidth]{\PathKExpInterp K_exp_interp-7.png}
\caption{Time 7}
\end{subfigure}
\begin{subfigure}[c]{0.24\textwidth}
\includegraphics[width=1.0\textwidth]{\PathKExpInterp K_exp_interp-8.png}
\caption{Time 8}
\end{subfigure}
\begin{subfigure}[c]{0.24\textwidth}
\includegraphics[width=1.0\textwidth]{\PathKExpInterp K_exp_interp-9.png}
\caption{Time 9}
\end{subfigure}
\begin{subfigure}[c]{0.24\textwidth}
\includegraphics[width=1.0\textwidth]{\PathKExpInterp K_exp_interp-10.png}
\caption{Time 10}
\end{subfigure}
\begin{subfigure}[c]{0.24\textwidth}
\includegraphics[width=1.0\textwidth]{\PathKExpInterp K_exp_interp-11.png}
\caption{Time 11}
\end{subfigure}
\begin{subfigure}[c]{0.24\textwidth}
\includegraphics[width=1.0\textwidth]{\PathKExpInterp K_exp_interp-12.png}
\caption{Time 12} \label{fig:VisualizePriceImpactSpikeExpInterp:ell}
\end{subfigure}
\caption{Shows the result of a flash limit order being placed near the current price in the order book and corresponds to Sec. \ref{ssec:priceimpactlimit} using the non-uniform sampling time algorithm from Sec. \ref{ssec:updateequationarrivaltimes}. Each sub-figure, Fig. \ref{fig:VisualizePriceImpactSpikeExpInterp:a} to Fig. \ref{fig:VisualizePriceImpactSpikeExpInterp:ell}, is a snapshot of the system at simulation sampling events $1$ through $12$. In Fig. \ref{fig:STVisualizePriceImpactMarketOrder:d}, an impluse arrives and shocks the order book out of equilibrium.  The time between these simulation update events is $\Delta t_m$, and is non-uniform. This results in different log-price differences $\Delta x_m$ than found in the background grid, and these are indicated by marking the two points $p\pm \Delta x_m$ with black stars. The non-uniform sampling algorithm stores off grid points which approximate the on grid points ($\varphi$, shown as a solid blue) and these approximations are shown as dashed, and dashed-dotted, blues lines given in the legend as $\varphi^{i\pm1}$. The green line is the orders which are about to be added, and the gold lines those about to be removed.}
\label{fig:VisualizePriceImpactSpikeExpInterp}
\end{figure*}

Numerically, at the chosen time, a limit order with height $V$ is placed at the grid price point to the right of the current price (or on the grid point if the current price lies exactly on the grid point). This is visualised in Figure \ref{fig:VisualizePriceImpactSpikeExpInterp:b}. The system then returns to equilibrium as shown over the remaining frames. We note that the result of the spines in Figure \ref{fig:VisualizePriceImpactSpikeExpInterp:b} is to create ripples about the spike when it is first placed. One might think these would cause problems in the price impact diagram but, surprisingly, they do not. Additionally, the behaviour of the points on the jump lattice may appear erratic but the system is out of equilibrium and this behaviour is not unexpected.

We are interested in the effect of $\gamma$ on price impact. We consider Figure \ref{fig:VisualizePriceImpactSpikeExpInterp}. Suppose Figure \ref{fig:VisualizePriceImpactSpikeExpInterp:a} occurs at time $n$ in $L_0$. The resulting price impact plot is shown in Figure \ref{fig:PI-LO-CI-NU-E_} as a function of the size of the limit order $Q$ which occurs at Figure \ref{fig:VisualizePriceImpactSpikeExpInterp:b} and where each different colour line represents a different choice for $\Delta n$. The x-axis on this plot is normalised by the total volume of bids $A$ in the order book. This normalisation was chosen because there is no unique normalisation for a flash limit order, as there would be for a market order. 

The non-uniform time algorithm does have an effect on the price impact. To see this, note that we have highlighted, on the y-axis, the set of values defined by $b_{\Delta n} \equiv \{1/2 + (\Delta n-1):\Delta n \in \{1..7\}\}$. This ordinarily represents an upper bound on line $\mathcal{J}_{\Delta n}$ (\ref{app:appendix:additional}). At first glance this appears to still be the case in \ref{fig:VisualizePriceImpactSpikeExpInterp:b}, however we argue that it is not. Line $\mathcal{J}_{2}$ should be bounded above by $b_{2} = 3/2$ and but reaches $5/2$, passing smoothly through the $3/2$ barrier. Similarly, for the other two lines corresponding to greater $\Delta n$. However, each line does become bounded eventually but we propose that there is no upper bound for price impact lines in general. 

The explanation for this is simple: if one considers the non-uniform time algorithm given in Section \ref{ssec:updateequationarrivaltimes}, together with  Figure \ref{fig:visualisationofexpscheme}, one notices that any point may gather information about points a distance $\Delta x_m$ away. Thus, for large enough $\Delta x_m$, a point which is any distance away from the position of the flash limit order spike, may learn of its existence instantaneously. However, larger values of $\Delta x_m$ are suppressed given its relation to $\Delta t_m$ which is drawn from an exponential distribution, as described in Section \ref{ssec:updateequationarrivaltimes}. The specific set of upper bounds encountered in Figure \ref{fig:visualisationofexpscheme} are a result of the specific realisation of $\Delta t_n$ used in this simulation.

\subsubsection{Simulated market order}\label{ssec:priceimpactmarket}

\begin{figure*}[ht!]
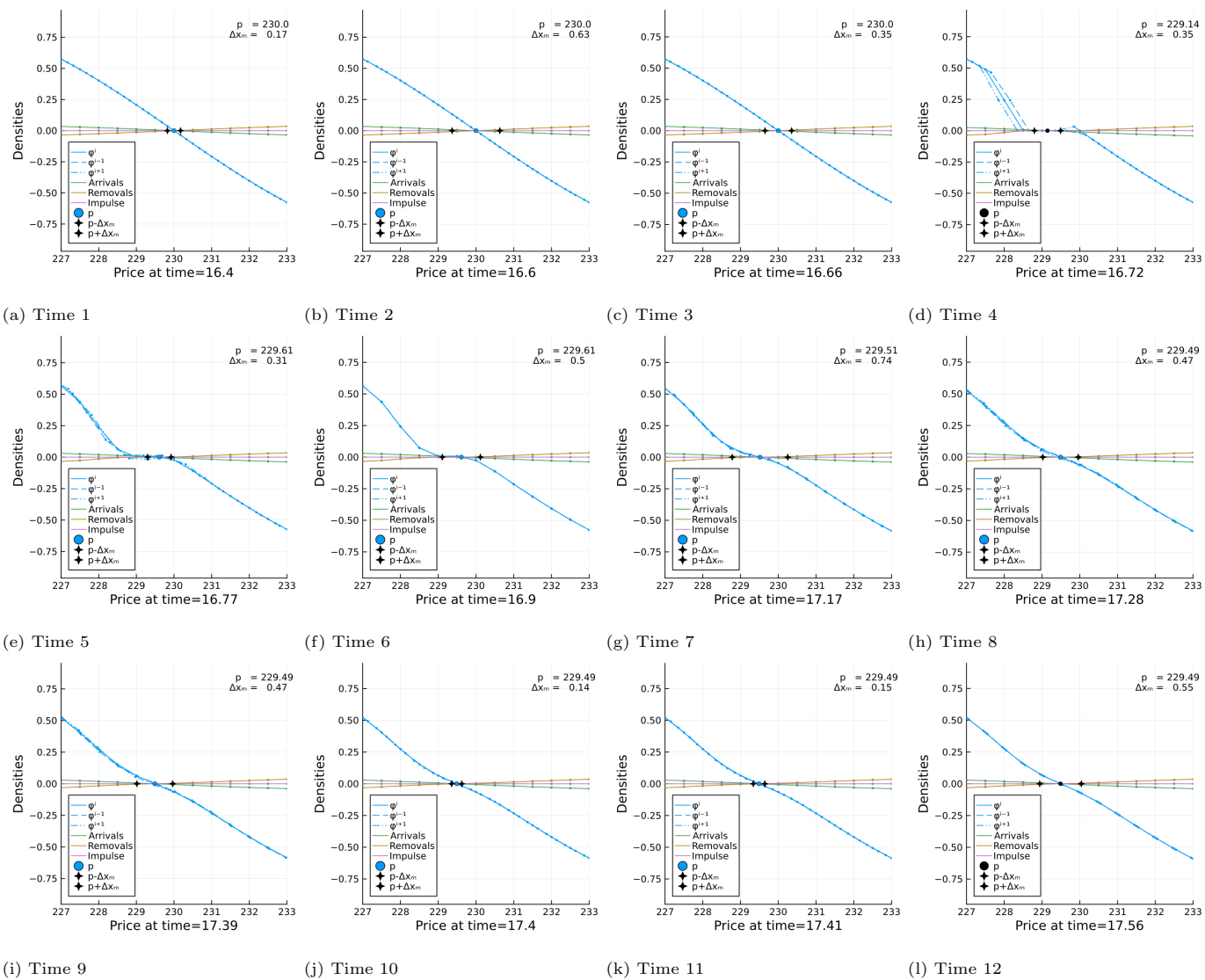

\centering
\begin{subfigure}[c]{0.24\textwidth}
\includegraphics[width=1.0\textwidth]{\PathKExpMO K_exp_MO-1.png}
\caption{Time 1}\label{fig:STVisualizePriceImpactMarketOrder:a}
\end{subfigure}
\begin{subfigure}[c]{0.24\textwidth}
\includegraphics[width=1.0\textwidth]{\PathKExpMO K_exp_MO-2.png}
\caption{Time 2}\label{fig:STVisualizePriceImpactMarketOrder:b}
\end{subfigure}
\begin{subfigure}[c]{0.24\textwidth}
\includegraphics[width=1.0\textwidth]{\PathKExpMO K_exp_MO-3.png}
\caption{Time 3}\label{fig:STVisualizePriceImpactMarketOrder:c}
\end{subfigure}
\begin{subfigure}[c]{0.24\textwidth}
\includegraphics[width=1.0\textwidth]{\PathKExpMO K_exp_MO-4.png}
\caption{Time 4} \label{fig:STVisualizePriceImpactMarketOrder:d}
\end{subfigure}
\begin{subfigure}[c]{0.24\textwidth}
\includegraphics[width=1.0\textwidth]{\PathKExpMO K_exp_MO-5.png}
\caption{Time 5}
\end{subfigure}
\begin{subfigure}[c]{0.24\textwidth}
\includegraphics[width=1.0\textwidth]{\PathKExpMO K_exp_MO-6.png}
\caption{Time 6}
\end{subfigure}
\begin{subfigure}[c]{0.24\textwidth}
\includegraphics[width=1.0\textwidth]{\PathKExpMO K_exp_MO-7.png}
\caption{Time 7}
\end{subfigure}
\begin{subfigure}[c]{0.24\textwidth}
\includegraphics[width=1.0\textwidth]{\PathKExpMO K_exp_MO-8.png}
\caption{Time 8}
\end{subfigure}
\begin{subfigure}[c]{0.24\textwidth}
\includegraphics[width=1.0\textwidth]{\PathKExpMO K_exp_MO-9.png}
\caption{Time 9}
\end{subfigure}
\begin{subfigure}[c]{0.24\textwidth}
\includegraphics[width=1.0\textwidth]{\PathKExpMO K_exp_MO-10.png}
\caption{Time 10}
\end{subfigure}
\begin{subfigure}[c]{0.24\textwidth}
\includegraphics[width=1.0\textwidth]{\PathKExpMO K_exp_MO-11.png}
\caption{Time 11}
\end{subfigure}
\begin{subfigure}[c]{0.24\textwidth}
\includegraphics[width=1.0\textwidth]{\PathKExpMO K_exp_MO-12.png}
\caption{Time 12} \label{fig:STVisualizePriceImpactMarketOrder:ell}
\end{subfigure}
\caption{Shows the result of a large market order arriving and corresponds to Sec. \ref{ssec:priceimpactmarket} and uses the non-uniform sampling time algorithm from Sec. \ref{ssec:updateequationarrivaltimes}. Each sub-figure, Fig. \ref{fig:STVisualizePriceImpactMarketOrder:a} to Fig \ref{fig:STVisualizePriceImpactMarketOrder:ell}, is a snapshot of the system at a certain time. In Fig. \ref{fig:STVisualizePriceImpactMarketOrder:d}, a large market order to sell arrives which is executed against the set of current offers to buy. The time between simulation updates $\Delta t_m$ and is a non-uniformly sampled. This results in a different $\Delta x_m$ than found in the background grid and these are indicated by marking the two points $p\pm \Delta x_m$ with black stars. The non-uniform algorithm stores off grid points which approximate the on grid points ($\varphi$, shown in blue) and these approximations are shown as the dashed, and dashed-dotted, blue lines, and in the legend as $\varphi^{i\pm1}$. The green line is the orders which are about to be added via the source term, and the gold line is the orders which are about to be removed.}
\label{fig:STVisualizePriceImpactMarketOrder}
\end{figure*}

We computed the price impact function from the market order impulse response. A market order is an order that some amount of volume be bought (or sold) at the prevailing best ask (bid). For larger market orders, this may cause one to ``walk the order book" as bids (offers) at higher (lower) prices are consumed in order to execute the order. This will allows us to model the price impact. 

Here a market order of volume $Q$ is executed against the order book at time $t$ as a single child order. This requires finding the depth $x_t(Q)$ at which the order will complete its execution given the prevailing mid-price $p_t$ and available liquidity $Q = \pm \int_{p_t}^{\pm x_t} \varphi(x,t) dx$
and to then remove (add) the required orders to the order book profile at time $t$ for the mid-price $p_t$ for buying (+1) or selling (-1):
\begin{equation}
   s_{_\text{MO}}(x,t) = \mp \int_{p_t}^ {\pm x_t(Q)} \varphi(x,t) dx.
\end{equation}
Numerically, at the corresponding time, an order with target volume $Q$ arrives. We then begin from the current value of numerical intercept defined by using straight line interpolation. Call the value of this intercept $p$. We then proceed to integrate numerically from $p$ in the desired direction until we find a position $p'$ such that the area between $p$ and $p'$ is equal to $Q$. 

Let $p'$ lie between grid point $x_{i}$ and $x_{i+1}$ where $x_{i}$ is between $p$ and $p'$, while $x_{i+1}$ is just outside this range. Set $\varphi$ to be zero at all grid points between $p$ and $p'$. Then consider the proportion $\xi=|(p'-x_{i+1})/\Delta x|$, which is at most unity and goes to zero as $p'$ approaches the grid point $x_{i+1}$. We then multiply $\varphi_{x_{i+1}}$ by $\xi$. This ensures that the furthest grid point goes smoothly to zero as the size of $Q$ increases, and is necessary to obtain a smooth price impact function. This is visualised in Figure \ref{fig:STVisualizePriceImpactMarketOrder:b}. The system then returns to equilibrium as shown over the remaining frames. 

\begin{figure*}[ht!]
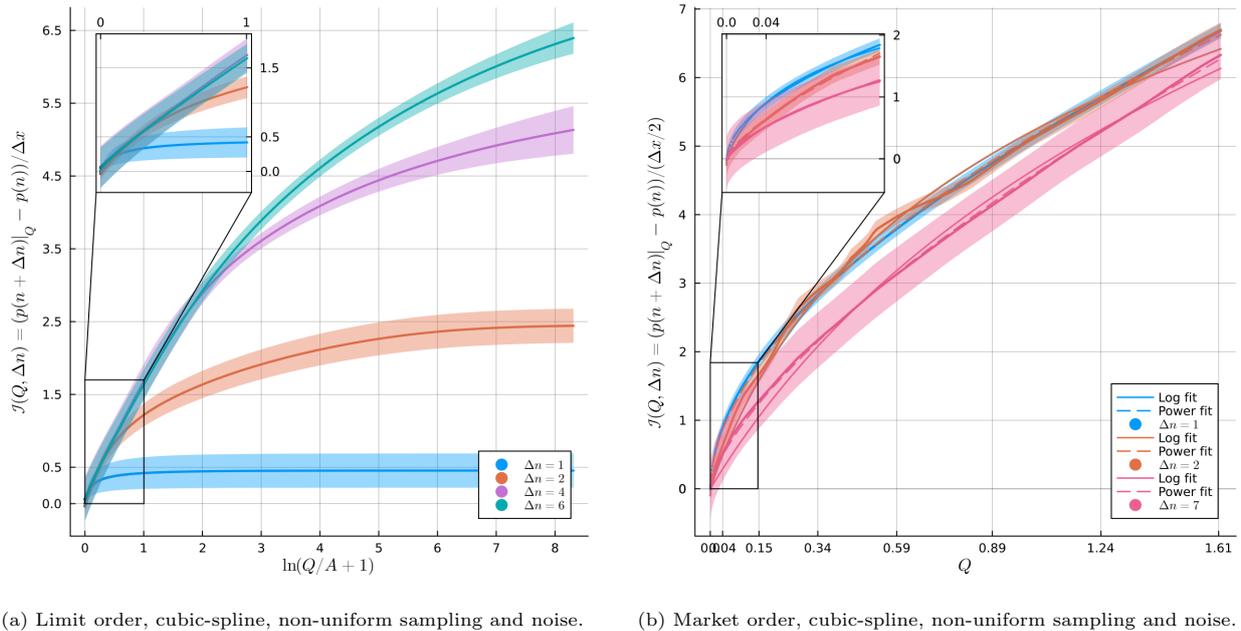

\centering
\begin{subfigure}[c]{0.45\textwidth}
\includegraphics[width=\textwidth]{\PathSingles PI_DD_Exp-Int-uncert.png}
\caption{Limit order, cubic-spline, non-uniform sampling and noise.}
\label{fig:PI-LO-CI-NU-E_}
\end{subfigure}
\begin{subfigure}[c]{0.45\textwidth}
\includegraphics[width=\textwidth]{\PathSingles PI_DD-MO_exp-Int-general-uncert.png}
\caption{Market order, cubic-spline, non-uniform sampling and noise.}\label{fig:PI-MO-CI-NU-E_}
\end{subfigure}
\caption{Fig. \ref{fig:PI-LO-CI-NU-E_} shows the price impact for a flash limit order (see Sec. \ref{ssec:priceimpactlimit}) with intercept $p$ estimated using cubic-spline interpolation when simulated using the non-uniform sampling scheme (see Eqn. \ref{eq:UpdateEquationExpArrival}). Fig. \ref{fig:PI-MO-CI-NU-E_} shows the price impact for a market order (see Sec. \ref{ssec:priceimpactmarket}) with the same configuration. In both plots, 10 replications were used with different noise paths $V_t$ to introduce uncertainty. These graphs show how the price impact plots for the two order types are affected by using non-uniform sampling (Sec. \ref{ssec:updateequationarrivaltimes}) and by the choice of the sampling rate free-parameter $\gamma_1$ (Sec. \ref{ssec:timescales}). Here we have $\gamma_1 = \Delta n$ where each line in each figure shows the price impact measured after a delay of $\Delta n \in \mathbb{N}^{+}$ time steps in $L_0$ (see Sec. \ref{ssec:timescales}) since the placement of the order. This line is defined as $\mathcal{J}_{\Delta n} \equiv \{(Q,\mathcal{J}(Q,\Delta n)):\forall Q\}$. The function $\mathcal{J}(Q,\Delta n)$ is different for the left and right figures above. Each $\mathcal{J}_{\Delta n}$ corresponds to the price impact one would observe for different choices of $\Delta n$. In Fig. \ref{fig:PI-MO-CI-NU-E_} 
we include a power fit and a log fit to each line $\mathcal{J}_{\Delta n}$. In Fig. \ref{fig:PI-LO-CI-NU-E_} each line $\mathcal{J}_{\Delta n}$ are plotted against the log of the volume of the flash limit order and normalised by the total area under the bid side $A$ (Sec \ref{ssec:priceimpactlimit}). Some $\mathcal{J}_{\Delta n}$ pass through one or more of the points defined by $b_{\Delta n} \equiv \frac{1}{2}+(\Delta n-1)$ and are not bounded above by $b_{\Delta n}$ (see $\Delta n = 2$). This is unlike the case for linear interpolation where there are kinks (for example see Fig. \ref{fig:PI-LO-LI-U_-NE} and Fig. \ref{fig:PI-LO-CI-U_-NE}). There is also some interval of volumes, starting from zero up to some maximum volume, for which $\mathcal{J}_{\Delta n}$ is approximately linear (on the log scale) and that the size of this interval increases for larger $\Delta n$. The inset shows this difference for $\Delta n=1$ and $\Delta n=2$. This demonstrates how the choice of the sampling rate from the underlying model will influence the ability of flash orders to change the price of the order book (see Sec. \ref{ssec:sampling}).  
}
\label{fig:PriceImpact-AllBellsAndWhistlesUncert}
\end{figure*}

We are again interested in the effect of $\gamma$ on price impact. We consider Figure \ref{fig:STVisualizePriceImpactMarketOrder}. Suppose Figure \ref{fig:STVisualizePriceImpactMarketOrder:b} occurred at time $n$ in $L_0$.  The resulting price impact plot is shown in \ref{fig:PI-MO-CI-NU-E_} as a function of the size of the market order $Q$ which occurs in Figure \ref{fig:STVisualizePriceImpactMarketOrder:d} and where each different colour line represents a different choice for $\Delta n$.

We then fit two functions to the resulting data in \ref{fig:PI-MO-CI-NU-E_}. First, we fit a power law of the form $\Delta p(Q) = a Q^b$ where $a$ and $b$ are fitted parameters. Then we fit a logarithmic function of the form $\Delta p(Q) = c \ln(1+dQ)$. By fitting the parameter $d$ we have removed the need to declare the normalisation of $Q$, but this presents its own problems. Most importantly, over any finite range from zero to some upper bound, a function of the form $c \ln(1+dQ)$, having two non-trivial fit parameters, can mimic the shape of the square root (or power law) to fitting accuracy. The log fit will intercept the square root fit (or power law) twice and will pass from below, to above, to below again. We include Figure \ref{fig:PI-MO-CI-NU-NE} in \ref{app:appendix:additional} to make this more clear, which is simply Figure \ref{fig:PI-MO-CI-NU-E_} without any uncertainty and was obtained by running the simulation once without $V_t$.

The most obvious difference between market orders in Figure \ref{fig:PI-MO-CI-NU-E_} compared to the price impacts of limit orders given in Figure \ref{fig:PI-LO-CI-NU-E_} is that each line in the price impact of market orders is inherently unbounded above. Indeed, the market order may consume as much of the order book as it wishes to, and it is not hard to see the instantaneous impact of this is going to be approximately square root. This is what we see in the line $\mathcal{J}_{1}$ in Figure \ref{fig:PI-MO-CI-NU-E_}. This is the line which makes the strongest case for not being able to be fit using a log function. In general, one can expect not to be able to tell the power law and log fits apart because of the noise in the data. It should be noted that for this $\mathcal{J}_1$, the diffusion dynamics have yet to have any effect on the recorded price and the entire price movement is due to the way the market order consumes limit orders. 

If one allows some time to pass in $L_0$ ($\Delta n = 2$) the price impact (orange) is initially different from the blue but aligns with it for sufficiently large volumes. This makes sense as it corresponds to the idea that the flat region in Figure \ref{fig:STVisualizePriceImpactMarketOrder:d} can become large enough that the new mid price will not be reached by the propagation of the unaffected prices (in a sense very similar to that explained in \ref{app:PriceImpactKinks}) for more than one step in $L_0$. The dynamics of the system do not reach the price for an entire step in $L_0$. If one waits for $\Delta n=7$ times steps, one sees an even further deviation from square-root behaviour with a fitted power law with exponent of approximately $0.7$. We further note the presence of kinks once again  \footnote{One can imagine modifying Fig.\ref{fig:KinkVisualisation} such that the blue bars have vanished. This then corresponds to the visualisation of market orders in Fig. \ref{fig:STVisualizePriceImpactMarketOrder:d} in the region where $\varphi$ has been zeroed. The red in Fig. \ref{fig:KinkVisualisation} then corresponds to the first non-zero value of $\varphi$ in Fig. \ref{fig:STVisualizePriceImpactMarketOrder:d}, to the left of the current price, spreading towards the current price. One then obtains kinks in a similar way as before.}. 

\subsection{Measuring the volatility and trading rate} \label{ssec:dailyvolume}

Consider a power law impact function \cite{ATHL2005,TLDLJB2011,MTBPRL2014,BILL2015,KO2016,BB2018}: 
\begin{equation}\label{eq:powerlawpriceimpact}
F(Q) = Y \sigma_D \left( \frac{Q}{V_D} \right)^{\delta} =  \left(Y \frac{\sigma_D}{(V_D)^\delta}\right) Q^{\delta}
\end{equation}
Here, one sees the daily volatility $\sigma_D$, the daily volume traded $V_D$, a power $\delta$ and a factor $Y$ which is of order unity. We write the equation in the second form above in order to make a comparison below.

To obtain data from which we can measure the model equivalent of the Daily Volume (DV) traded $V_D$, and then the associated mid-price volatility, over the same time period, $\sigma_D$, we run our model with the default parameters for simulated equivalent of 8 hours of trading, as described in Section \ref{ssec:stylisedfacts}. However, we do this four more times, with different seeds. Combining this with the data we already have, this means we have the equivalent (on a volume traded basis) of 5 ``days" worth of 8 hour trading sessions of typical equity market data\footnote{This was generated using five seeds: \{3535956730, 4898384128, 5554355463, 0586258657, 3453348462\}}. 

The instantaneous trading rate is needed to measure total volume  $V_D$ traded during the simulation session that represents 8 hours of trading. The trading rate is integrated over the desired period, where the number of trades taking place per unit time is given by $D \partial_x \varphi(x,t)|_{x=p_t}$ which leads to a total daily volume traded given by Equation \ref{eq:volumetraded} \cite{DBMB2014,BB2018}. We compute this expression numerically. In equilibrium, the derivative is constant for all time and across all days, and so together with the diffusion $D$, one need only declare the amount of time over which one wishes to measure. The value of $D \partial_x \varphi(x,t)|_{x=p_t}$ was found to be $0.0957$. This then means that the total volume traded in a simulated 8 hour trading day is given by $V_D = 0.0957 \times 25000 = 2392.5$.

Now we divide our five simulated trading days of data up into single slices equivalent to simulated hours, which gives us $32$ ``one hour" samples. For each sample we compute the log price difference from the beginning of the period to the end. We plot the results in Figure \ref{fig:volatilitydata}. From this we get an hourly sampled price fluctuation variance $\sigma^2_{h}$. To compute daily price fluctuation variance $\sigma^2_D$ we use: $\sigma_D=\sqrt{8} \sigma_{h}$. We note that this is an approximation because we have anomalous diffusion with $\alpha=0.8$; however, we only wish to check whether the proportionality constant $Y$, in Equation [\ref{eq:powerlawpriceimpact}] is approximately of order unity, and so we will use this approximation. The value of $\sigma_{h}$ was found to be 52.02 and so we obtain $\sigma_D\approx147.15$.

From Section \ref{ssec:priceimpactmarket} and table \ref{tab:fitforpriceimpact}, we know the values of the scale and power parameters in the power law fit of the market impact shown in Figure \ref{fig:PI-MO-LI-U_-NE}. This corresponds to $\delta$ and the number in large brackets in equation [\ref{eq:powerlawpriceimpact}]. 
This means we may use equation [\ref{eq:powerlawpriceimpact}] to solve for $Y$ such that the numerical fit matches this equation. Upon doing so we obtain $Y=4.58712\pm0.00096$. 

\section{Conclusions} \label{sec:conclusion}

Starting with the Discrete Time Random Walks (DTRW) numerical scheme proposed by \citet{ADHJL2015} and extending this to include a forcing term we then demonstrate that our version of the numerical scheme generates the correct diffusion behaviour over time. This is shown in tables \ref{tab:measuredspikevarparameters} and \ref{tab:measuredspikevarparameterssmalldx} - visually in figures \ref{fig:IncreasingVarianceOfSpike} and \ref{fig:IncreasingVarianceOfSpikeSmallDx}. We show the variety of stylised facts which can result from the scheme using different model parameters; this can be seen in Figure \ref{fig:StylizedFactsDifferentDx} (with additional detail in Figure \ref{fig:StylizedFactsDifferentDiffusions}). This highlights the limitations of the model.

The simulations are a discrete numerical approximation (equations \ref{eq:FinalUpdateEquation} and \ref{eq:UpdateEquationExpArrival}) for a continuum of order book events (Equation \ref{eq:MainPDE}) where each order book event results in a mid-price (Equation \ref{eq:midprice}). Trading in the model is itself unobservable because the model represents the order book as bids and offers where the annihilation interface boundary instantaneously has zero bids or offers at the prevailing mid-price. Another view on this problem is that the trading rate is constant in time, and given by Equation \ref{eq:volumetraded}, which means there are no distinct trade events.

The solution of Equation \ref{eq:MainPDE} is conceptually a continuous object. The simulation generates approximate solution of this PDE via equations \ref{eq:FinalUpdateEquation} and \ref{eq:UpdateEquationExpArrival}, but model still needs to be map to observable events which are themselves a discrete process. Although the discretised update equation is consistent with the continuous time representation in appropriate limits there is a model ambiguity under sampling. This sampling problem is discussed in Section \ref{ssec:sampling}. The exact set of points any numerical scheme gives are arbitrary, here dependant on the choice of $\Delta t$, and therefore on an unknown scale, and thus one must choose the sampling that will index and extract events out of the scheme. This freedom of interpretation was discussed in Section \ref{ssec:timescales}. This freedom introduces an implicit hyper parameter that should be carefully considered within a calibration framework. 

The interpretation we adopt in this paper is that we are sampling the order book at externally imposed trade event times. This is because we want to compute price impact curves where the volume of trade events are associated with mid-price changes. These trade events are labelled $\ell$ and sampled as integer values from the underlying simulation time $t$. This then requires that we fix the average ratio of the waiting times between the sampling times representing trade events $\delta t$ (See Equation \ref{eq:gamma2}) with the time increments $\Delta t$ in the background order book simulation model. 

To ensure that the resultant mid-prices preserve the correct diffusion properties, Equation \ref{eq:latticesize} must hold. This means that the choice of time increments in the background, $\Delta t$, fixes the log price increments $\Delta x$, and that we need to ensure that the sampling preserves the properties of this diffusion process. For this reasons our scheme has two types of sampling. First, we demonstrate the simulation using {\it uniform sampling} (Equation \ref{eq:FinalUpdateEquation}). Second, we extend the scheme to accommodate {\it non-uniform time sampling} (Equation \ref{eq:UpdateEquationExpArrival}). 

In this setting we can use random walks with differing diffusion scaling parameter values, $\alpha$, and compare them by sampling each appropriately. This results in distributions with differing return variances and auto-correlation strengths, but consistent diffusions (this is shown in Figure \ref{fig:StylizedFactsDifferentDiffusions}). We then show that we naturally obtain auto-correlations that can match those seen in real markets when made in conjunction with the introduction of auto-correlations into the random walk, $V_t$, that represents exogenous information arrival. This is reported here and can be seen in the stylised fact plots in Figure \ref{fig:StylizedFactsDifferentDx} (and more generally in \ref{app:appendix:additional}, and Figure \ref{fig:StylizedFactsDifferentDiffusions}).

The desired behaviour is that order flow correlations tend to die off much slower than auto correlations in the absolute value of the log returns. One might also expect that the information arrival self-correlation could be used to produce more extreme events however they do not. We show that, despite using a strong self-correlation of $\rho=0.9$, the tails of all the return distributions are very light\footnote{One might think that making the distribution of $V_t$ heavy tailed will result in the returns being so too. This is not unexpected given the underlying diffusive nature of the numerical scheme.}. This is not due to sub-diffusion (as this remains true in Figure \ref{fig:StylizedFactsDifferentDiffusions} where one of the lines represents $\alpha = 1.0$). The model does not appear to contain enough self-dependence to introduce volatility clustering into the returns time series despite having an information arrival, driving potential with auto-correlations of $\rho=0.9$ and $\alpha=0.8$ (even $\alpha = 0.6$ in Figure \ref{fig:StylizedFactsDifferentDiffusions}). This implies that although we can weakly recover the slow decay in trade-sign auto-correlations relative to those in the absolute value of the returns, this is insufficient and we do not recover the appropriate level of extreme events because of the strongly diffusive nature of the numerical scheme.

Despite this, the key contribution of this work is to demonstrate the reasonableness of the price impact functions for different model parameter configurations when sub-sampling the simulation lattice in the presence of: 1.) a flash limit order, and 2.) a market order in a simulated lit order book. These are shown in Figure \ref{fig:PI-LO-CI-NU-NE}. In this model these two case can be difficult to tell apart when they both have small parent orders, but the choice of which to use to push the system out of equilibrium can have strong numerical effects because the first case is bounded, and the second case is not. This is important, because both power law and log fits seem reasonable for the simulated models. Here the fitting of a power law, relative to a log price impact functional form, is dependent on the sampling scale hyper-parameter, which follows from the requirement to map the continuous time model to discretely sampled data, and to numerically simulate this on a discrete grid. One must say more about the scales involved. 

Although we can provide a heuristic check of the commonly cited price impact formula, Equation \ref{eq:powerlawpriceimpact}, and find that the proportionality factor of $Y$ is of order unity -approximately $4$. Which means that we have a reasonable fit. Significantly more work is required. The key point here is that this highlights the importance of the impact of different sampling choices, when sampling the underlying model. We can find smooth and empirically acceptable price impact curves when sampling at rates lower than that of the underlying model.   

Concretely, we investigate the impact of the choice of the ratio of trade sampling events to underlying simulated event times {\it i.e.} the impact of the choice of $\gamma_2$ (Equation \ref{eq:gamma2}) on the observable price impact curves for the two types of parent orders that have immediate execution. We find, for flash limit orders that, the more simulation steps you allow between every trade, the higher the bound on the spread of the limit order and the longer the time for which it is linear on a log scale. For market orders, this effect is lessened. 

In all cases we observe kinks in the price impact, that are a result of the sampling inherent in any numerical scheme needed to solve PDEs. We observed kinks in the price impact plots that resulted from flash limit orders (Section \ref{ssec:priceimpactlimit}) and we were able to explain analytically why the kinks occur (Section \ref{app:PriceImpactKinks}) and the associated boundedness in price impact plots that results from the uniform sampling numerical scheme. Further, we proposed a new scheme which allows for time to become a statistical process in and of itself, and then showed that this scheme (with cubic spline interpolation) was able to remove the kinks and boundedness of flash limit orders, and produced comparable results to the uniform sampling scheme in the case of market orders. The resampling of the uniform process defines a new stochastic process, we speculate that this new process should itself have a well defined continuum limit for certain choices of the non-uniform sampling but do not show that this is in fact the case. 

The computation cost of the method is important for large scale simulation work. We provide a formula for how the complexity of computation grows with decreasing $\alpha$ (Equation \ref{eq:computecomplexity}). From this we find that, even using a finite memory kernel, there is a rapid jump in complexity from $\alpha=1.0$ to anything less, then moderate growth in complexity until $\alpha=0.6$. Beyond this, computation times quickly become intractable for sub-diffusion's. Using the full kernel is generally infeasible for long running times with most $\alpha$ values.

With this computational cost in mind, we recommend using an $\alpha=1.0$ model because there is no clear advantage provided in terms of the sub-diffusion's contribution to the stylised facts. In addition, we recommend the use of the non-uniform time algorithm as it conceptually simplifies the sampling requirements while retaining the desired features of the stylised facts. Together, these may both be able to be used for future calibration and simulation work. We again note, however, that neither will be able to produce effects which rely on memory, such as required to recover appropriate volatility clustering. As such, the model in its current form needs additional features to generate extreme events and volatility clustering.

\section{*Acknowledgements} \label{sec:acknowledgements}
We thank Salahuddeen Ahmad and Etienne Pienaar for thoughtful comments. We thank Chris Angstmann for many helpful conversations, and Byron Jacobs for encouragement. 


\bibliography{Anomalous}

%
%
\appendix
\section{Anomalous diffusion} \label{app:appendix:diffusion}

\tocless\subsection{Anomalous diffusion update equation}
\label{app:AnomalousDerivation}

Reaction-diffusion systems' can be modelling in terms of creation and annihilation processes \cite{ANGSTMANN2016508}. First, we assume the order annihilation process on a discrete time and price grid is an in-homogeneous Poisson process with rate parameter\footnote{Here $a(x,t) = \mlm a_{\Delta}(x_i,t_n)$}: $a_\Delta(x,t)$. This means the probability of no annihilation, or order cancellation, between times $t_m=m \Delta t$ and time $t_{n-1}=(n-1) \Delta t$ prior to the update time $t_n$, at price point $x_i - i \Delta x$ on the lattice is given by the survival probability function $\theta^{i}_{m,n-1}$:
\begin{equation}
    \theta^{i}_{m,n-1} = \exp \left( - \int_{t_m}^{t_{n-1}} 
 a_\Delta (x_i, t') dt' \right). \label{eq:survivalfn}
\end{equation}
To make it to time $t_n$, the order (particle) must survive one more time step. The probability it will be annihilated between time $t_{n-1}$ and time $t_n$ is then given by: $A^{i}_{n-1} = 1 - \theta^i_{n-1,n}$.

We now consider the entire system. There are orders that arrive (creation), orders that survive removal (annihilation), and orders that move from different price positions on the surrounding order-book lattice, to the current price $x_i$, due to random effects over the time increment $\Delta t$ ({\it i.e.} from time $t_{n-1}$ to time $t_n$). From these we subtract orders that have survived from prior times up to time $t_{n-1}$, at price $x_i$ and then find the overall remaining order density. Using such a discrete master equation formulation of the anomalous reaction-diffusion equation at time $t_n$ on a time and price grid, we can consider the overall density of orders $\varphi^i_n$ at price $x_i$ \cite{ANGSTMANN2016508}:
\begin{align}
    \varphi^{i}_{n} &=  C^{i}_{n-1} + \theta^i_{n-1,n} \varphi^{i}_{n-1} \nonumber \\ 
    &+\sum_{j=-1}^{j=+1} \lambda^{i|i+j}_{n-1} \sum_{m=0}^{n-1} K_{n-m} \theta^{i+j}_{m,n-1} \varphi^{i+j}_{m} \nonumber \\
    &-\sum_{m=0}^{n-1} K_{n-m} \theta^{i}_{m,n-1} \varphi^i_m.
    \label{eq:Update1}
\end{align}
Here $K_{n-m}$ is a memory kernel for a process with Sibuya waiting times \cite{ANGSTMANN2016508} which defines the memory due to the fractional diffusion. 

Random effects are described by right, self and left transition rates, back to state $i$ from price states $i+j$ for $j \in [-1,0,+1]$. These are probabilities on the lattice $\lambda^{i|i+j}_{n}= P^{(-j)}_{i,n}$ so that:
\begin{equation}
    \lambda^{i|i+j}_{n} = P^{(+1)}_{i,n} \delta_{i-1,i+j}+P^{(0)}_{i,n} \delta_{i,i+j}+P^{(-1)}_{i,n} \delta_{i+1,i+j}. \label{eq:TransitionProb1}
\end{equation}
Here $P^{(+1)}_{i,n}$, $P^{(0)}_{i,n}$, and $P^{(-1)}_{i,n}$ are the probabilities of making a right jump, self jump, and a left jump, respectively: 
\begin{equation}
\sum_{k=-1}^{+1} P^{(k)}_{i,n}=1. \label{eq:DiffusionProbNorm}
\end{equation}
We allow the probabilities to depend on the time $t_n$, but are independent of the the lattice price location, $x_i$; so we can drop the price index $i$. The time dependence is necessary because the forcing function will be time dependent. 

Substituting Equation [\ref{eq:TransitionProb1}] into Equation [\ref{eq:Update1}]:
\begin{equation} 
\begin{aligned} \label{eq:UpdateRuleFinal}
    \varphi^{i}_{n}& = \sum_{m=0}^{n-1} K_{n-m} \biggl[P^{(+1)}_{n-1} \theta^i_{m,n-1} \varphi^{i-1}_{m} 
    +P^{(-1)}_{n-1} \theta^i_{m,n-1} \varphi^{i+1}_{m} \\ 
    &+(P^{(0)}_{n-1}-1) \theta^i_{m,n-1} \varphi^i_m \biggr] 
    +\theta^i_{n-1,n} \varphi^{i}_{n-1} + C_{i,n-1}.
\end{aligned} 
\end{equation}
This is the update equation scheme proposed in \cite{ANGSTMANN2016508}. We can also write the density of created orders $C^i_n$ as a creation rate $c^i_{n-1}$ on the grid: $C^i_n=c^i_{n-1} \Delta t$.

\tocless\subsection{Deriving the Partial Differential Equation} 
\label{app:derivePDE}

Starting with the update Equation [\ref{eq:UpdateRuleFinal}] we subtract $\varphi^i_{n-1}$ from both sides and use that the density of annihilated orders at time $t_{n-1}$: $A^{i}_{n-1} \varphi^i_{n-1} = (1 - \theta^i_{n-1,n})\varphi^i_{n-1}$:
\begin{align}
    \varphi^{i}_{n} - \varphi^{i}_{n-1} &= \sum_{m=0}^{n-1} K_{n-m}  \biggr[P^{(+1)}_{n-1} \theta^{i-1}_{m,n-1} \varphi^{i-1}_{m} \nonumber \\
    &+P^{(-1)}_{n-1} \theta^{i+1}_{m,n-1} \varphi^{i+1}_{m} +(P^{(0)}_{n-1}-1)\theta^{i}_{m,n-1} \varphi^{i}_{m}\biggr] \nonumber \\
    &-A^{i}_{n-1} \varphi^{i}_{n-1} + C^{i}_{n-1}.
\label{eq:WalkForNumericalScheme} 
\end{align}
The survival function  of the cancelled orders from Equation [\ref{eq:survivalfn}] can be factored:
\begin{equation}
\theta^{i}_{m,n-1} =  \theta^i_{m,k} \theta^i_{k,n-1} = \frac{\theta^{i}_{k,n-1}}{\theta^{i}_{k,m}}. \label{eq:factorsurvival}
\end{equation} 
Here $m \leq k \leq n-1$. The numerator no longer depends on $m$. Substituting this and factoring the terms that don't depend on $m$ out of the summations:
\begin{align}
    \varphi^{i}_{n} - \varphi^{i}_{n-1}& = P^{(+1)}_{n-1} \theta^{i-1}_{k,n-1}\sum_{m=0}^{n-1} K_{n-m}  \frac{\varphi^{i-1}_{m}}{\theta^{i-1}_{k,m}} \nonumber \\
    & +P^{(-1)}_{n-1} \theta^{i+1}_{k,n-1} \sum_{m=0}^{n-1} K_{n-m}  \frac{\varphi^{i+1}_{m}}{\theta^{i+1}_{k,m}} \nonumber \\
    &+(P^{(0)}_{n-1}-1) \theta^{i}_{k,n-1} \sum_{m=0}^{n-1} K_{n-m}  \frac{\varphi^{i}_{m}}{\theta^{i}_{k,m}} \nonumber \\
    &-A^{i}_{n-1} \varphi^{i}_{n-1} + C^{i}_{n-1}. \label{eq:derivePDE1}
\end{align}
At this point we can introduce the forcing functions, which are defined as: 
\begin{equation}
f_{\Delta}(t_n) = \tfrac{1}{r \beta \Delta x}(P^{(+1)}_{n} - P^{(-1)}_{n}).
\end{equation}
Here $r$ is a constant representing the probability to change position due to diffusion, $\beta$ is an inverse temperature, and $\Delta x$ the price (spatial) grid size. This means that the force opposes the diffusion jumps when they occur (See [\ref{app:force}]). We choose a system of probabilities such that the probability of a self jump is always $1-r$. We then write 
\begin{equation}\label{eq:deltaforce}
    F_n = r \beta \Delta x f_{\Delta} = P^{(+1)}_{n} - P^{(-1)}_{n}. 
\end{equation}
Combining this information with the forcing function definition we obtain expressions for each jump probability where we see that stochastic force will positively (negatively) bias the jumps to the current price grid location $x_i$ depending whether the jump is to the right (left) of the current price: 
\begin{align}
    P^{(+1)}_{n} &= \tfrac{1}{2}(r + F_{n}), \\
    P^{(0)}_{n} &= 1 - r, \mbox{ and } \\
     P^{(-1)}_{n} &= \tfrac{1}{2}(r - F_{n}). 
\end{align}
Thus the probability of self jumps never changes, but the the probability for jumps left or right may be exchanged with one another which introduces the affect of the forcing function. 

It is convenient to define:
\begin{equation} 
\Phi^{i}_{k,n-1} \equiv \sum_{m=0}^{n-1} K_{n-m}  \frac{\varphi^{i}_{m}}{\theta^{i}_{k,m}}.
\end{equation}
Substituting this into Equation [\ref{eq:derivePDE1}] to find:
\begin{align}
\varphi^{i}_{n} - &\varphi^{i}_{n-1} = C^{i}_{n-1} -A^{i}_{n-1} \varphi^{i}_{n-1} \nonumber \\ 
&+ \tfrac{r}{2} \left[{ \theta^{i-1}_{k,n-1} \Phi^{i-1}_{k,n-1} - 2 \theta^{i}_{k,n-1} \Phi^{i}_{k,n-1} + \theta^{i+1}_{k,n-1} \Phi^{i+1}_{k,n-1}}\right] \nonumber \\
&+ \tfrac{1}{2} \left[{F_{n-1} \theta^{i-1}_{k,n-1} \Phi^{i-1}_{k,n-1} - F_{n-1}  \theta^{i+1}_{k,n-1}  \Phi^{i+1}_{k,n-1}}\right].
\end{align}
The approach is to now implement a discrete Laplace transform over the Dirac combs on the uniform and discrete price and time lattice, and to then inverse Laplace transform this back to the continuous order volume and time domain on the real line. This will then recover a partial differential equation \cite{ANGSTMANN2016508}. We first take the unilateral star transform: $\{Y^i_m\}^*_s= \mcl{Z}_n^{*}\{ {Y^i_m |s, \Delta t}\}$ (unilateral on $t \in \R^+$) with respect to the time lattice points $t_n$ indexed by $n$. Then another unilateral star transform: $\{ Y^i_m \}^*_q = \mcl{Z}_i^* \{ { Y^i_m |q,\Delta x} \}$ (unilateral on $x \in \R^+$) with respect to price (space) lattice points $x_i$ indexed by $i$. The prices are on $[0,+\infty]$ we do not need to use a bilateral transform for the spatial domain. For all variables $Y$ on the lattice: $e^{-k s \Delta t} {Y^i_m}^* =  {Y^i_{m-k}}^*$, and $e^{-k q \Delta x} {Y^i_m}^* = {Y^{i-k}_{m}}^*$. We write the removed and created orders in terms of annihilation and creation rates over the increment: $C^i_{n-1} = c^{i}_{n-1} \Delta t$ and $A^i_{n-1} = a^i_{n-1} \Delta t$. We shift all the variables onto the same price and time points on the lattice to then obtain:
\begin{widetext}
\begin{align}
\left({1- e^{- s \Delta t}}\right)& \biggl\{ \biggl\{ {\varphi^i_n} \biggr\}^*_{s} \biggr\}^{*}_{q} 
=  e^{- s \Delta t}\biggl\{ \biggl\{ {c^{i}_{n} \Delta t} \biggr\}^*_{s} \biggr\}^{*}_{q}  - e^{- s \Delta t} \biggl\{ \biggl\{ {a^{i}_{n} \Delta t \varphi^{i}_{n}} \biggr\}^*_{s} \biggr\}^{*}_{q} \negthickspace \nonumber \\
&+ \frac{r}{2}e^{- s \Delta t} \left( {e^{-q \Delta x} - 2 + e^{+q \Delta x}} \right) \biggl\{ \biggl\{ {\theta^{i}_{k,n} \Phi^{i}_{k,n} } \biggr\}^*_{s} \biggr\}^{*}_{q} 
+\frac{1}{2} e^{- s \Delta t} \left( { e^{-q \Delta x} - e^{+q \Delta x} } \right) \biggl\{ \biggl\{ {F_{n} \theta^{i}_{k,n} \Phi^{i}_{k,n}} \biggr\}^*_{s} \biggr\}^{*}_{q}. \label{eq:SameSpaceTimePoints}
\end{align}
\end{widetext}
We write the transformed variables, for some $Y$ as $\{ Y \}^*$ and drop the redundant lattice indices; these are functions of $s$ and $q$. Taylor expand the exponentials\footnote{$e^{x} = \sum_{n=0}^{\infty} \frac{x^n}{n!}$.} to second order, and keep leading terms of the order of ${\Delta x}^2$ and ${\Delta t}$ where the functions on the lattice are at least ${\cal O}(\Delta x)$:
\begin{align}
&\left[{ s \Delta t }\right] \{{\varphi}\}^* = (\{c \}^*- \{a \varphi\}^*) \Delta t  \label{eq:secondorderSSTP}\\
& +\tfrac{r}{2} \left[{ q^2 {\Delta x}^2  }\right]\{\Theta \Phi \}^* - \left[{q \Delta x}\right] \{F \Theta \Phi\}^*
+ \mcl{O}({\Delta t}^2,{\Delta x}^3). \nonumber 
\end{align}
Take the inverse Laplace transform and define: 
\begin{equation}
    y_{\Delta}(x,t) \equiv \mcl{L}_q^{-1} \{ \mcl{L}_s^{-1} \{ \{ Y \}^* \} \}.
\end{equation}
We have by construction that:
\begin{equation}
   y_{\Delta}(x,t) z_{\Delta}(x,t) \equiv \mcl{L}_q^{-1} \{ \mcl{L}_s^{-1} \{ \{ Y Z \}^* \} \}.
\end{equation}
From the properties of the inverse Laplace transform consider $s$ and hence $t$ (similarly for $q$ and $x$):
\begin{align}
\frac{d^n}{dt^n} y_{\Delta}(x,t) = \mcl{L}_q^{-1} \left\{ \mcl{L}_s^{-1} \left\{ s^n  \{ Y \}^* \right\} \right\}.
\end{align}
Putting this all together into Equation [\ref{eq:secondorderSSTP}] and divide by $\Delta t$, to find that at $ \mcl{O}({\Delta t})$ and $\mcl{O}({\Delta x}^2)$:
\begin{equation}
\de_t \varphi_{\Delta} = c_{\Delta} - a_{\Delta} \varphi_{\Delta} 
+ \tfrac{r}{2} \tfrac{\Delta x^2}{\Delta t} \de_{xx} \theta_{\Delta} \Phi_{\Delta}  
- \tfrac{\Delta x}{\Delta t} \de_x f_{\Delta} \theta_{\Delta} \Phi_{\Delta}. \nonumber 
\end{equation}
We can now use that $\hat F = r \beta \Delta x f_{\Delta}$ from Equation [\ref{eq:deltaforce}], and that the partial derivatives commute with the grid differentials to find:
\begin{align}
\de_t \mlm \varphi_{\Delta} & =  \mlm c_{\Delta}- \mlm a_{\Delta} \varphi_{\Delta} \\ 
&+ \de_{xx}  \mlm \frac{r}{2} \frac{\Delta x^2}{\Delta t} \left({\theta_{\Delta} \Phi_{\Delta}}\right) \nonumber \\
&- 2 \beta  \de_x  \mlm \frac{r}{2} \frac{\Delta x^2}{\Delta t} \left({f_{\Delta} \theta_{\Delta} \Phi_{\Delta}}\right).
\nonumber 
\end{align}
Now, there is some function $y(x,t)$ that $y_{\Delta}(x,t)$ is trying to approximate on the lattice: $Y^{i}_{n} = y_\Delta(i \Delta x,n \Delta t)$:
\begin{equation}
y(x,t) 
= \mlm y_{\Delta}(x,t).
\label{eq:capital_delta_rule}
\end{equation}
From the product of limits for well behaved $y_{\Delta}$ and $z_{\Delta}$:
\begin{equation}
y(x,t) z(x,t) = \mlm y_{\Delta} z_{\Delta} \label{eq:product_rule}
\end{equation} 
The stochastic force is only time dependent, and we can factor the time increments to include a fractional time dependence:
\begin{align}
\de_t \varphi &=c  - a \varphi \\ 
&+  \de_{xx} \left[{\mlm \frac{r}{2} \frac{\Delta x^2}{\Delta t^{\alpha}}\frac{1}{\Delta t^{1-\alpha}} \theta_{\Delta} \Phi_{\Delta}} \right] \nonumber \\
&- 2 \beta f(t) \de_x \left[{\mlm \frac{r}{2} \frac{\Delta x^2}{\Delta t^{\alpha}} \frac{1}{\Delta t^{1-\alpha}} \theta_{\Delta} \Phi_{\Delta}}\right]
\nonumber \label{eq:Transformed1}
\end{align}
Now, where $D^{1-\alpha}_{t}$ is the Riemann-Liouville derivative of order $1-\alpha$ and $D_\alpha$ the associated diffusion parameter \cite{ANGSTMANN2016508}:
\begin{align}
D_{\alpha} \theta D_{t}^{1-\alpha} \left( {\frac{\varphi}{\theta}} \right) = \mlm \frac{r \Delta x^2}{2 \Delta t^{\alpha}} \theta_{\Delta} \left[{\frac{1}{\Delta t^{1-\alpha}} \Phi_{\Delta}}\right]. 
\end{align}
Putting this into Equation [\ref{eq:Transformed1}] gets the required equation:
\begin{align}
\de_t \varphi &= c - a \varphi + \de_{xx} \left[{D_{\alpha} \theta D_{t}^{1-\alpha} \left( {\frac{\varphi}{\theta}} \right) } \right] \nonumber \\
&-2 \beta f(t) \de_x \left[{D_{\alpha} \theta D_{t}^{1-\alpha} \left( {\frac{\varphi}{\theta}} \right) }\right]
\end{align}
We can now substitute in our chosen potential (See Equation [\ref{eq:forcepotential}]) to then find:
\begin{align}
\de_t \varphi &= c - a \varphi + \de_{xx} \left[{D_{\alpha} \theta D_{t}^{1-\alpha} \left( {\frac{\varphi}{\theta}} \right) } \right] \nonumber \\
&+ V_t \de_x \left[{\theta D_{t}^{1-\alpha} \left( {\frac{\varphi}{\theta}} \right) }\right]. \label{app:eq:MainPDE}
\end{align}
This approach can be compared to that provided by \citet{BB2018}. This is instructive because it high-lights some important points. Rather than starting with a discrete microscopic detailed balance equation in the Kolmogorov form (Equation \ref{eq:Update1}) and then transforming to the dual space and transforming back to allow for the continuum formulation to be consistently found using only the diffusion limit to find Equation \ref{app:eq:MainPDE}, here following \citet{ADHJL2015}. \citet{BB2018} start with a balance like-equation in the dual-space\footnote{$\phi(k,p)=\Phi(p) \phi_0(k) + \Lambda(j) \Psi(p) \phi(k,p) + \Phi(p) s(k,p)$ for jump length $\Lambda(t)$, waiting-times $\Psi(t)$, density of states $\phi(x,t)$ and survival function $\Psi(t)$; all in the dual-space with initial condition $\phi_0(k)$ using that $p \Phi(p) = 1 - \Psi(p)$ \cite{BB2018}.}, approximate the diffusion limit\footnote{$\Lambda(k) \approx 1- \sigma^2 k^2 - \eta$ for some non-normalised jump-lengths with some small probability bias $\eta$ \cite{BB2018} and RMS jump-lengths $\sigma$.}, and take a truncated waiting-time distribution\footnote{$\Psi(p)\approx 1 -\tau^{\alpha}\left[{(p+\epsilon)^{\alpha} - \epsilon^{\alpha}}\right]$ for some $t_c = \epsilon^{-1}$ and a power law exponent $\alpha<1$.} to find a propagator equation\footnote{$\phi(k,p) = G_{\alpha,\epsilon}(k,p)[\phi_0(k) + s(k,p)]$ for a propagator $G_{\alpha,\epsilon}(k,p)$ \cite{BB2018}.} whose waiting time distribution is necessarily truncated in time. To then motivate small time-scale limit $t \ll t_c$ equations that can then, using the inverse Fourier-Laplace method, be used to find a reaction-diffusion equation\footnote{$\partial_t \phi = K D^{1-\alpha}_t (\partial_{xx} \phi - \varphi \phi) + s(x,t)$ for diffusion constant $K$ and fractional Riemann-Liouville operator $D_t^{1-\alpha}$; to again recover, for $t\gg t_c$ a diffusive reaction-diffusion equation.}. The key requirement here is the need for a consistent numerical scheme rather than an approximated interpretative model.

\tocless\subsection{Stochastic force jump probability consistency} \label{app:force}

The jump probabilities $\frac{r}{2}$ are now biased by an external force \cite{HLS2010}. Consider an order restricted in movement to only two new price positions within a potential $V(x,t)$ that is immediately adjacent to the orders current price position. We can then pick the probabilities of right and left jumps ($\pm 1)$ with some thermodynamic constant $C$:
\begin{equation}
p^\pm(x,t)= C Z^{-1} e^{-\beta V(x_{i \pm 1},t)}. \nonumber
\end{equation}
The partition function is: $Z =\sum_{k = \pm 1} e^{-\beta V(x_{i+k},t)}$. The jump probabilities are found at the end of the time increment so that $p^+ +p^-=1$. If the stochastic force was the only source of randomness in the system this would then define $C=1$. However, in this system the stochastic force is chosen to oppose the jumps due to the diffusion (Equation [\ref{eq:TransitionProb1}]). This means that the normalisation is $(\frac{1}{2}r + \frac{1}{2}r)=C$ because probabilities of left, right and self jumps due to the diffusion are also normalised (Equation [\ref{eq:DiffusionProbNorm}]), with the probablity of jumping being $r$, this choice sets $C=r$. 

We can find the potential in terms of the probabilities:
\begin{equation}
p^+(x,t) - p^-(x,t) = C \frac{e^{-\beta V(x_{i+1},t)} - e^{- \beta V(x_{i-1},t)}}{e^{-\beta V(x_{i+1},t)} + e^{-\beta V(x_{i-1},t)}}. \label{eq:probabilitydiff}
\end{equation}
Taylor expand this to find: 
\begin{equation}
p^+(x,t) - p^-(x,t) = - C \beta \Delta x \de_x V(x,t) + {\cal O}(\Delta x^3).
\end{equation}
A forcing function can be found from the gradient of a potential {\it i.e.} $f(x,t)=-\de_x V(x,t)$:
\begin{equation}
f_{\Delta}(x,t) =  \frac{1}{C \beta \Delta x }(p^+(x,t) - p^-(x,t)).
\end{equation}
Here $f(x,t) = \lim_{\Delta x \to 0} f_{\Delta}(x,t)$ and $p^+$ and $p^-$ are written in terms of the right and left transition probabilities, $P^{(+1)}$ and $P^{(-1)}$, associated with the diffusion process that the stochastic force opposes once a left or right jump occurs. 

If we now make the choice of potential as:
\begin{equation}
V(x,t)=\frac{V_t x}{2 D_{\alpha} \beta}
\end{equation}
Substitute the potential into Equation [\ref{eq:probabilitydiff}] and Taylor expand  the potential to first order:
\begin{equation}
f(t) = \frac{V_t}{2 D_{\alpha} \beta}. \label{eq:forcepotential}
\end{equation}
Which is consistent to first order with finding the force directly from the gradient of potential. 

\tocless\subsection{Diffusion limit update equation} \label{app:simplediffusionupdate}

Starting with the update equation in Equation [\ref{eq:UpdateRuleFinal}] at time $t_n$ with the convolution from time $t_{m}$:
\begin{align}
\varphi^{i}_{n}=& C^i_{m,n} + \theta^i_{n-1,n} \varphi^i_{n-1} +\sum_{m=0}^{n-1} K_{n-m}  \biggl[ P^{(+1)}_{i,n} \theta^{i-1}_{m,n-1} \varphi^{i-1}_{m} \nonumber \\
+ & P^{(-1)}_{i,n} \theta^{i+1}_{m,n-1} \varphi^{i+1}_{m} +(P^{(0)}_{i,n}-1)\theta^i_{m,n-1} \varphi^i_m
\biggr].\label{eq:DiffDeriv1}
\end{align} 
For the case where $\alpha=1$, we have $K_{n-m}=\delta_{n-m,1}$, which eliminates the sum and sets $m=n-1$, and ignoring the source term and using that $\theta^k_{n-1,n-1} =1$:
\begin{align}
\varphi^{i}_{n}= P^{(+1)}_{i,n} \varphi^{i-1}_{n-1} 
+ P^{(-1)}_{i,n} \varphi^{i+1}_{n-1} 
+ (P^{(0)}_{i,n}-1-\theta^i_{n-1,n}) \varphi^i_{n-1}
 \nonumber
\end{align} 
If the survival function is independent of price, then $\theta^{i}_{n-1,n}=e^{-\nu \Delta t}$ and with jump probability is $r$ (so that the probability of self jump is $1-r$). With no stochastic force:
\begin{equation}
\varphi^{i}_{n} = \tfrac{r}{2} \varphi^{i-1}_{n-1} 
+ \tfrac{r}{2} \varphi^{i+1}_{n-1} -(r-e^{-\nu \Delta t}) \varphi^i_{n-1}. \nonumber
\end{equation}
When Taylor expanded to first order in $\Delta t$ this is the simple diffusion update equation used in prior work \cite{Gant2022a,Gant2022b}.

\tocless\subsection{Boundary and initial condition constraints} \label{app:boundarycons}

Consider the boundary conditions for Equation [\ref{eq:MainPDE}]:
\begin{align}
\de_t \varphi &= c - a \varphi + \left[{D_{\alpha} \theta D_{t}^{1-\alpha} \left( {\frac{\varphi}{\theta}} \right) } \right]_{xx} \negthickspace  \negthickspace  \negthickspace + V_t \left[{\theta D_{t}^{1-\alpha} \left( {\frac{\varphi}{\theta}} \right) }\right]_x. \nonumber 
\end{align}
For the case when the survival function $\theta$ is only a function of time -- this is the case we consider in this paper, then the derivatives commute with respect to spacial past of the operator $D^{1-\alpha}_t$ and with $\theta$, to obtain:
\begin{align}
\de_t \varphi &= c - a \varphi +  { \theta D_{t}^{1-\alpha} \left( {\frac{1}{\theta}  D_{\alpha} \varphi_{xx}} \right) }  + {\theta D_{t}^{1-\alpha} \left(  {\frac{1}{\theta} V_t \varphi_x} \right) }. \nonumber 
\end{align}
Here the inner-most parts of the final two terms have the same operations performed on them; we can factor these two terms to obtain:
\begin{align}\label{eq:boundaryconditions3}
\de_t \varphi &= c - a \varphi + \theta D_{t}^{1-\alpha} \left[ {\frac{1}{\theta} \left( D_{\alpha} \varphi_{xx} + V_t \varphi_x \right)} \right].
\end{align}
The spatial integral of $\varphi$ is defined as $\Phi(t)$: 
\begin{equation}
    \Phi(t) = \int_{L_1}^{L_2} \varphi(x,t) dx. \label{eq:totalmass}
\end{equation}
This measures the net volume of limit order bids and offers in the system. We integrate both sides of Equation [\ref{eq:boundaryconditions3}] and use Equation [\ref{eq:totalmass}]to obtain:
\begin{align}
\de_t \Phi &= \int_{L_1}^{L_2}  \negthickspace  c \, dx - a \Phi + {\theta D_{t}^{1-\alpha} \left[ {\frac{1}{\theta} \left( D_{\alpha} \de_{x} \varphi + V_t \varphi \right)} \right]_{L_1}^{L_2} }. \nonumber
\end{align}
For the  net volume of limit order bids and offers
in the system to be in a steady state we want $\de_t \Phi=0$ for both the lit, and latent order book. This implies three conditions. \\

\begin{statement}{Steady-state sufficient conditions }

If $\de_t \Phi=0$, then the order book is in a steady-state. There are three sufficient conditions that ensure that order book is in a pre-equilibrium steady-state:
\begin{enumerate}[label=\roman*.)]
\item $\Phi(0)=0$,
\item $\int_{L_1}^{L_2} c \, dx = 0$ for all time, and 
\item $\left[D_{\alpha} \de_{x} \varphi + V_t \varphi \right]_{L_1,t}^{L_2,t} = 0$ for all time. 
\end{enumerate}  \label{thm:sufficientconds}
These are respectively conditions on the initial conditions, the order imbalance, and order density conservation. 
\end{statement}
The first condition in Statement \ref{thm:sufficientconds} is achieved by our choice of initial conditions. The second and third conditions were trivially achieved in this paper by the choice of an anti-symmetric source (See Section \ref{ssec:mass} and Equation [\ref{eq:source}]) and appropriately distant (from each other) boundaries such that $\varphi$ and $\de_x \varphi$ are zero at the boundaries. 

However, one could enforce the third condition, by instead declaring that it is true at each boundary separately, which then amounts to enforcing, separately, at each boundary:
\begin{equation}
    D_{\alpha} \de_{x} \varphi = - V_t \varphi.
\end{equation}
This can be achieved using ghost points \cite{ANGSTMANN2016508} to give:
\begin{equation}
    \varphi_{i \pm 1} = \varphi_i \mp \frac{\Delta t \, V_t}{D_\alpha} \varphi_i.
\end{equation}
Here $i\pm1$ refers to the right (left) ghost point respectively. 

These conditions are sufficient but not necessary. If one tries to simply solve Equation [\ref{eq:boundaryconditions3}] for the ghost point directly, but using discretisation of the operators involved, one should be able to get a general relationship. The integral inside the operator $D_t^{1-\alpha}$ can be approximated as a sum multiplied by $\Delta t$. and all time and space derivatives, using a first order difference in time, to then obtain an approximation scheme that can be solved. This gives:
\begin{align}
    \left[{D_\alpha \de_{x} \varphi + V_{t_n} \varphi}\right]_{L_1,t_{n-1}}^{L_2,t_{n-1}} 
     =  \Gamma (\alpha)\frac{\Delta t_{n}^{1-\alpha}}{\theta_{n-1,n}} \sum_{i=L_1}^{L_2} c(x_i,t_n). \nonumber
\end{align}
Here we have also used equations [\ref{eq:survivalfn}] and [\ref{eq:factorsurvival}]. One immediately sees that if area under the source $c$ is zero, then one gets condition iii.) from Statement \ref{thm:sufficientconds}. Recall that the spacial derivative of $\varphi$ at each boundary implicitly contains the ghost point there. This  means that both ghost points appear in the above expression on the left hand side. When the right hand side is zero we can then turn this equation into two conditions and solve for each ghost point separately; by assuming the expression holds at each boundary independently. However, when the right hand side is not zero, we cannot reduce this into two conditions without additional constraints.

Next we consider the initial conditions. The key problem is that they are difficult to define for this system as it relates to the experiment. This is because the steady state of Equation [\ref{eq:boundaryconditions3}] is likewise difficult to define in the presence of a combination of path dependent memory and stochasticity. If one removes the term containing $V_t$, then the steady state of the system can be solved analytically, as is done by \citet{ADHJL2015}. Usually, this would be the full solution since one can simply change co-ordinates into the frame of the system as is done in \citet{DBMB2014} - this removes the $V_t$ term. The analytic solution from \citet{ADHJL2015} can then be used, to then change back to the original coordinate system, to get the complete steady state solution. 

Specifically, one makes the transform from $x$ to $y=x-\hat p_t$ where $\hat p_t = \int_0^t ds V_s$ which centers the intercept at $y=0$ in this moving frame. One then computes the analytic solution in this frame, then changes back to $x$, which sets the price moving again. However, anomalous diffusion and the operator $D^{1-\alpha}_t$ makes this problematic. This is because the operator includes the entire history of $p_t$. In the simple diffusion case this is thus a trivial problem, in the general setting of anomalous diffusion this approach is challenging because there is a dense set of necessary frame changes required. 

Thus, even though the intercept in the frame at time $t$ is always at $y=0$, at some previous time $t'$, it will have been at $p_t-\hat p_{t'}$ - even in this moving frame! Thus $V_t$ re-appears once more. It also remains unclear how to define the steady state of this system in way which would allow one to simulate the system. If one did use the steady state analytic solution in some way, or used some solution which is averaged over $V_t$, one would then run into the problem of what it means to be in a steady state. 

In general for this system, thanks to the operator $D^{1-\alpha}_t$, this means an infinite amount of time has already passed and so the numeric kernel that this operator implies would already have an infinite history to contend with at the very start of the simulation. What is clear that, for this to be tractable, some sort of truncation is necessary.

\section{Implementation} \label{app:appendix:implementation}

\begin{algorithm}[hp]
\caption{Diffusion Simulation Algorithm} \label{alg:SFDRW}
\begin{algorithmic}
\Require{$\{M, x_0, x_M, r,\lambda_t, D_{\alpha}, \alpha, \nu, \beta, \lambda, \mu\}$ (Table \ref{tab:parameters})} 
\State{1: Compute Time Steps: $\{\Delta t_k\}_{k=1}^M$ and $\Delta t$}
\State{2: Compute State Steps: $\{ \Delta x_k \}_{k=1}^M$ and $\Delta x$ (Eqn. [\ref{eq:latticesize}]) }
\State{3: Initialise Background Lattice: $(x_i,t_n)$}
\State{4: Compute Initial Conditions: $\varphi^i_0$}
\State{\ForAll{$n$}
    \State{5: Set Time and State Steps: $\Delta t_n, \Delta x_n$}
    \State{6: Compute Boltzmann Potentials: $V_t$}
    \State{7: Compute Jump Probabilities: $r$, $F_n$}
    \State{8: Update Sources: $s(x_i,t_n,p_n)$}
    \State{9: Update Boundary Conditions: $\varphi^0_n$, $\varphi^M_n$}
    \State{10: Update Interior Points: $\varphi^i_n$ (Eqn. [\ref{eq:UpdateEquationExpArrival},\ref{eq:onjumplattice},\ref{eq:MemKerRecursive}])} 
    \State{11: Find mid-price: $p_n=\{ x_i : \min \{|\varphi^i_n|\} \}$}
    \EndFor} \\
\State{\Return{$\varphi^i_n$ and $p_n$ on lattice $(x_i,t_n)$}}
\end{algorithmic}
\end{algorithm} 

\begin{figure}
\centering
\includegraphics[width=0.40\textwidth]{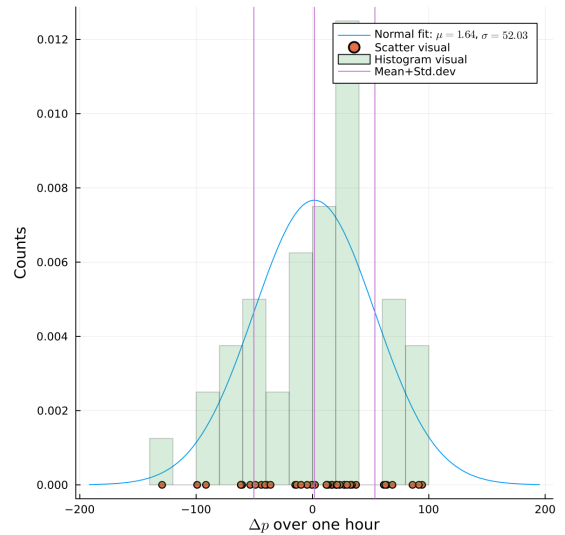}
\caption{Distribution of price changes over one hour for 5 simulated, 8 hour days of trading - calculated for Sec. \ref{ssec:dailyvolume}. The horizontal axis shows the 40 data points obtained. We include a histogram and normal fit of this data. We are interested in the variance of this distribution and so we label that as well.}
\label{fig:volatilitydata}
\end{figure}

\tocless\subsection{Price impact kinks} \label{app:PriceImpactKinks}

\begin{figure}
\begin{center}
\resizebox{0.45\textwidth}{!}{
\input{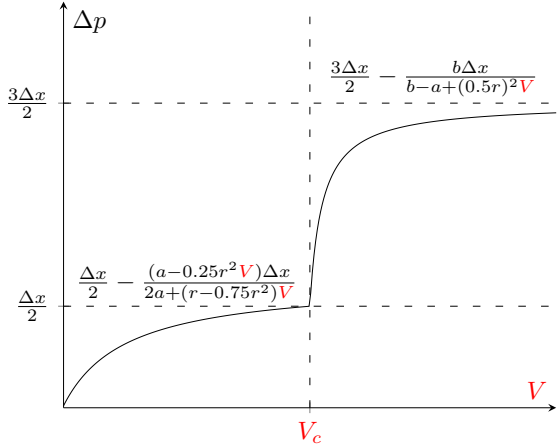}}
\end{center}
\caption{We show a schematic plot of the volume $V$ against the price impact which explains the kinks observed in Fig. \ref{fig:PI-LO-LI-U_-NE} and Fig. \ref{fig:PI-LO-CI-U_-NE} (see Sec. \ref{ssec:priceimpactlimit}) as discussed in  \ref{app:PriceImpactKinks}. The central vertical stippled line represents the critical value of the volume. We highlight the form of the price impact to the left of the critical value and to the right of the critical value. We see that the left and right forms agree at the critical value and that the form has the correct upper bounds to correspond to $\Delta n=2$ in Fig. \ref{fig:PI-LO-LI-U_-NE}.}
\label{fig:KinkPlot}
\end{figure}

\begin{figure*}[ht!]
\centering
\begin{subfigure}[c]{0.45\textwidth}
\includegraphics[width=\textwidth]{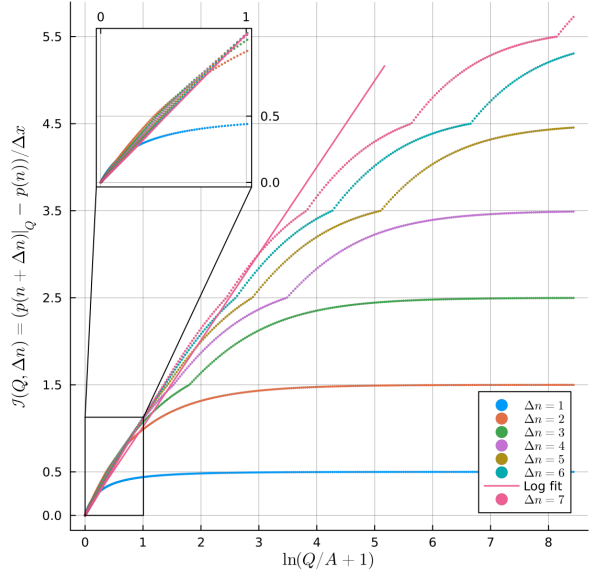}\label{fig:LimitOrderPriceImpact}
\caption{Limit order, linear interpolation and uniform sampling}\label{fig:PI-LO-LI-U_-NE}
\end{subfigure}
\begin{subfigure}[c]{0.45\textwidth}
\includegraphics[width=\textwidth]{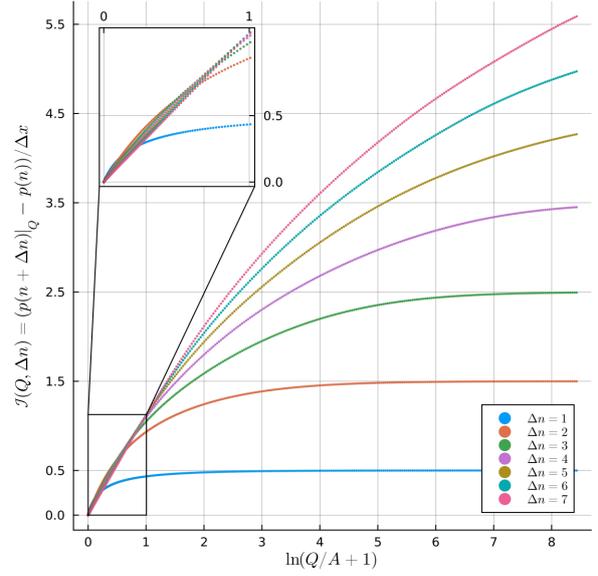}
\caption{Limit order, cubic-spline interpolation and uniform sampling}\label{fig:PI-LO-CI-U_-NE}
\end{subfigure}
\caption{Fig. \ref{fig:PI-LO-LI-U_-NE} shows the price impact for a limit order with intercept $p$ estimated using linear interpolation when simulating using the uniform sampling scheme (see Eqn. \ref{eq:FinalUpdateEquation}). Fig. \ref{fig:PI-LO-CI-U_-NE} gives the price impact simulated using cubic-spline interpolation while keeping the configuration otherwise unchanged. These graphs show how the price impact plots of limit orders are affected by using cubic-spline interpolation and by the choice of $\gamma_1$ (Sec. \ref{ssec:timescales}). Here we have $\gamma_1 = \Delta n$ where each line in both figures shows the price impact measured after $\Delta n \in \mathbb{N}^{+}$ time steps in $L_0$ (see Sec. \ref{ssec:timescales}) since the placement of the market order. For additional discussion, see Sec. \ref{ssec:priceimpactlimit}.
}
\label{fig:}
\end{figure*}

\begin{figure*}[ht!]
\centering
\begin{subfigure}[c]{0.45\textwidth}
\includegraphics[width=\textwidth]{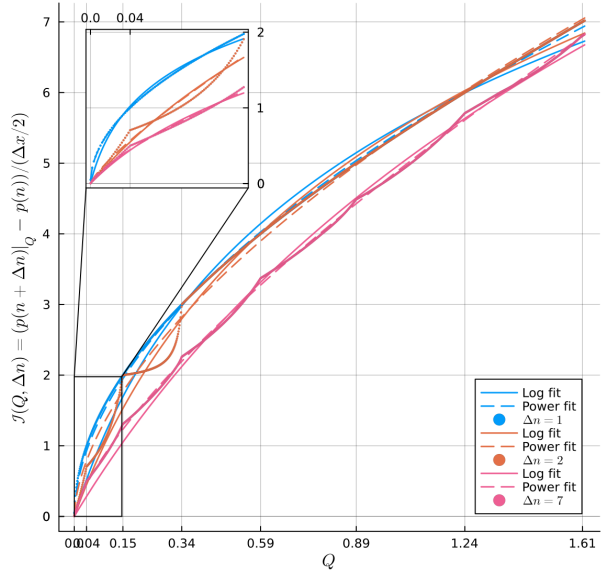}
\caption{Market order, linear interpolation and uniform sampling.}\label{fig:PI-MO-LI-U_-NE}
\end{subfigure}
\begin{subfigure}[c]{0.45\textwidth}
\includegraphics[width=\textwidth]{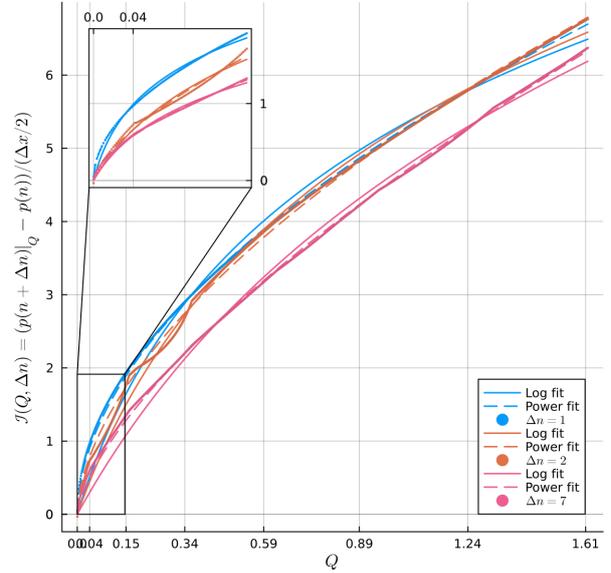}
\caption{Market order, linear interpolation non-uniform sampling.}\label{fig:PI-MO-LI-NU-NE}
\end{subfigure}
\caption{Fig. \ref{fig:PI-MO-LI-U_-NE} gives the price impact for a market order (see Sec. \ref{ssec:priceimpactmarket}) with intercept $p$ estimated using linear interpolation when simulating using the uniform sampling scheme (see Eqn. \ref{eq:FinalUpdateEquation}). Fig. \ref{fig:PI-MO-LI-NU-NE} gives the price impact simulated using the non-uniform sampling scheme (Eqn. \ref{eq:UpdateEquationExpArrival}) while keeping the same configuration otherwise unchanged. These graphs show how the price impact of market orders is affected by using non-uniform sampling as well as by the choice of the sampling rate free-parameter $\gamma_1$ (Sec. \ref{ssec:timescales}). Here we have $\gamma_1 = \Delta n$ where each line in both figures shows the price impact measured after $\Delta n \in \mathbb{N}^{+}$ time steps in $L_0$ (see Sec. \ref{ssec:timescales}) since the placement of the market order. For additional discussion, see Sec. \ref{ssec:priceimpactmarket}.
}
\label{fig:fig:PriceImpactMarket-Linear}
\end{figure*}

In this section we explain the ``kinks" (see Section \ref{ssec:priceimpactlimit}) which appear in Figure \ref{fig:PI-LO-LI-U_-NE} (and by extension also explain the kinks in Figure \ref{fig:PI-MO-LI-U_-NE}.). Throughout this section we assume that we are using the same conditions as in Section \ref{ssec:priceimpactlimit} i.e. $\alpha=1.0$ and $V_t=0$ for all time. 

\begin{figure*}
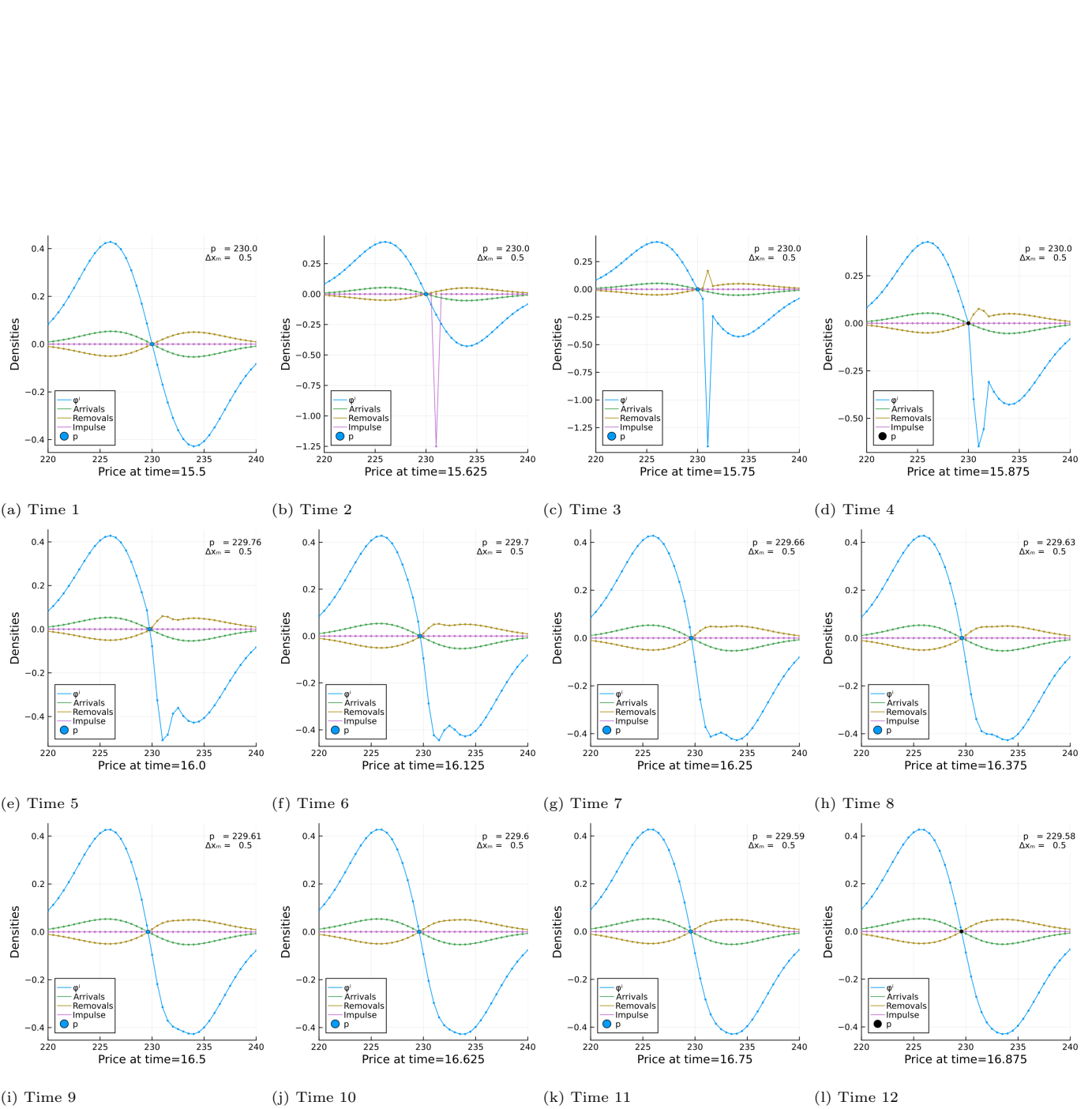

\centering
\begin{subfigure}[c]{0.24\textwidth}
\includegraphics[width=1.0\textwidth]{\PathVK K-1.png}
\caption{Time 1}\label{fig:VisualizeImpactSpike:a}
\end{subfigure}
\begin{subfigure}[c]{0.24\textwidth}
\includegraphics[width=1.0\textwidth]{\PathVK K-2.png}
\caption{Time 2}\label{fig:VisualizeImpactSpike:b}
\end{subfigure}
\begin{subfigure}[c]{0.24\textwidth}
\includegraphics[width=1.0\textwidth]{\PathVK K-3.png}
\caption{Time 3}\label{fig:VisualizeImpactSpike:c}
\end{subfigure}
\begin{subfigure}[c]{0.24\textwidth}
\includegraphics[width=1.0\textwidth]{\PathVK K-4.png}
\caption{Time 4}
\end{subfigure}
\begin{subfigure}[c]{0.24\textwidth}
\includegraphics[width=1.0\textwidth]{\PathVK K-5.png}
\caption{Time 5}
\end{subfigure}
\begin{subfigure}[c]{0.24\textwidth}
\includegraphics[width=1.0\textwidth]{\PathVK K-6.png}
\caption{Time 6}
\end{subfigure}
\begin{subfigure}[c]{0.24\textwidth}
\includegraphics[width=1.0\textwidth]{\PathVK K-7.png}
\caption{Time 7}
\end{subfigure}
\begin{subfigure}[c]{0.24\textwidth}
\includegraphics[width=1.0\textwidth]{\PathVK K-8.png}
\caption{Time 8}
\end{subfigure}
\begin{subfigure}[c]{0.24\textwidth}
\includegraphics[width=1.0\textwidth]{\PathVK K-9.png}
\caption{Time 9}
\end{subfigure}
\begin{subfigure}[c]{0.24\textwidth}
\includegraphics[width=1.0\textwidth]{\PathVK K-10.png}
\caption{Time 10}
\end{subfigure}
\begin{subfigure}[c]{0.24\textwidth}
\includegraphics[width=1.0\textwidth]{\PathVK K-11.png}
\caption{Time 11}
\end{subfigure}
\begin{subfigure}[c]{0.24\textwidth}
\includegraphics[width=1.0\textwidth]{\PathVK K-12.png}
\caption{Time 12}
\end{subfigure}
\caption{Shows the result of a flash limit order being placed near the current price in the order book and corresponds to Sec. \ref{ssec:priceimpactlimit}. The flash limit order shocks the order book out of equilibrium. Each sub-figure in the above is a snapshot of the system in simulation event time. In each sub-figure, the plot in blue is the density we are modelling at the current time, the plot in green is the orders which are about to be added via the source term and the gold plot is the orders which are about to be removed via the removal rate. The purple line which appears only in Fig. \ref{fig:VisualizeImpactSpike:b} indicates the shock that is about to be added to the system.}
\label{fig:VisualizeImpactSpike}
\end{figure*}

Consider Figure \ref{fig:KinkVisualisation}. Here we show in blue the general shape of the order book that may occur near the current price which we place at $0$ on the x-axis (log price). We show four discretized bars which represent the order book in its discretized form; this can be found in our code and can be seen in Figure \ref{fig:VisualizeImpactSpike} - which this section deals with closely. We have included on these diagrams the result of straight line interpolations between the grid points which then allows one to determine the price. All items in blue represent the state of the order book before the placement of a flash limit order and are unchanged throughout Figure \ref{fig:kt1}-Figure \ref{fig:kt32}. 

Now, place the flash limit order at a distance of $1.5 \times \Delta x$ from the current price (Section \ref{ssec:priceimpactlimit}) - this is shown in red, noting that the current price is between grid points. This modifies both the on grid values of $\varphi$ and the resulting straight line interpolations; all of which is shown in red in Figure \ref{fig:kt1}. After one time step in simulation time, we get Figure \ref{fig:kt2}. To arrive here, we have manually allowed the red bar to diffuse (the blue bars do not diffuse as we assume they were already in equilibrium before this time). To have the red bar diffuse, we allow a proportion $r/2$ of it to step left, $r/2$ of it to step right, and $1-r$ to remain where it is (see Section \ref{ssec:simulatinglimitordrebook} for the motivation). We now notice that the new straight line interpolation has shifted the current price, and we label the distance of this shift $\Delta p$. This would correspond to the line for $\Delta n = 1$ in Figure \ref{fig:PI-LO-LI-U_-NE}. No matter how large the red bar at $0.5\Delta x$ gets, $\Delta p$ cannot move past $-0.5\Delta x$; this means that $\Delta p$ is bounded above by $\Delta x / 2$ at this point. 

Now, allow another step to pass in order book event time. We then obtain two possibilities for what happens: Figure \ref{fig:kt31} and Figure \ref{fig:kt32}. First, consider Figure \ref{fig:kt31}. In this case, upon allowing particles to diffuse in the same way described above, the particles which crossed zero  were not sufficient to completely annihilate with the blue orders that lie to the left of the zero intercept at $-0.5 \Delta x$. Second, the other case which could occur is shown in Figure \ref{fig:kt32}. In this case the particles which cross the zero intercept are large enough to completely annihilate the order represented by the blue bar at $-0.5 \Delta x$ and will replace it with a positive red bar that is above the x-axis (price axis). This then allows the price intercept to move into the region $[-0.5 \Delta x, -1.5 \Delta x ]$. Since, as one can see in each diagram, it is only the order volume $V$ which sets the scale of the red orders, we can infer that there must be some value for $V$ at one which crosses over from Figure \ref{fig:kt31} to Figure \ref{fig:kt32}. We call this value the critical volume $V_c$. 

The entire system is algebraic and every point on Figure \ref{fig:kt1} to Figure \ref{fig:kt32} is labelled, and with simple algebra the impact on price that the flash order of size $V$ will have can be determined. We have done this, and plot the resulting analytic function in Figure \ref{fig:KinkPlot}. In this figure, one sees the sought after ``kink" occurring at the critical volume, in addition to some analytic expressions for the shape of the price impact elsewhere. One can see that this fully explains the orange line corresponding to $\Delta n = 2$ in Figure \ref{fig:PI-LO-LI-U_-NE} -- taking note that Figure \ref{fig:PI-LO-LI-U_-NE} is on log scale. The explanation for the other lines can, similarly, be inferred from here.


We now comment on the nature of limit orders when not using cubic-spline interpolation or non-uniform times to see what these have added. The price impact for the computationally least demanding case, where one uses straight line interpolation and uniformly distributed times, is shown in Figure \ref{fig:PI-LO-LI-U_-NE} (with no error bars). Here there are ``kinks" in the line $\mathcal{J}_{\Delta n}$ where is crosses any $b_{\Delta n}$ and in addition, each $\mathcal{J}_{\Delta n}$ is bounded above by $b_{\Delta n}$. We explain both of these observations in detail in \ref{app:PriceImpactKinks}. 

\begin{figure*}
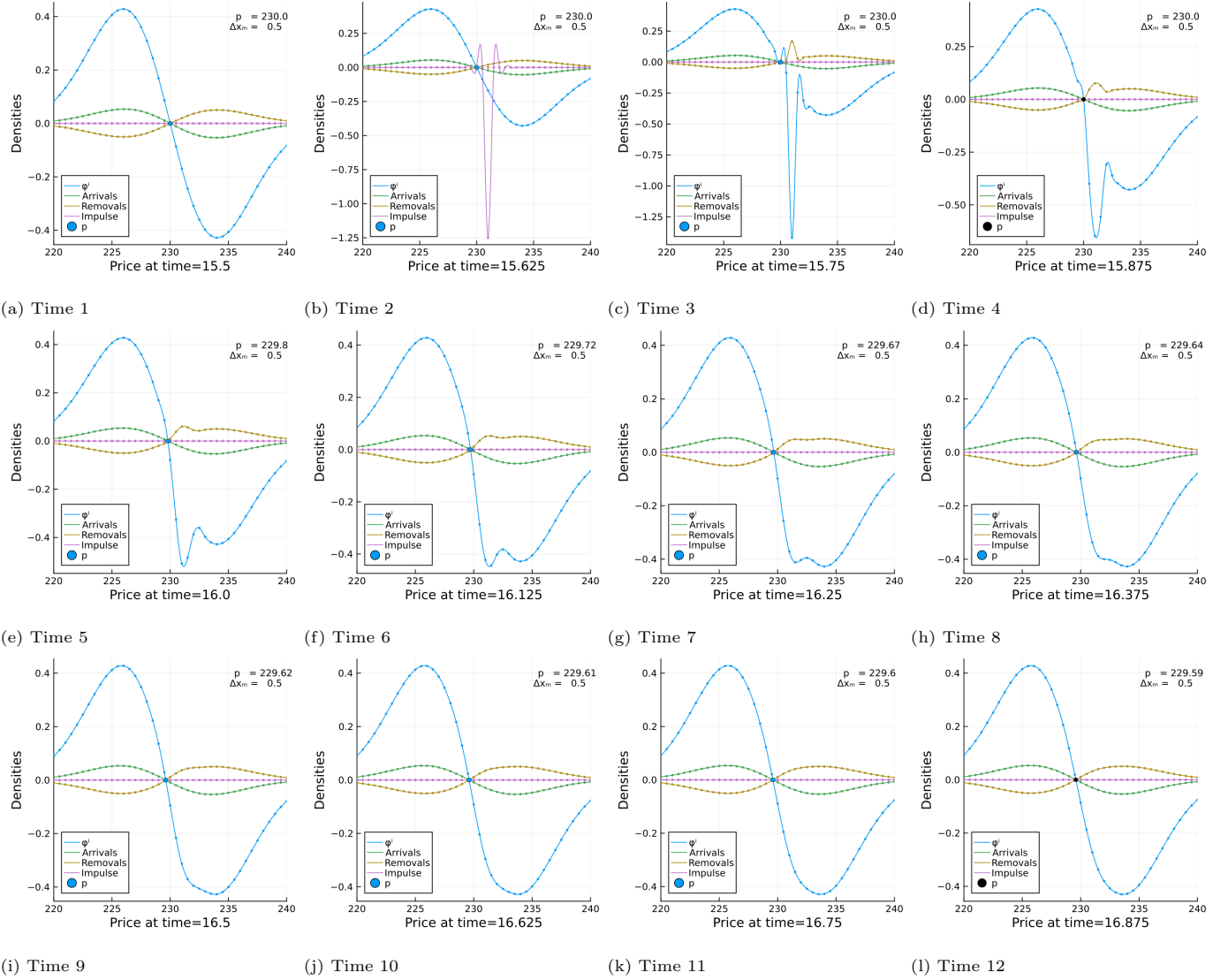

\centering
\begin{subfigure}[c]{0.24\textwidth}
\includegraphics[width=1.0\textwidth]{\PathKInterp K_interp-1.png}
\caption{Time 1}\label{fig:VisualizePriceImpactSpikeInterpolated:a}
\end{subfigure}
\begin{subfigure}[c]{0.24\textwidth}
\includegraphics[width=1.0\textwidth]{\PathKInterp K_interp-2.png}
\caption{Time 2}\label{fig:VisualizePriceImpactSpikeInterpolated:b}
\end{subfigure}
\begin{subfigure}[c]{0.24\textwidth}
\includegraphics[width=1.0\textwidth]{\PathKInterp K_interp-3.png}
\caption{Time 3}\label{fig:VisualizePriceImpactSpikeInterpolated:c}
\end{subfigure}
\begin{subfigure}[c]{0.24\textwidth}
\includegraphics[width=1.0\textwidth]{\PathKInterp K_interp-4.png}
\caption{Time 4}
\end{subfigure}
\begin{subfigure}[c]{0.24\textwidth}
\includegraphics[width=1.0\textwidth]{\PathKInterp K_interp-5.png}
\caption{Time 5}
\end{subfigure}
\begin{subfigure}[c]{0.24\textwidth}
\includegraphics[width=1.0\textwidth]{\PathKInterp K_interp-6.png}
\caption{Time 6}
\end{subfigure}
\begin{subfigure}[c]{0.24\textwidth}
\includegraphics[width=1.0\textwidth]{\PathKInterp K_interp-7.png}
\caption{Time 7}
\end{subfigure}
\begin{subfigure}[c]{0.24\textwidth}
\includegraphics[width=1.0\textwidth]{\PathKInterp K_interp-8.png}
\caption{Time 8}
\end{subfigure}
\begin{subfigure}[c]{0.24\textwidth}
\includegraphics[width=1.0\textwidth]{\PathKInterp K_interp-9.png}
\caption{Time 9}
\end{subfigure}
\begin{subfigure}[c]{0.24\textwidth}
\includegraphics[width=1.0\textwidth]{\PathKInterp K_interp-10.png}
\caption{Time 10}
\end{subfigure}
\begin{subfigure}[c]{0.24\textwidth}
\includegraphics[width=1.0\textwidth]{\PathKInterp K_interp-11.png}
\caption{Time 11}
\end{subfigure}
\begin{subfigure}[c]{0.24\textwidth}
\includegraphics[width=1.0\textwidth]{\PathKInterp K_interp-12.png}
\caption{Time 12}
\end{subfigure}
\caption{Shows the result of a flash limit order being placed near the current price in the order book and corresponds to Sec. \ref{ssec:priceimpactlimit}, here using cubic spline interpolation to estimate the off grid points, and therefore the mid-price. The flash limit order shocks the order book which is then out of equilibrium in Fig. \ref{fig:VisualizePriceImpactSpikeInterpolated:c}. One can see at this time that the interpolation causes the order book to undulate about the point at which it is shocked. Each sub-figure in the above is a snapshot of the system in simulation event time. In each sub-figure, the plot in blue is the density we are modelling at the current time, the plot in green is the orders which are about to be added via the source term and the gold plot is the orders which are about to be removed via the removal rate. The purple line which appears only in Fig. \ref{fig:VisualizeImpactSpike:b} indicates the shock that is about to be added to the system.}
\label{fig:VisualizePriceImpactSpikeInterpolated}
\end{figure*}

As a next step, one may try to smooth these kinks by using cubic splines to determine the intercept $p(t)$ \footnote{This was implemented using the ``Interpolations" library in Julia} while still using uniformly distributed times. This combination is visualised in Figure \ref{fig:VisualizePriceImpactSpikeInterpolated} and the resulting price impact is shown in Figure \ref{fig:PI-LO-CI-U_-NE}. We see that cubic splines fix the ``kinks" where a $\mathcal{J}_{\Delta n}$ crosses a $b_{\Delta n}$ but not the upper bounds on each $\mathcal{J}_{\Delta n}$. This is then finally fixed by including non-uniform sampling as described above. Although it is not shown explicitly, the price impact lines in Figure \ref{fig:PI-LO-CI-U_-NE}, at the points where they pass through the critical displacements $b_{\Delta n}$, pass exactly through the kinks in Figure \ref{fig:PI-LO-LI-U_-NE}. These points are the most accurate in the sense that they are stable under both interpolation schemes.


We now comment on the nature of market orders when not using cubic splines or non-uniform times. To do this we include Figure \ref{fig:PI-MO-LI-U_-NE} which shows the price impact for market orders with uniform times and linear interpolation (no errors bars this time) and subsequently Figure \ref{fig:PI-MO-LI-NU-NE} which keeps the linear interpolation but uses non-uniform times instead. We notice that in \ref{fig:PI-MO-LI-U_-NE}, with the additional complexity removed, we are able to determine both co-ordinates of every kink exactly with the highlighted horizontal tick values being the volume which needs to traded be to change the price by $\Delta x/2$ and is determined by numerically integrating. After introducing non-uniform time in \ref{fig:PI-MO-LI-NU-NE}, we can no longer exactly determine the kinks' co-ordinates beforehand.

\begin{figure*}[ht!]
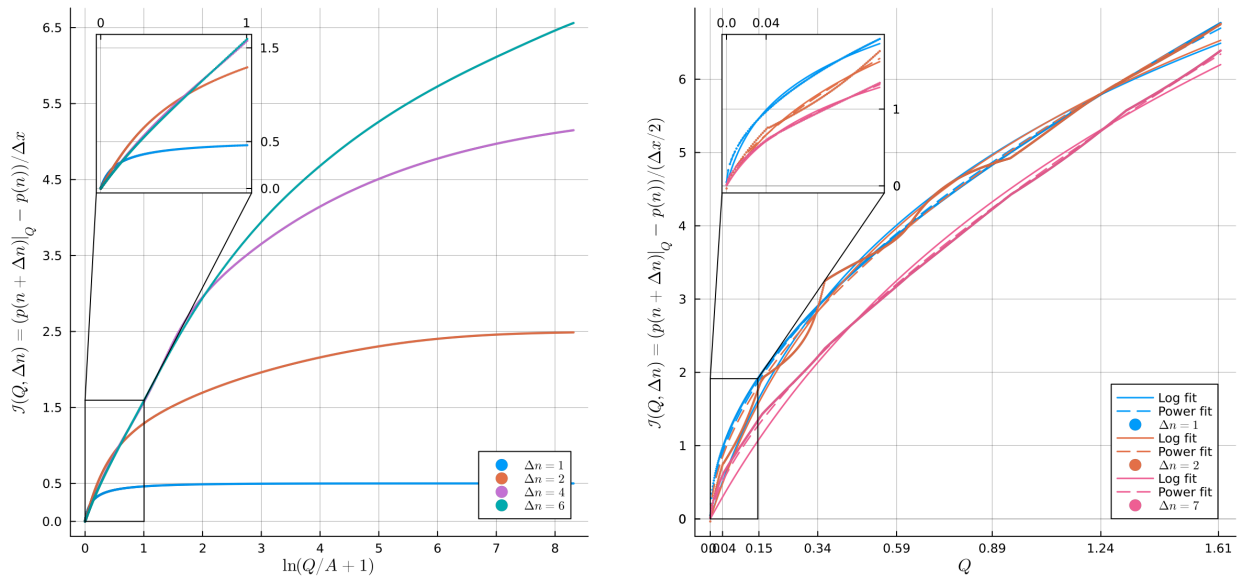

\centering
\begin{subfigure}[c]{0.45\textwidth}
\includegraphics[width=\textwidth]{\PathSingles PI_DD_Exp-Int.png}
\caption{Limit order, cubic-spline interpolation and non-uniform sampling.}
\label{fig:PI-LO-CI-NU-NE}
\end{subfigure}
\begin{subfigure}[c]{0.45\textwidth}
\includegraphics[width=\textwidth]{\PathSingles PI_DD-MO_exp-Int-general.png}
\caption{Market order, cubic-spline interpolation and non-uniform sampling.}\label{fig:PI-MO-CI-NU-NE}
\end{subfigure}
\caption{Fig. \ref{fig:PI-LO-CI-NU-NE} gives the price impact for a limit order (see Sec. \ref{ssec:priceimpactlimit}) with intercept $p$ estimated using linear interpolation when simulating using the non-uniform sampling scheme (see Eqn. \ref{eq:UpdateEquationExpArrival}). Fig. \ref{fig:PI-MO-CI-NU-NE} gives the price impact for a market order (see Sec. \ref{ssec:priceimpactmarket}) with the same configuration.  These graphs show how the price impact plots for the two order types are affected by using non-uniform sampling (Sec. \ref{ssec:updateequationarrivaltimes}) and by the choice of the sampling rate free-parameter $\gamma_1$ (Sec. \ref{ssec:timescales}). Here we have $\gamma_1 = \Delta n$ where each line in each figure shows the price impact measured after a delay of $\Delta n \in \mathbb{N}^{+}$ time steps in $L_0$ (see Sec. \ref{ssec:timescales}) since the placement of the flash limit order. For additional discussion, see Sec. \ref{ssec:priceimpactmarket}.
}
\label{fig:PriceImpact-AllBellsAndWhistles}
\end{figure*}

Concretely, Figure \ref{fig:PI-LO-CI-NU-NE} shows the price impact measured after differing time steps, and is plotted against the log of the volume of the flash limit order and this is normalised by the total area under the bid side $A$ as described in Section \ref{ssec:priceimpactlimit}. We highlight values on the price axis defined by $\frac{1}{2}+(\Delta n-1)$. Compared with Figure \ref{fig:PI-LO-LI-U_-NE}, and even Figure \ref{fig:PI-LO-CI-U_-NE}, here one uses cubic splines to find the intercept and the non-uniform time algorithm (see Section \ref{ssec:updateequationarrivaltimes}). One observes this results in a function with no kinks: each line passes through the set of highlighted prices smoothly. But in addition to this, some lines labelled with $\Delta n$ have surpassed their expected bound of $\frac{1}{2}+(\Delta n-1)$ such as the $\Delta n = 2$ line. One also observes that there is some interval of volumes starting from zero up to some maximum volume for which the price impact is approximately straight (on this log scale) and that the size of this interval increases for larger $\Delta n$. Most notably, as seen in the inset, the difference between $\ell=1$ and $\ell=2$. Thus, one's choice of $\gamma$ (see Section \ref{ssec:sampling} and equations \ref{eq:gamma1} through \ref{eq:gamma12}). will influence the ability of flash limit orders to change the price of the order book.

\begin{table*}
    \centering
    \begin{tabular}{|c|c|c|c|c|c|}
        \hline
        &                   &                                     & $\Delta n=1$           & $\Delta n=2$          & $\Delta n=7$           \\
        \hline
     \multirow{4}{*}{Figure \ref{fig:PI-MO-LI-U_-NE}} 
                            & \multirow{2}{*}{$ax^b$}         & a &  $1.33702 \pm 0.00032$ & $1.3277 \pm 0.0010$ & $1.19982 \pm 0.00033$ \\
                            &                                & b &   $0.54045 \pm 0.00049$ & $0.5888 \pm 0.0017$ & $0.72762 \pm 0.00066$ \\ 
                            \cline{2-6}
                            & \multirow{2}{*}{$c\log(1+dQ)$}  & c &  $0.7868  \pm 0.0068$  & $0.9239 \pm 0.0057$ & $1.548 \pm 0.012$     \\
                            &                                & d &   $4.619 \pm 0.082$     & $3.314   \pm 0.039$ & $1.197 \pm 0.014$     \\
                            \cline{2-6}
        \hline
        \hline
     \multirow{4}{*}{Figure \ref{fig:PI-LO-CI-NU-NE}}  
                            & \multirow{2}{*}{$ax^b$}         & a &  $1.29129 \pm 0.00029$  &  $1.28288 \pm 0.00056$  & $1.14056 \pm 0.00029$\\
                            &                                & b &   $0.53823 \pm 0.00047$  &   $0.58065 \pm 0.00094$ & $0.67964 \pm 0.00060$\\
                            \cline{2-6}
                            & \multirow{2}{*}{$c\log(1+dQ)$}  & c &  $0.7537 \pm 0.0064$    &   $0.8696 \pm 0.0047$   & $1.189 \pm 0.012$ \\
                            &                                & d &   $4.698 \pm 0.083$      &   $3.484 \pm 0.036$     & $1.65 \pm 0.026$   \\
                            \cline{2-6}
        \hline
    \end{tabular}
    \caption{Fit parameters corresponding to Fig. \ref{fig:PI-MO-LI-U_-NE} and Fig. \ref{fig:PI-LO-CI-NU-NE} (see Sec. \ref{ssec:priceimpactmarket}). The first column labels which figure the fits in its row correspond to. The second column shows the functional forms which are to be fit and just to their right, the free parameters in those functional forms. Finally, in the rows corresponding to those free parameters, are their fitted values and their uncertainties. Uncertainties are shown to two significant figures. The last 3 columns show which lines each fit belongs to and correspond to legend labels in Fig. \ref{fig:PI-MO-LI-U_-NE} and Fig. \ref{fig:PI-LO-CI-NU-NE}.}
    \label{tab:fitforpriceimpact}
\end{table*}

\tocless\subsection{Computational complexity and cost}\label{app:complexity}

\begin{figure}[H]
\centering
\includegraphics[width=0.45\textwidth]{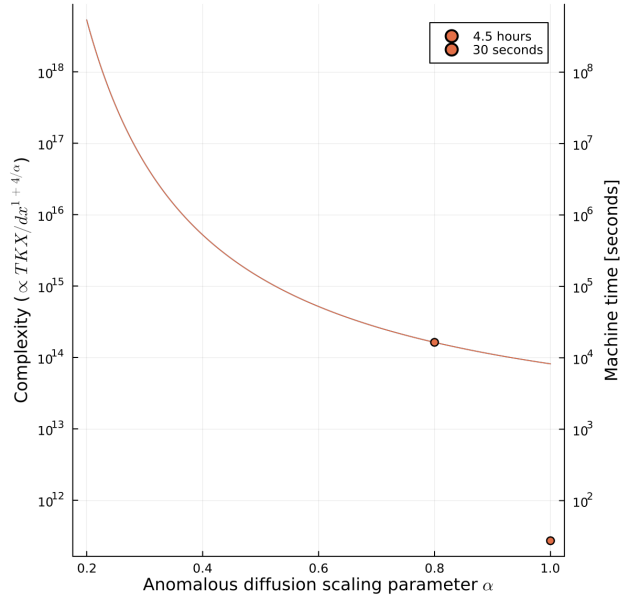}
\caption{The growth in computation time as a function of $\alpha$ (\ref{app:complexity}). On the left axis the growth has computational steps, and we map this to the right axis by assuming that a computer can do around $10^{10}$ steps per second. The horizontal axis has varying $\alpha$. We note that when $\alpha=1$, while one could in principle carry out K computations in this case, the kernel will be $0$ at all but one value.} 
\label{fig:computcomplex}
\end{figure}

We given an indication of the computational complexity of this scheme. Suppose one wishes to model prices over a range $X$, over a time $T$ (this time could refer to order book event time or calendar time) and wishes to have the kernel remember for a time $K$ (in the same units as $T$) into the past. The number of grid points one ends up using depends on the value of $\Delta x$ that one would like to have. This would mean one ends up with a grid of $M=X/\Delta x$ spacial points and $N=T/\Delta t$ time points. But we have that $\Delta t = (\Delta x)^{2/\alpha}$ (up to scaling) from equation [\ref{eq:latticesize}]. So the number of steps needed will be proportional to 
\begin{equation} \label{eq:computecomplexity}
\mbox{steps} \propto \frac{T}{\Delta t} \times \frac{K}{\Delta t} \times \frac{X}{\Delta x} = \frac{T \times K \times X}{(\Delta x)^{1+\sfrac{4}{\alpha}}}.
\end{equation}
Equation [\ref{eq:computecomplexity}] holds for $K \ll T$, which is the case considered in this paper. If one requires that the kernel remember its entire past, then the complexity grows as:
\begin{equation}
\mbox{steps} \propto XT^2(\Delta x)^{-(1+\sfrac{4}{\alpha})}.
\end{equation}
A plot of this can be seen in Figure \ref{fig:computcomplex}. The compute time as measured in machine time (seconds) is proportional to the computational complexity measured in computation steps or cycles. The actual machine time performance for two comparative computational tasks on the same hardware configuration were found for two different diffusion scaling parameter choices. First, for $\alpha = 1.00$ with a compute time of 30 seconds to generate 28,000 trade events. Second, for $\alpha=0.80$ with a compute time of 4 hours and 35 minutes for the same number of trade events. This makes heuristic sense because for $\alpha=0.80$ with the choice of $m_0$ the kernel remembers $600$ previous simulation steps as compared to none for $\alpha=1$ so that one expects a compute time of 5 hours = 600 $\times$ 30s.  Both comparative simulations find 28,000 trade events running on a Core i5 (11800H) with 24GB RAM. 

\section{Additional plots} \label{app:appendix:additional}

\tocless\subsection{Additional Stylised Fact}
The descriptions and generic features described in \ref{ssec:sfuniform} also hold true for the figures \ref{fig:StylizedFactsDifferentDiffusions}, \ref{fig:StylizedFactsDifferentCorrelation} and \ref{fig:StylizedFactsDifferentVariances} in which we vary the default parameters themselves. This is discussed in what follows.

\tocless\subsubsection{Varying the diffusion rate\texorpdfstring{: $\alpha$}{}}\label{ssec:sfchangealpha}

\begin{figure*}
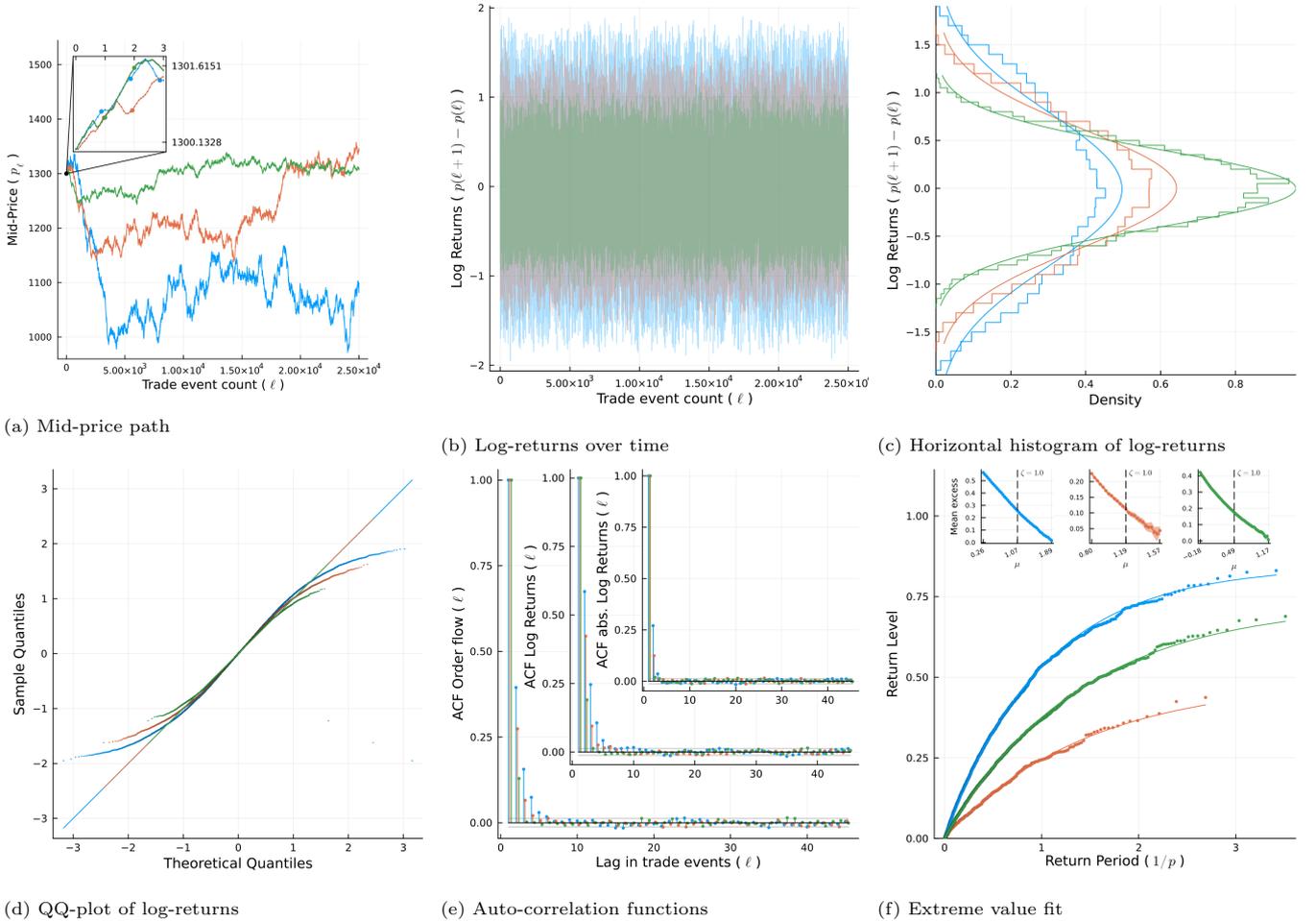

\centering
\begin{subfigure}[c]{0.32\textwidth}
\includegraphics[width=1.0\textwidth]{\PathSFDD SF_DD_1-mp.png} 
\caption{Mid-price path}\label{fig:StylizedFactsDifferentDiffusions:a}
\end{subfigure}
\begin{subfigure}[c]{0.32\textwidth}
\includegraphics[width=1.0\textwidth]{\PathSFDD SF_DD_2-lr.png}
\caption{Log-returns over time}\label{fig:StylizedFactsDifferentDiffusions:b}
\end{subfigure}
\begin{subfigure}[c]{0.32\textwidth}
\includegraphics[width=1.0\textwidth]{\PathSFDD SF_DD_3-hlr.png}
\caption{Horizontal histogram of log-returns}\label{fig:StylizedFactsDifferentDiffusions:c}
\end{subfigure}
\begin{subfigure}[c]{0.32\textwidth}
\includegraphics[width=1.0\textwidth]{\PathSFDD SF_DD_4-qqlr.png}
\caption{QQ-plot of log-returns}\label{fig:StylizedFactsDifferentDiffusions:d}
\end{subfigure}
\begin{subfigure}[c]{0.32\textwidth}
\includegraphics[width=1.0\textwidth]{\PathSFDD SF_DD_5-acfs.png}
\caption{Auto-correlation functions}\label{fig:StylizedFactsDifferentDiffusions:e}
\end{subfigure}
\begin{subfigure}[c]{0.32\textwidth}
\includegraphics[width=1.0\textwidth]{\PathSFDD SF_DD_6-expl.png}
\caption{Extreme value fit}\label{fig:StylizedFactsDifferentDiffusions:f}
\end{subfigure}
\caption{
The impact of different diffusion parameters $\alpha$ on the stylised facts are shown with free parameter tuple: $(\alpha,\rho=0.9,\sigma=1.0)$ (see \ref{ssec:stylisedfacts}). That is, each path has a different value of $\alpha$ (corresponding to the discussion in \ref{ssec:sfchangealpha}). As usual, the red path has the default value $\alpha=0.8$ appearing in all stylised facts plots for comparison (see figs. \ref{fig:StylizedFactsDifferentVariances}, \ref{fig:StylizedFactsDifferentCorrelation} and \ref{fig:StylizedFactsDifferentDx}) The blue path has $\alpha=1.0$ (no fractional diffusion), and the green has $\alpha=0.6$. In all paths, random kicks $V_t$ are drawn from a random walk defined as $V_t = \rho V_{t-1} + \epsilon_t$ with $\epsilon_t \sim \mathcal{N}(0,\sigma)$. Time is measured in terms of trade events $\ell$ (see level $L_2$ in Fig. \ref{fig:TimeScales}) but simulation events ($L_0$) can be seen in the inset of Fig. \ref{fig:StylizedFactsDifferentDiffusions:a}. Here each path has 25,000 trade events. For additional discussion, see \ref{ssec:sfchangealpha}.
See Tab. \ref{tab:parameters} for the full set of model and hyper-parameters.
}
\label{fig:StylizedFactsDifferentDiffusions}
\end{figure*}

In Figure \ref{fig:StylizedFactsDifferentDiffusions} we consider three values for $\alpha$. The red path shows the default $\alpha=0.8$ path and this path occurs in all stylised facts plots for comparison. The green shows $\alpha=0.6$, and blue shows $\alpha=1.0$ (ordinary diffusion). Each of these plot share the same generic features as described in \ref{ssec:sfuniform}. 

The most apparent affect of reducing the sub-diffusion is that it also reduces the variance of the price path. This can be seen in Figure \ref{fig:StylizedFactsDifferentDiffusions:c} and Figure
Figure \ref{fig:StylizedFactsDifferentDiffusions:a}. 
Figure \ref{fig:StylizedFactsDifferentDiffusions:e} shows that the larger the value of $\alpha$, the more auto-correlations are suppressed. In addition, the paths shown in Figure \ref{fig:StylizedFactsDifferentDiffusions:a}, and the histograms shown in Figure \ref{fig:StylizedFactsDifferentDiffusions:c}, are not merely scaled versions of one another (see stylised facts plot Figure \ref{fig:StylizedFactsDifferentVariances} to see how this could occur), but rather diffusion has change the dynamics such that all paths look to have been drawn from a different seed. Finally, it may be worth noting that all three of the plots in Figure \ref{fig:StylizedFactsDifferentDiffusions:c} intercept at near $\pm 0.5=\pm \Delta x$ in this formulation.  

\tocless\subsubsection{Varying the force variance\texorpdfstring{: $\sigma$}{}}\label{ssec:sfchangesigma}

\begin{figure*}
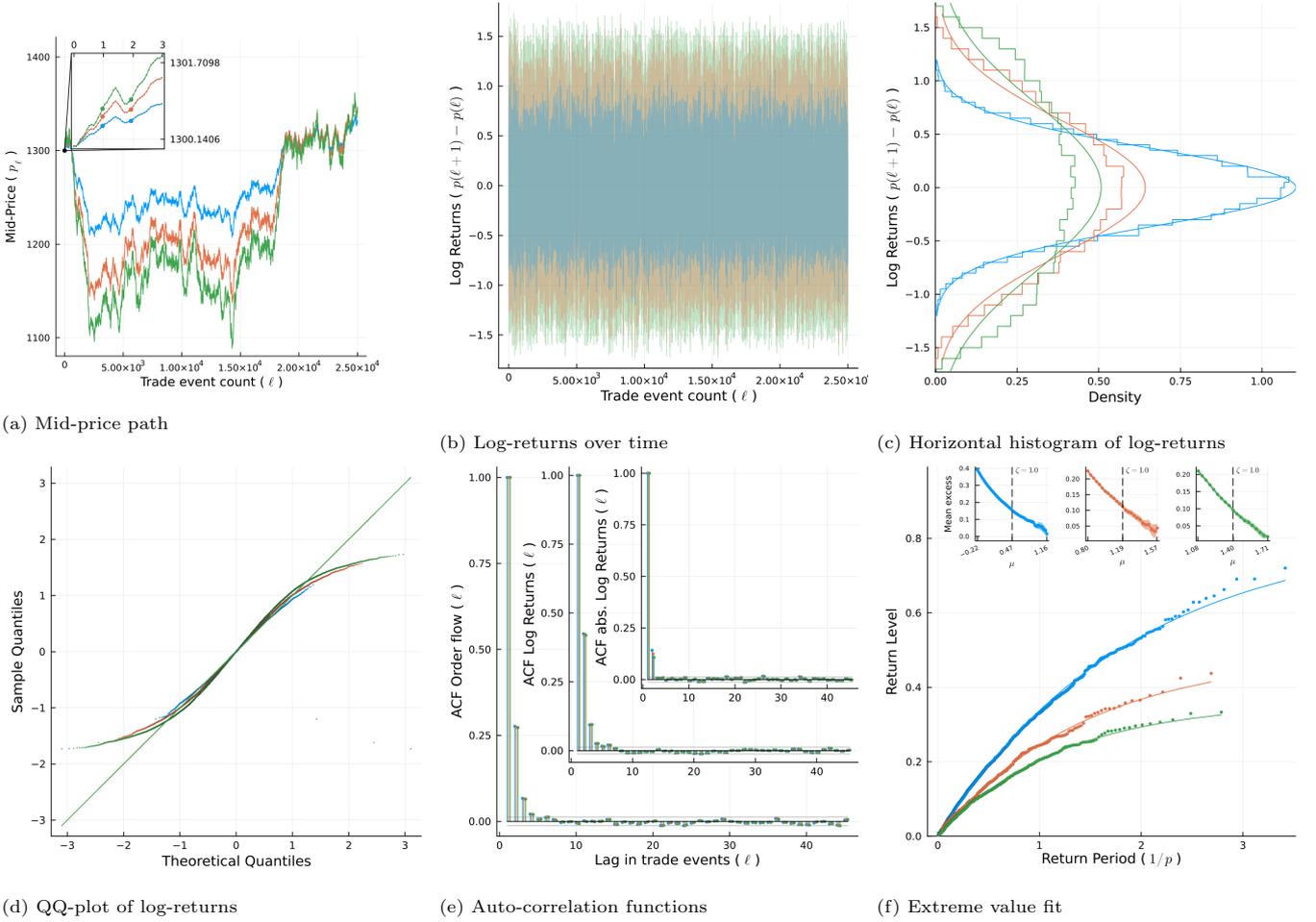

\centering
\begin{subfigure}[c]{0.32\textwidth}
\includegraphics[width=1.0\textwidth]{\PathSFDV SF_DV_1-mp.png} 
\caption{Mid-price path}\label{fig:StylizedFactsDifferentVariances:a}
\end{subfigure}
\begin{subfigure}[c]{0.32\textwidth}
\includegraphics[width=1.0\textwidth]{\PathSFDV SF_DV_2-lr.png}
\caption{Log-returns over time}\label{fig:StylizedFactsDifferentVariances:b}
\end{subfigure}
\begin{subfigure}[c]{0.32\textwidth}
\includegraphics[width=1.0\textwidth]{\PathSFDV SF_DV_3-hlr.png}
\caption{Horizontal histogram of log-returns}\label{fig:StylizedFactsDifferentVariances:c}
\end{subfigure}
\begin{subfigure}[c]{0.32\textwidth}
\includegraphics[width=1.0\textwidth]{\PathSFDV SF_DV_4-qqlr.png}
\caption{QQ-plot of log-returns}\label{fig:StylizedFactsDifferentVariances:d}
\end{subfigure}
\begin{subfigure}[c]{0.32\textwidth}
\includegraphics[width=1.0\textwidth]{\PathSFDV SF_DV_5-acfs.png}
\caption{Auto-correlation functions}\label{fig:StylizedFactsDifferentVariances:e}
\end{subfigure}
\begin{subfigure}[c]{0.32\textwidth}
\includegraphics[width=1.0\textwidth]{\PathSFDV SF_DV_6-expl.png}
\caption{Extreme value fit}\label{fig:StylizedFactsDifferentVariances:f}
\end{subfigure}
\caption{
The impact of different driving force variances $\sigma$ on the stylised facts are shown with free parameter tuple: $(\alpha=0.8,\rho=0.9,\sigma)$. (see \ref{ssec:stylisedfacts}). Each path has a different variance $\sigma$ (corresponding to the discussion in \ref{ssec:sfchangesigma}.) As usual, the red path has the default value of $\sigma=1.0$ appearing in all stylised facts plots for comparison (see figs. \ref{fig:StylizedFactsDifferentDiffusions},  \ref{fig:StylizedFactsDifferentCorrelation} and \ref{fig:StylizedFactsDifferentDx}). The blue path then has $\sigma=0.5$ and the green has $\sigma=1.5$. In all paths, random shocks $V_t$ are drawn from a random walk defined as $V_t = \rho V_{t-1} + \epsilon_t$ with $\epsilon_t \sim \mathcal{N}(0,\sigma)$. Time is measured in terms of trade events $\ell$ (see level $L_2$ in Fig. \ref{fig:TimeScales}) but simulation events ($L_0$) can be seen in the inset of Fig. \ref{fig:StylizedFactsDifferentVariances:a}. Here each path has 25,000 trade events. For additional discussion, see \ref{ssec:sfchangesigma}.
See Tab. \ref{tab:parameters} for the full set of model and hyper-parameters.
 }
\label{fig:StylizedFactsDifferentVariances}
\end{figure*}

In Figure \ref{fig:StylizedFactsDifferentVariances}, we consider three values for $\sigma$. The red path shows the default $\sigma=1.0$ path, and this path occurs in all stylised facts plots for comparison. The green shows $\sigma=1.5$, and blue shows $\sigma=0.5$. Each of these plot share the same generic features as described in \ref{ssec:sfuniform}. 

The most apparent effect of changing the variance of the force term, that represents random information arrival, is that one scales the variance of the overall path without changing its dynamics in a more complex way. This can be seen in Figure \ref{fig:StylizedFactsDifferentVariances:a} from which it is clear that all paths are drawn from the same seed. This is comforting, because in the non-anomalous diffusion case, $\alpha$=1.0, one may time transform to the frame of the order book \cite{DBMB2015QF} by noticing that:
\begin{equation}
    p(t) = \int_0^t V_{t'} dt'.
\end{equation}
Since, in this case, $V_{t'}$ is Brownian noise, one would expect that that variance of the process grows proportional to $\sigma$. 

Finally, it may be worth noting that all three of the plots in Figure \ref{fig:StylizedFactsDifferentVariances:c} intercept at near $\pm 0.5=\pm \Delta x$.  

\tocless\subsubsection{Varying the self-correlation\texorpdfstring{: $\rho$}{}}\label{ssec:sfchangerho}

\begin{figure*}
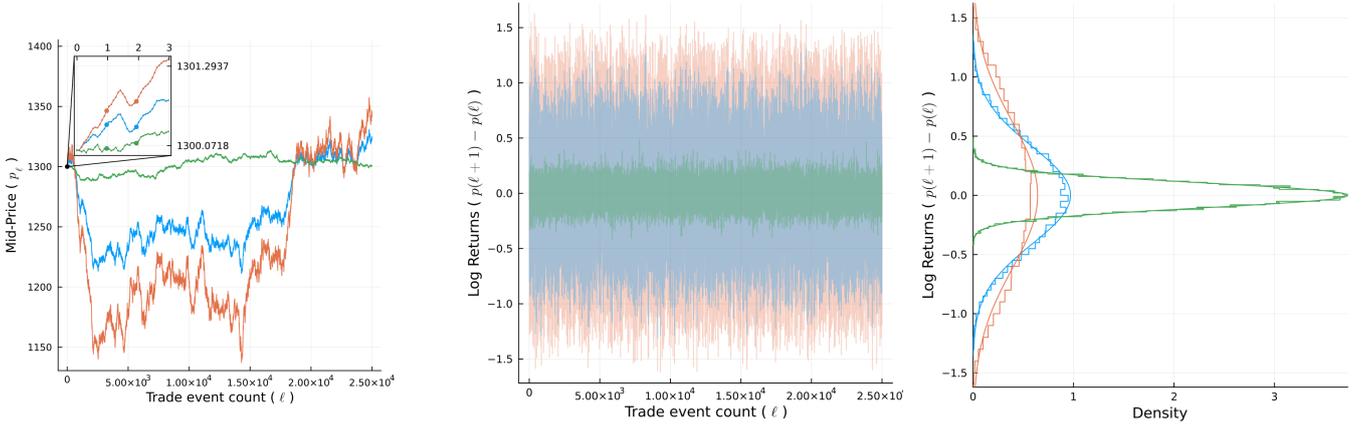
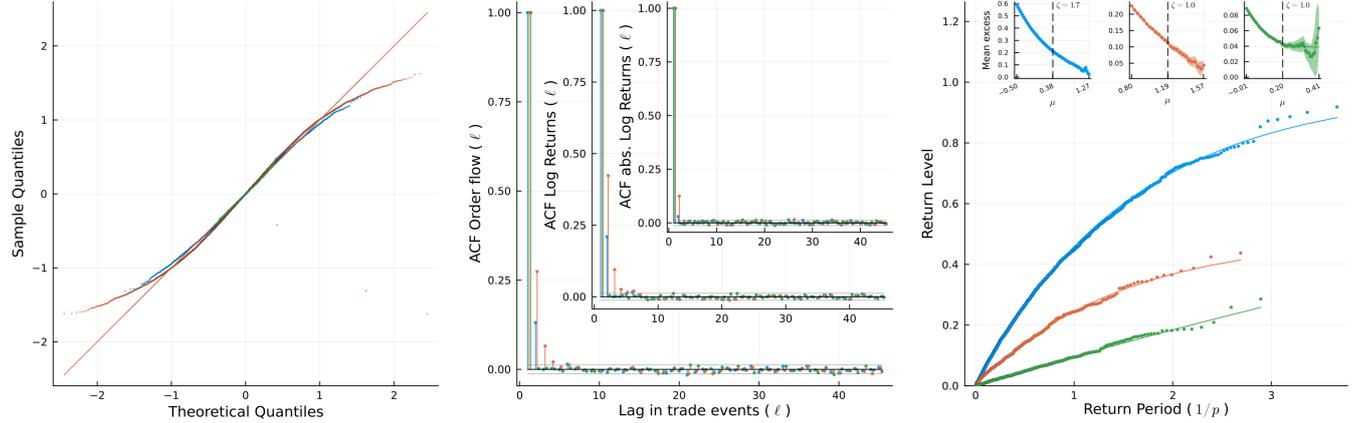

\centering
\begin{subfigure}[c]{0.32\textwidth}
\includegraphics[width=1.0\textwidth]{\PathSFDC SF_DC_1-mp.png}
\caption{Mid-price path}\label{fig:StylizedFactsDifferentCorrelation:a}
\end{subfigure}
\begin{subfigure}[c]{0.32\textwidth}
\includegraphics[width=1.0\textwidth]{\PathSFDC SF_DC_2-lr.png}
\caption{Log-returns over time}\label{fig:StylizedFactsDifferentCorrelation:b}
\end{subfigure}
\begin{subfigure}[c]{0.32\textwidth}
\includegraphics[width=1.0\textwidth]{\PathSFDC SF_DC_3-hlr.png}
\caption{Horizontal histogram of log-returns}\label{fig:StylizedFactsDifferentCorrelation:c}
\end{subfigure}
\begin{subfigure}[c]{0.32\textwidth}
\includegraphics[width=1.0\textwidth]{\PathSFDC SF_DC_4-qqlr.png}
\caption{QQ-plot of log-returns}\label{fig:StylizedFactsDifferentCorrelation:d}
\end{subfigure}
\begin{subfigure}[c]{0.32\textwidth}
\includegraphics[width=1.0\textwidth]{\PathSFDC SF_DC_5-acfs.png}
\caption{Auto-correlation functions}\label{fig:StylizedFactsDifferentCorrelation:e}
\end{subfigure}
\begin{subfigure}[c]{0.32\textwidth}
\includegraphics[width=1.0\textwidth]{\PathSFDC SF_DC_6-expl.png}
\caption{Extreme value fit}\label{fig:StylizedFactsDifferentCorrelation:f}
\end{subfigure}

\caption{
The impact of different values of the self-correlation $\rho$ (of the driving force $V_t$) on the stylised facts are shown with free parameter tuple: $(\alpha=0.8,\rho,\sigma=1.0)$ (see Sec. \ref{ssec:stylisedfacts}). That is, the force $V_t$, which is related to itself in time as $V_t = \rho V_{t-1} + \epsilon_t$ with $\epsilon_t \sim \mathcal{N}(0, \sigma)$, has a different value for $\rho$ in each path (corresponding to the discussion in \ref{ssec:sfchangerho}.) As usual, the red path has the default $\rho=0.9$ appearing in all stylised facts for comparison (see figs. \ref{fig:StylizedFactsDifferentDiffusions}, \ref{fig:StylizedFactsDifferentVariances} and \ref{fig:StylizedFactsDifferentDx}.) The blue path has $\rho = 0.8$ and the green has $\rho=0$ ($V_t$ is therefore not a random walk but is now random noise). Time is measured in terms of trade events $\ell$ (see level $L_2$ in Fig. \ref{fig:TimeScales}) but simulation events ($L_0$) can be seen in the inset of Fig. \ref{fig:StylizedFactsDifferentCorrelation:a}. Here each path has 25,000 trade events. For additional discussion, see \ref{ssec:sfchangerho}.
See Tab. \ref{tab:parameters} for the full set of model and hyper-parameters.
}
\label{fig:StylizedFactsDifferentCorrelation}
\end{figure*}

In Figure \ref{fig:StylizedFactsDifferentCorrelation}, we consider three values for self correlations $\rho$ between the random information shocks represented by $V_t$. The red path shows the default $\rho=0.9$ path and this is the common path that occurs in all stylised facts plots for comparison. The green shows $\rho=0$ where $V_t$ is uncorrelated random noise, and blue shows $\rho=0.8$. These plots mostly share the same generic features as described in \ref{ssec:sfuniform} except for the auto-correlations of the green path.

The most apparent interpretation of these plots is that it is the self-correlation parameter $\rho$ alone that creates auto-correlation in the ACF plots (at least for $\alpha=0.8$). This can be seen in Figure \ref{fig:StylizedFactsDifferentCorrelation:e} where the green path has no significant auto-correlation anywhere despite having $\alpha = 0.8$. 
This seems surprising, but there is auto-correlation in the green path at the level of $L_0$ (not shown), but this dies too quickly to reach the next time step in $L_1$. This captures the importance of sampling. 
Finally, it may be worth noting that in Figure \ref{fig:StylizedFactsDifferentCorrelation:c}, the green path seems to have a hard upper and lower bound at $\pm 0.5=\pm \Delta x$, while the blue and red plots intercept at these points. 


\begin{figure*}
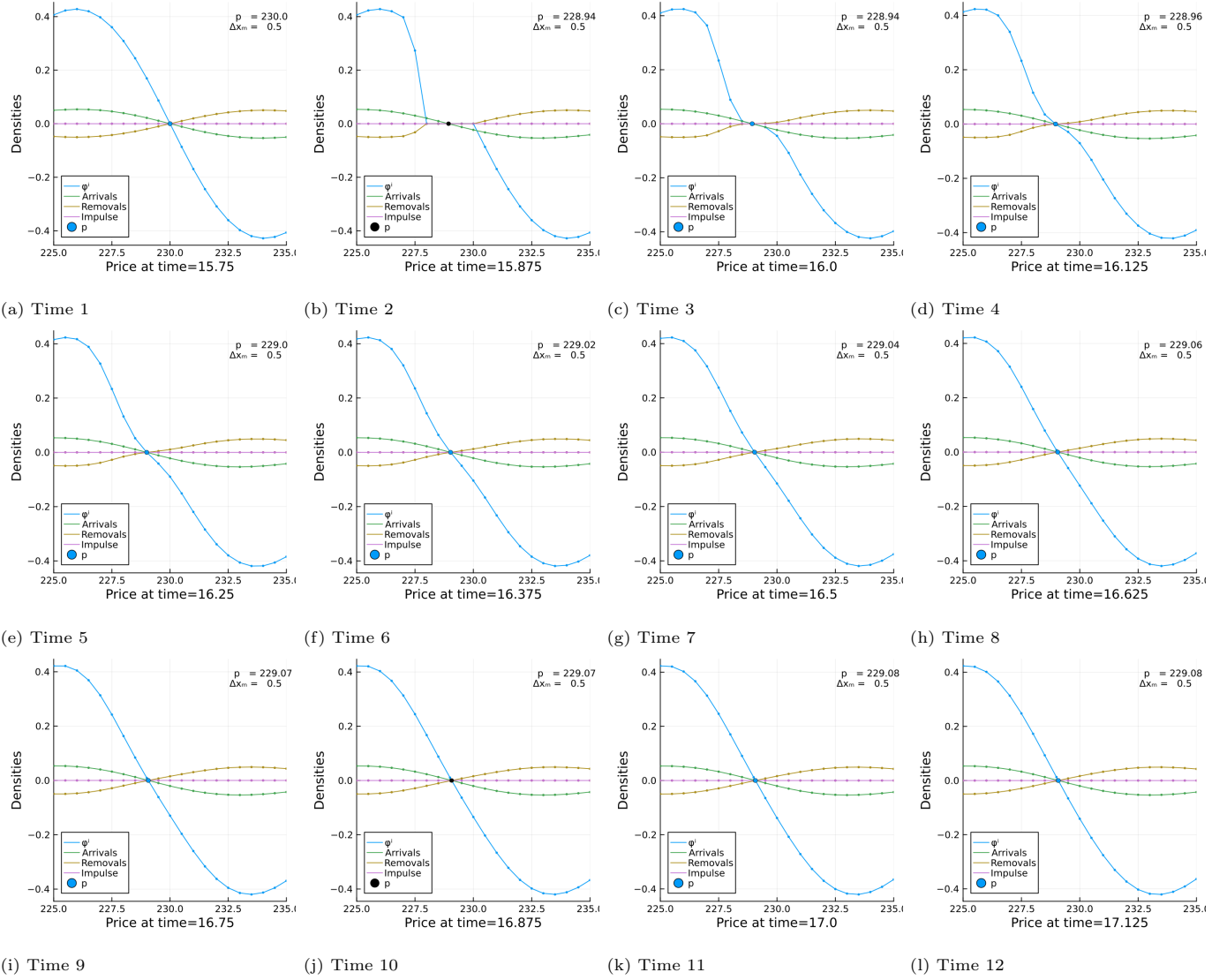

\centering
\begin{subfigure}[c]{0.24\textwidth}
\includegraphics[width=1.0\textwidth]{\PathKMO K_MO-1.png}
\caption{Time 1}\label{fig:VisualizePriceImpactMarketOrder:a}
\end{subfigure}
\begin{subfigure}[c]{0.24\textwidth}
\includegraphics[width=1.0\textwidth]{\PathKMO K_MO-2.png}
\caption{Time 2}\label{fig:VisualizePriceImpactMarketOrder:b}
\end{subfigure}
\begin{subfigure}[c]{0.24\textwidth}
\includegraphics[width=1.0\textwidth]{\PathKMO K_MO-3.png}
\caption{Time 3}\label{fig:VisualizePriceImpactMarketOrder:c}
\end{subfigure}
\begin{subfigure}[c]{0.24\textwidth}
\includegraphics[width=1.0\textwidth]{\PathKMO K_MO-4.png}
\caption{Time 4}\label{fig:VisualizePriceImpactMarketOrder:d}
\end{subfigure}
\begin{subfigure}[c]{0.24\textwidth}
\includegraphics[width=1.0\textwidth]{\PathKMO K_MO-5.png}
\caption{Time 5}\label{fig:VisualizePriceImpactMarketOrder:e}
\end{subfigure}
\begin{subfigure}[c]{0.24\textwidth}
\includegraphics[width=1.0\textwidth]{\PathKMO K_MO-6.png}
\caption{Time 6}\label{fig:VisualizePriceImpactMarketOrder:f}
\end{subfigure}
\begin{subfigure}[c]{0.24\textwidth}
\includegraphics[width=1.0\textwidth]{\PathKMO K_MO-7.png}
\caption{Time 7}\label{fig:VisualizePriceImpactMarketOrder:g}
\end{subfigure}
\begin{subfigure}[c]{0.24\textwidth}
\includegraphics[width=1.0\textwidth]{\PathKMO K_MO-8.png}
\caption{Time 8}\label{fig:VisualizePriceImpactMarketOrder:h}
\end{subfigure}
\begin{subfigure}[c]{0.24\textwidth}
\includegraphics[width=1.0\textwidth]{\PathKMO K_MO-9.png}
\caption{Time 9}\label{fig:VisualizePriceImpactMarketOrder:i}
\end{subfigure}
\begin{subfigure}[c]{0.24\textwidth}
\includegraphics[width=1.0\textwidth]{\PathKMO K_MO-10.png}
\caption{Time 10}\label{fig:VisualizePriceImpactMarketOrder:j}
\end{subfigure}
\begin{subfigure}[c]{0.24\textwidth}
\includegraphics[width=1.0\textwidth]{\PathKMO K_MO-11.png}
\caption{Time 11}\label{fig:VisualizePriceImpactMarketOrder:k}
\end{subfigure}
\begin{subfigure}[c]{0.24\textwidth}
\includegraphics[width=1.0\textwidth]{\PathKMO K_MO-12.png}
\caption{Time 12}\label{fig:VisualizePriceImpactMarketOrder:l}
\end{subfigure}
\caption{Shows the result of a large market order arriving and corresponds to Sec. \ref{ssec:priceimpactmarket}. In Fig. \ref{fig:VisualizePriceImpactMarketOrder:b}, the market order to sell arrives and is executed against the set of current offers to buy. Each sub-figure in the above is a snapshot of the system in simulation event time. In each sub-figure, the plot in blue is the density we are modelling at the current time, the plot in green is the orders which are about to be added via the source term and the gold plot is the orders which are about to be removed via the removal rate.}
\label{fig:VisualizePriceImpactMarketOrder}
\end{figure*}

\end{document}